\documentclass[preprint]{iucr}              

     \journalcode{J}              
\usepackage{graphicx}
\usepackage{chemformula}
\usepackage{amssymb}
\usepackage{xspace}
\usepackage{siunitx}
\usepackage{url} 
\usepackage{makecell}\setcellgapes{1.5pt}

\definecolor{tagcolor}{HTML}{21a889}

\definecolor{dgreen}{HTML}{008000}

\newcommand{\be}{\begin{enumerate}[wide, labelwidth=!, labelindent=0pt,
        label=\textbf{\textcolor{blue}{\arabic*}.}]}
    \newcommand{\bei}{\begin{enumerate}}
        \newcommand{\ee}{\end{enumerate}}
    \newcounter{saveenumi}

\newcommand{\pdfgetxthree}{\textsc{PDFgetX3}\xspace}
\newcommand{\xpdfsuite}{\textsc{xPDFsuite}\xspace}
\newcommand{\diffpy}{\textsc{diffpy-cmi}\xspace}
\newcommand{\pdfitc}{\textsc{PDFitc}\xspace}
\newcommand{\strumin}{\textsc{structureMining}\xspace}

\newcommand{\simmap}{\textsc{similarityMapping}\xspace}
\newcommand{\nmfmap}{\textsc{nmfMapping}\xspace}
\newcommand{\dawn}{\textsc{DAWN}\xspace}
\newcommand{\topas}{\textsc{TOPAS Academic V6}\xspace}
\newcommand{\pyfai}{\textsc{pyFAI}\xspace}  
\newcommand{\sibe}{\begin{enumerate}[wide, labelwidth=!, labelindent=0pt,
        label=\textbf{\textcolor{blue}{A{\arabic*}.}}]}
    \newcommand{\sibei}{\begin{enumerate}}
        \newcommand{\siee}{\end{enumerate}}

\newcommand{\fig}[1]{Fig.~\ref{fig:#1}}
\newcommand{\figs}[1]{Figs.~\ref{fig:#1}}
\newcommand{\sect}[1]{Section~\ref{sec:#1}}

\newcommand{\tabl}[1]{Table~\ref{table:#1}}

\newcommand{\latonea}{$a\;[\mathrm{\AA}]$}

\newcommand{\latoneb}{$b\;[\mathrm{\AA}]$}

\newcommand{\latonec}{$c\;[\mathrm{\AA}]$}

\newcommand{\latonebeta}{$\beta\;[^{\circ}]$}

\newcommand{\uisoone}[1]{$u_{\mathrm{iso,#1}}\;\Bigl[\mathrm{\AA}^{2}\Bigr]$}

\newcommand{\deltatwoone}{$\delta_{2}\;\Bigl[\mathrm{\AA}^{-2}\Bigr]$}

\newcommand{\cdsone}{$\mathrm{cds}\;\bigl[\mathrm{\AA}\bigr]$}

\newcommand{\atomposone}[4]{$#1_{\mathrm{#2,#3}}\;\bigl[#4\bigr]$}

\newcommand{\rw}{$R_{\mathrm{w}}$}
\newcommand{\rwvalue}[1]{$R_{\mathrm{w}}=#1$}
\newcommand{\rwp}{$R_{\mathrm{wp}}$}

\newcommand{\rbragg}{$R_{\mathrm{Bragg}}$}

\newcommand{\gobs}{$G_{\mathrm{obs}}$}
\newcommand{\gcalc}{$G_{\mathrm{calc}}$}
\newcommand{\gdiff}{$G_{\mathrm{diff}}$}

\newcommand{\ewe}{$E_{\mathrm{we}}$}
\begin{document}                  


\title{
\textit{Operando} pair distribution function analysis of nanocrystalline functional materials: the case of \ch{TiO2}-bronze nanocrystals in Li-ion battery electrodes
}
     
\author[a]{Martin A.}{Karlsen}
\author[b]{Jonas}{Billet}
\author[c]{Songsheng}{Tao}
\author[b]{Isabel}{Van Driessche}
\author[c]{Simon J. L.}{Billinge}
\cauthor[a]{Dorthe B.}{Ravnsbæk}{dorthe@chem.au.dk}

\aff[a]{Department of Chemistry, Aarhus Univeristy, DK-8000 Aarhus C, \country{Denmark}}
\aff[b]{Department of Chemistry, Ghent University, 9000 Gent, \country{Belgium}}
\aff[c]{Department of Applied Physics and Applied Mathematics with Materials Science and Engineering, Columbia University, New York City, NY, 10027, USA}
%





\keyword{operando}\keyword{pair distribution function analysis}\keyword{nanocrystalline functional materials}\keyword{\ch{TiO2}-bronze}\keyword{Li-ion battery}


%
\maketitle                        
\includegraphics[width=\columnwidth]{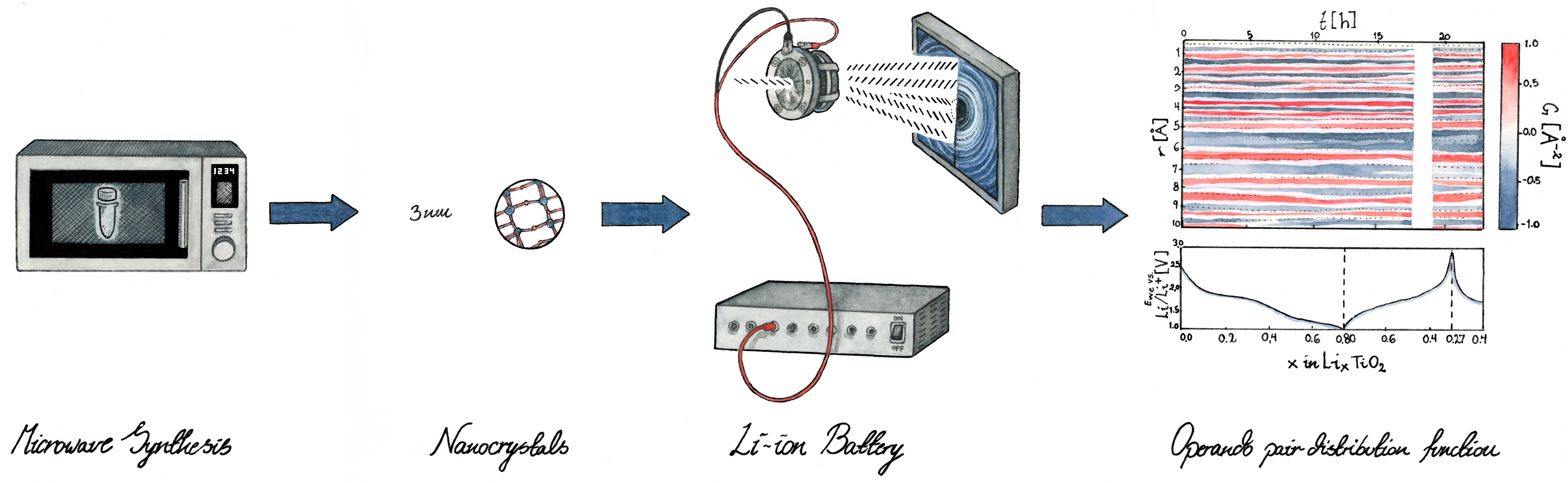}
\begin{synopsis}
Demonstration of tools for \textit{operando} pair distribution function analysis of nanocrystalline functional materials. The particular case being \SI{3}{nm} \ch{TiO2}-bronze nanocrystals as active electrode material in a Li-ion battery.
\end{synopsis}
\newpage
\begin{abstract}
Structural modelling of \textit{operando} pair distribution function (PDF) data of functional materials can be highly complex.
To aid the understanding of complex \textit{operando} PDF data, we here demonstrate a toolbox for PDF analysis.
The tools include the \strumin, \simmap, \nmfmap apps available through the online service 'PDF in the cloud' (\pdfitc, \url{www.pdfitc.org}), as well as noise-filtering using principal component analysis (PCA).
The tools are applied to both \textit{ex situ} and \textit{operando} PDF data for \SI{3}{nm} \ch{TiO2}-bronze nanocrystals, which function as the active electrode material in a Li-ion battery.
The tools enable structural modelling of the \textit{ex situ} and \textit{operando} PDF data, revealing two pristine \ch{TiO2} phases (bronze and anatase) and two lithiated \ch{Li_{x}TiO2} phases (lithiated versions of bronze and anatase), and the phase evolution during Galvanostatic cycling is characterized.
\end{abstract}
%

%
\section{Introduction}
\label{sec:intro}
For many electrode materials for rechargable batteries, crystallinity, i.e., long-range structural order, has been thought of as a prerequisite~\cite{Whittingham2004, Goodenough2010, Christensen2021}.
However, in recent years, it has been realized that crystalline defects, nanosizing, amorphization, etc., may be beneficial for electrochemical performance~\cite{Uchaker2014, Sheng2014, Chae2014, Hua2017, Luo2017, Wang2018, Christensen2019, Christensen2021}.
Insights into structural transformations of battery electrodes are obtainable through \textit{operando} experiments~\cite{Chianelli1978, Chianelli1979, Latroche1992}.
For crystalline phases, \textit{operando} powder x-ray diffraction (PXRD) and Rietveld analysis~\cite{Rietveld1969} are also applicable for electrode materials for batteries~\cite{Tarascon1999, Bak2018}.
If the length of structural coherence of a phase shortens, PXRD and Rietveld analysis are no longer ideal tools for extracting information on the atomic structure.
Instead, information on the atomic structure may be extracted through x-ray total scattering (XTS) and atomic pair distribution function (PDF) analysis~\cite{Billinge2004, Billinge2007, Billinge2009, Billinge2012}.
Using PDF analysis, phase transitions involving non-crystalline phases under dynamic conditions may be explored through \textit{operando} XTS combined with PDF analysis~\cite{Hua2017, Christensen2018, Christensen2019, Christensen2019b}.
As battery electrodes are multicomponent systems containing both active material, conductive carbon, and polymeric binder, PDF data for battery electrodes are usually highly complex and therefore hard to model. To assist structural modelling, we demonstrate multiple types of model-free analyses to gain insights into \textit{operando} PDF data for nanocrystalline battery electrodes.

\begin{figure}
	\center
	\includegraphics[width=\columnwidth]{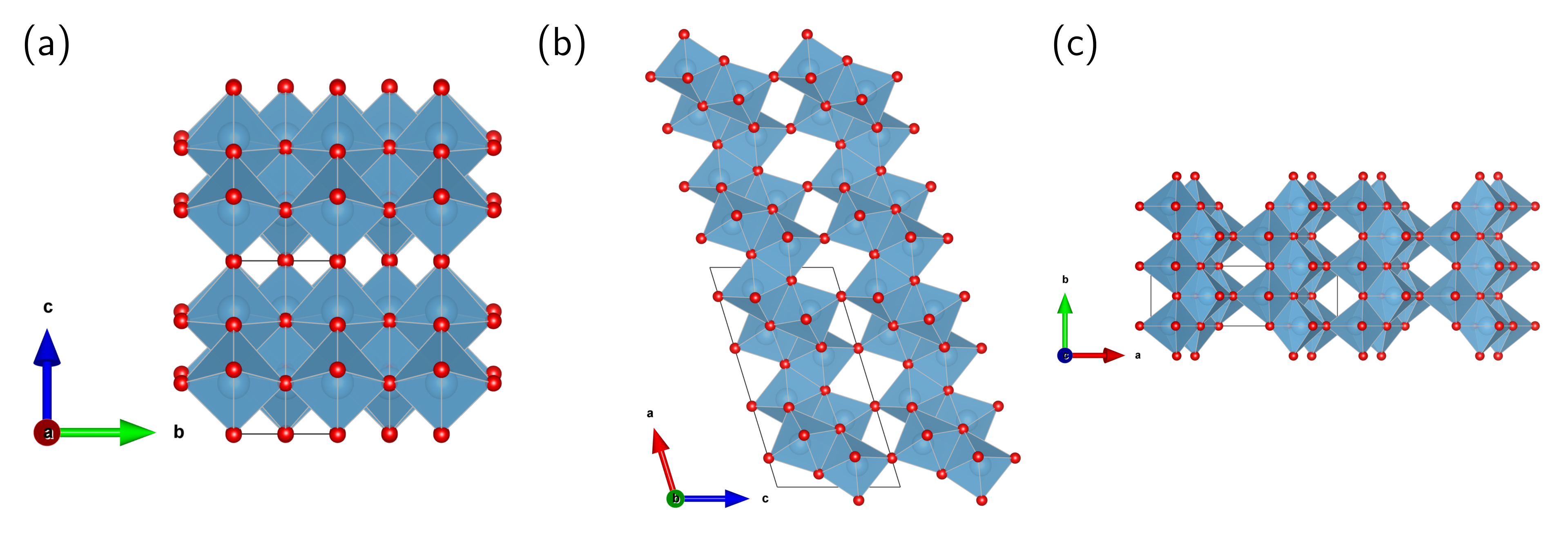}
	\caption{
Visualization of the \ch{TiO2}-bronze structure. $2\times2\times2$ monoclinic unit cells have been displayed along the crystallographic (a) $a$, (b) $b$, and (c) $c$ axes. Ti atoms are shown in light blue and O atoms are shown in red.
	}
	\label{fig:tio2_bronze_stru}
\end{figure}

This study is concerned with \ch{TiO2}-based electrode materials for rechargeable Li-ion batteries.
The family of titanium dioxide, \ch{TiO2}, polymorphs is large and diverse. The family members share the common building block of \ch{TiO6}-octahedra, which are connected in different ways giving rise to the various polymorphs~\cite{Liu2013, Aravindan2015}.
Examples on \ch{TiO2} polymorphs include
anatase~\cite{Cromer1955},
rutile~\cite{Cromer1955},
brookite~\cite{Pauling1928},
bronze~\cite{Marchand1980},
columbite~\cite{Simons1967},
hollandite~\cite{Latroche1989},
ramsdellite~\cite{Akimoto1994},
baddeleyite~\cite{Sato1991},
\ch{TiO2}-O-I~\cite{Dubrovinskaia2001},
and \ch{TiO2}-O-II~\cite{Dubrovinsky2001},
where the former four polymorphs are common at ambient temperatures and pressures.
The diversity of the \ch{TiO2} poymorphs results in versatile use as functional materials, such as wide band-gap semiconductors ($\sim$\SI{3.2}{eV})~\cite{Elmouwahidi2018} with spectral activity in the ultraviolet (UV) domain~\cite{Concalves2008}.
\ch{TiO2} materials also find use as photovoltaics, e.g., dye sensitized photovoltaic modules~\cite{Kay1996} and solar cells~\cite{Concalves2008} photocatalysts~\cite{Fujishima2008, Fresno2014}, supercapacitors~
\cite{Elmouwahidi2018}, and electrochemical storage. 
Regarding electrochemical storage, \ch{TiO2} materials have been widely explored as intercalation-type electrode materials for Li-ion batteries. The theoretical gravimetric capacity of \ch{TiO2} materials in Li-ion batteries reaches \SI{335}{mA\cdot h\cdot g^{-1}} for intercalation of one equivalent of \ch{Li+}, which make \ch{TiO2}  materials promising alternatives to the commercial anode material \ch{Li4Ti5O12} with a gravimetric capacity of \SI{175}{mA\cdot h\cdot g^{-1}} and carbon-based anodes, where graphite offers of a gravimetric capacity of \SI{372}{mA\cdot h\cdot g^{-1}}.
In addition to a high theoretical gravimetric capacity, \ch{TiO2} materials are also attractive as electrode materials due to low production cost and low environmental impact~\cite{Deng2009, Yang2009, Froschl2012, Christensen2019}.
Among the \ch{TiO2} polymorphs, the bronze polymorph has received additional attention due its high operation power and capacity performances~\cite{Gao2019}.
Compared to commercial graphite anodes, \ch{TiO2}-bronze also offers higher operation safety through its higher discharge voltage plateau ($>$\SI{1.7}{V}~vs. Li/\ch{Li+})~\cite{Liang2022}.
Its monoclinic unit cell, depicted along the three crystallographic axes in \fig{tio2_bronze_stru}, belongs to the $C2/m$ space group. The network of edge- and corner-sharing \ch{TiO6}-octahedra has channels along the $b$ axis, i.e., the [010] direction, suitable for ion intercalation~\cite{Arrouvel2009, Pham2021}.
\subsection{Recent tools for PDF analysis: PDF in the cloud}
\label{sec:intro_pdfitc}
To assist scientists in their PDF analyses, the service 'PDF in the cloud' (\pdfitc) is available at \url{www.pdfitc.org}~\cite{Yang2021}, which offers a number of different apps for PDF analysis.
Herein, we demonstrate the use of the \strumin, \simmap, and \nmfmap apps available at \pdfitc. 
Furthermore, we also make use of principal component analysis (PCA), for noise-filtering of the \textit{operando} data, where the severe dampening from both the instrument and the nanosized sample results in a low signal-to-noise ratio already at relatively low $r$-values in the \textit{operando} PDF data.
The novel tools greatly assist our PDF refinement of \textit{ex situ} data for pristine and chemically lithiated \ch{TiO2}-bronze nanocrystals. 
However, the value of the tools is even greater, when it comes to \textit{operando} investigation of the structural evolution during Li-ion intercalation, when using the \ch{TiO2}-bronze nanocrystals as electrodes in rechargeable Li-ion batteries.
The tools will be described briefly below.
\subsection{\strumin}
\label{sec:intro_strumin}
The \strumin app~\cite{Yang2020} offers phase identification for PDFs. Phase identification or 'fingerprinting' is common for PXRD data in many laboratories, since the establishment of the Hanawalt File~\cite{Hanawalt1938} in the first part of the twentieth century, which has been followed by a number of descendants, including the Powder Diffraction File~\cite{Gates-Rector2019}.
To run a search query, the user uploads an experimental (or simulated) PDF together with relevant metadata like compositional and experimental details.
Completing the search query, the \strumin app returns a list a of crystallographic information files (.cif)~\cite{Hall1991}. The rank of each .cif file is based on the weighted residual value, $R_{\mathrm{w}}$, when refining the user-uploaded PDF data using a structural model based on the .cif file. This allows the user to base a quantitative analysis, i.e., PDF refinement, on one or more of the .cif files returned by the app.
\subsection{\simmap}
\label{sec:intro_simmap}
The \simmap app is meant for probing similarity for a series of PDFs uploaded by the user.
During a query, the app runs a Pearson correlation analysis~\cite{Pearson1895} for the set of PDFs uploaded by the user.
The Pearson correlation coefficient of two PDFs becomes a measure of similarity between the PDFs.
Then, the similarity of two PDFs can be interpreted as the similarity between the phase contents of the materials from which the PDFs originate.
The analysis is purely statical in nature and therefore model-free, quick, and straightforward.
For \textit{operando} data, the \simmap can be used to identify the onset of phase transitions but also whether phase evolution appears to possess solid-solution or two-phase characteristics.
\subsection{\nmfmap}
\label{sec:intro_nmfmap}
The \nmfmap app~\cite{Liu2021, Thatcher2022} offers non-negative matrix factorization (NMF) analysis for a series of PDFs. The app decomposes a series of PDFs and describes the trends in the data with as few components as possible.
Being an unsupervised machine-learning (ML) technique, NMF analysis share some characteristics with PCA. The differences between NMF and PCA include the nature of the constraints of the matrix decompomsition.
The non-negative constraint of the matrix decomposition for the NMF analysis means that the (normalized) NMF weights can be interpreted as fractions of the total scattering signal, which is directly related to the phase fractions of the material.
Therefore, the number of components from NMF analysis can provide a hint on the number of phases present during an \textit{operando} experiment and the behavior of the NMF weights offer insights into evolution of the phases that the NMF components present.
Being a purely statistical analysis, the output NMF components do not necessarily represent actual PDFs of the physical phases. However, if the behavior of the corresponding NMF weights can be interpreted in a physically meaningful way, this can be of truly high value, when doing quantitative analysis through refinement of highly complex \textit{operando} PDF data.
\subsection{Principal component analysis}
\label{sec:intro_pca}
Inspired by the \nmfmap app, we also make use of PCA (not a part of \pdfitc) to filter noise from experimental data, to enable qualitative inspection of \textit{operando} PDF data, which suffer from low signal-to-noise ratio at higher $r$-values. 
For case of nanocrystals presented herein, the PDF signal is heavily damped by the instrument and the sample such that the signal-to-noise ratio becomes a challenge already at relatively low $r$-values.
PCA is used for denoising instead of NMF, because the PDFs need to be shifted on the G-scale both before and after NMF analysis. In contrase, PCA is directly applicable to the PDF data.
\section{Methods}
\label{sec:methods}
\subsection{Nanocrystal synthesis}
\label{sec:method_nanocrystal_synthesis}
Two batches of \ch{TiO2}-nanocrystals were studied. 
The first batch was used for \textit{ex situ} characterization, including chemical lithiation (see \sect{method_chemical_lithiation} below). 
The second batch was used for \textit{ex situ} and \textit{operando} characterization.
The \ch{TiO2}-bronze nanocrystals were synthesized using a microwave setup as previously described by Billet \textit{et al.}~\cite{Billet2018} The molar concentration of Ti was \SI{0.244}{M}. The molar concentration of glycolic acid was \SI{0.25}{M} and that of sulphuric acid was \SI{0.72}{M}.  The reaction mixture was treated at \SI{130}{\celsius} for \SI{5}{min}. Finally, the nanocrystals were washed three times with water.
\subsection{Chemical lithiation}
\label{sec:method_chemical_lithiation}
For the first batch of \ch{TiO2} nanocrystals, the material was dried over night under vacuum at \SI{60}{\celsius}. \SI{50}{mg} of nanocrystals were suspended in anhydrous heptane. To ensure complete lithiation, three equivalents (\SI{7}{mL}) of $N$-butyllithium (\SI{2.7}{M} in heptane, Sigma-Aldrich) were added dropwise to the suspension under magnetic stirring in an Ar-filled atmosphere. The mixture was left to react for two days. The remaining liquid was removed and the powder was washed in heptane three times and dried. To obtain a fine powder, the chemically lithiated materials was mortared using an agate mortar and pestle.
\subsection{Electrode fabrication}
\label{sec:method_electrode_fabrication}
For a \SI{200}{mg} cathode pellet mixture, \SI{60}{wt\%} active material (\SI{120}{mg} \ch{TiO2}-bronze nanocrystals, batch two), \SI{30}{wt\%} conductive carbon (\SI{30}{mg} SuperP C45 (Imerys) and \SI{30}{mg} Acetylene Black (VXC72, Cabot Corp. )), and \SI{10}{wt\%} polymeric binder (\SI{20}{mg} polyvinylidene fluoride (PVDF, Kynar, Arkena)) were used. The active material and the conductive carbon were weighed separately, whereas the polymeric binder was obtained from a \SI{4}{wt\%} $N$-methyl-2-pyrrolidone (NMP, 99.5 \%, anhydrous, Sigma-Aldrich) solution. The active material, conductive carbon, and PVDF/NMP solution were mixed in a plastic vial with a teflon ball using a vortex mixer to obtain a slurry. The slurry was poured onto a sheet of aluminum foil. The slurry was spread on the aluminum foil using the 'doctor blade' method. The coated aluminum foil was left to dry in the fumehood over night at \SI{60}{\celsius}. The drying ended with one hour at \SI{90}{\celsius} to ensure complete evaporation of the NMP.
The dry cathode composite was scraped off the aluminum foil using a plastic spatula and mortared using an agate mortar and pestle to obtain a fine powder. 8-12~mg of the composite were uniaxially pressed into \SI{7}{mm}~$\varnothing$ pellets at \SI{1.8}{ton} for \SI{1}{min}.
\subsection{Electrochemical cell assembly}
\label{sec:method_echem_cell_assembly}
For the \textit{operando} x-ray total scattering studies, the AMPIX electrochemical cell~\cite{Borkiewicz2012} was used. The half-cell was assembled in an Ar-filled glovebox with a \SI{11.259}{mg} cathode pellet, i.e., \SI{6.755}{mg} of \ch{TiO2}-bronze nanocrystals, bottommost. A \SI{12}{mm}~$\varnothing$ Whatman GF/B separator was put on top of the cathode pellet. The separator was wetted with 7 drops of \SI{1}{M} \ch{LiPF6} in ethylene carbonate : dimethyl carbonate, 1:1 v/v (99.9\%, Solvionic) using a \SI{1}{mL} Pasteur pipette. Topmost, a metallic Li anode was placed. The Li anode was obtained by rolling lithium foil using a stainless steel rod. From the thinly rolled lithium foil, a \SI{10}{mm}~$\varnothing$ disk was punched out.
\subsection{Galvanostatic cycling}
\label{sec:method_gc}
During the \textit{operando} x-ray total scattering experiment, the electrochemical cell was Galvanostatically cycled using a current density of \SI{0.151}{mA}, corresponding to a C-rate of C/15.
\subsection{Measurements}
\label{sec:method_measurements}
Pristine and chemically lithiated powders were characterized through \textit{ex situ} powder x-ray diffraction and x-ray total scattering. The electrode with \SI{3}{nm} \ch{TiO2}-bronze nanocrystrals as the active material was characterized through \textit{operando} x-ray total scattering.
The synchrotron x-ray scattering experiments were conducted at beamline P02.1, PETRA III, DESY~\cite{Dippel2015}, using a Perkin Elmer XRD1621 area detector. 
Experiments were conducted for two batches of \ch{TiO2}-bronze nanocrystals.
For the first batch, \textit{ex situ} experiments for pristine and chemically lithiated material were conducted using an x-ray wavelength of 0.20721~Å.
For the second batch, \textit{ex situ} experiments were conducted for pristine material and for a cathode mixture containing the active material, polymeric binder, and conductive carbon. Also, for the second batch, an \textit{operando} experiment was conducted for the \SI{3}{nm} nanocrystals using an x-ray wavelength of 0.20739~Å. 
The \textit{ex situ} PXRD and XTS experiments were conducted using kapton polyimide capillaries (\SI{1.0}{mm} inner diameter, Cole-Parmer). 
An empty capillary was used for the background measurement and \ch{CeO2} was used for calibration.
For the \textit{operando} XTS experiment, the AMPIX electrochemical cell was used. For the background measurement, an AMPIX cell containing separator wetted with electrolyte was used and \ch{CeO2} was used for calibration, including experimental geometry and instrumental contributions.
\subsection{Data processing}
\label{sec:method_data_processing}
For the \textit{ex situ} data, the scattering data were processed using the \dawn software package~\cite{Filik2017}. 
The beamstop arm, dead pixels, and over-exposed pixels were masked using the 'fast masking' tool.
The 'mask by coordinate' feature, where lower and upper $Q$-limits for the mask are stated by the user, was used to mask the beamstop and to mask incomplete Debye-Scherrer rings at high $Q$, effectively setting the range of azimuthal integration.
For the \textit{operando} data, the Python fast azimuthal integration (\pyfai) software~\cite{Ashiotis2015} was used. 
The calibration was done for a crystalline \ch{CeO2} standard. A static mask was created by masking beamstop, beamstop arm, dead pixels, and over-exposed pixels.
To mask out single-crystal spots originating from the Li-anode, a dynamic mask was created for each of the \ch{TiO2}-bronze \textit{operando} frames, using a Python-based automasking routine based on image analysis.
To account for x-ray intensity fluctuations due to fluctuating current in the storage ring during the \textit{operando} experiment, the \textit{operando} data were scaled in reciprocal space.
For the \textit{ex situ} and \textit{operando} XTS data, background subtraction, normalization to obtain the total scattering structure function, $S(Q)$, further reduction to obtain the reduced total scattering structure function, $F(Q)$, and inverse Fourier transformation to obtain the reduced atomic pair distribution function, $G(r)$, were done using the \pdfgetxthree algorithm~\cite{Juhas2013} through the \xpdfsuite~\cite{Yang2015} program.
\subsection{\textit{Ex situ} PXRD and Rietveld analysis}
\label{sec:method_rietveld}
For Rietveld analysis of the of the PXRD data for the pristine and chemically lithiated materials of batch one, the \topas software~\cite{Coelho2018} was used. 
Crystallite sizes were estimated based on the Scherrer method~\cite{Scherrer1918}, based on the volume-weighted column height~\cite{Dinnebier2019}.
The instrumental contribution to the peak broadening was determined by refining the powder profile of the crystalline \ch{CeO2} standard.
For the structural modelling of the pristine material, a monoclinic \ch{TiO2}-bronze structure (space group $C2/m$, \ch{VO2}(B) structure type~\cite{Theobald1976}, ICSD~\cite{Belsky2002} collection code 41056~\cite{Feist1992}) was used.
For the structural modelling of the chemically lithiated material through Rietveld analysis, a single phase of lithiated \ch{TiO2}-bronze (space group $C2/m$, ICSD-180011~\cite{Armstrong2010}) with composition \ch{Li_{0.5}TiO2} was used.
\begin{figure}
	\center
	\includegraphics[width=0.95\columnwidth]{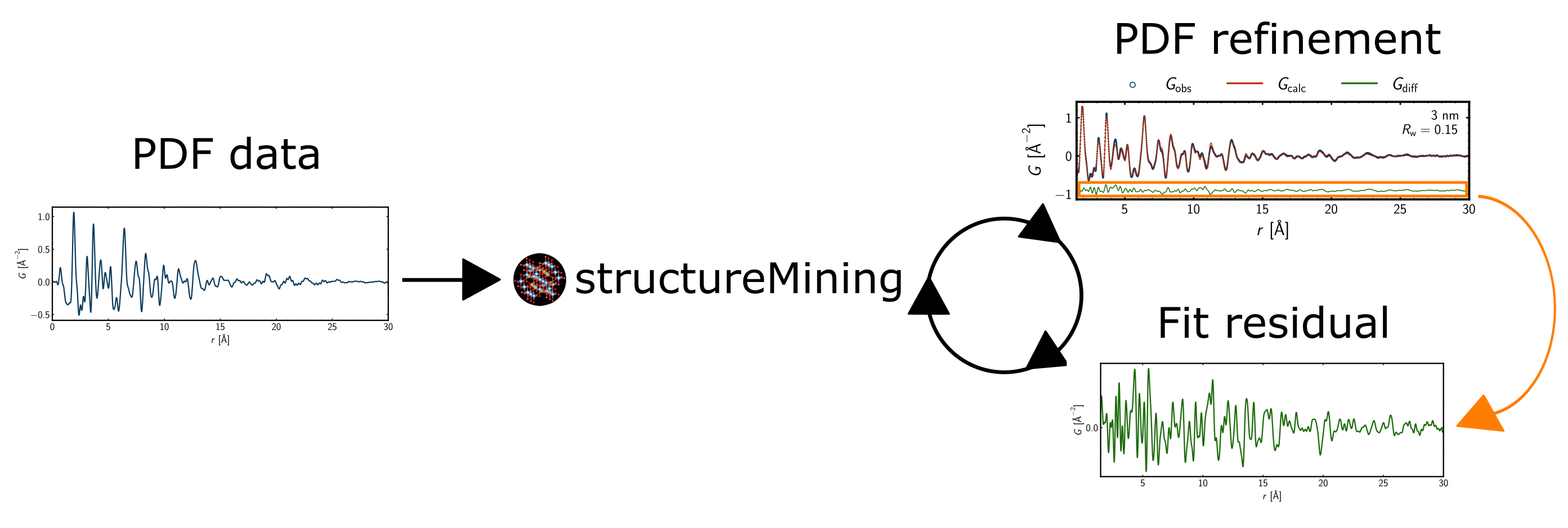}
	\caption{
Workflow when using the \strumin app at \pdfitc. 
PDF data is uploaded to the \strumin app and a model is build based on the output from the app and refined against the experimental data. 
If the model does not seem to describe the data sufficiently, the fit residual may be extracted and uploaded to the \strumin app, as if it were a PDF itself, i.e., with the same metadata as the original PDF. 
From the output of the \strumin app, a two-phase model is build and refined against the experimental data.
	}
	\label{fig:strumin_diffpy_workflow}
\end{figure}
\subsection{\textit{Ex situ} PDF modelling}
\label{sec:method_exsitu_pdf_modelling}
For analysis of the \textit{ex situ} PDF data, the \diffpy software~\cite{Juhas2015} was used. 
A crystalline \ch{CeO2} standard was refined to obtain instrumental dampening and broadening parameters, $Q_{\mathrm{damp}}$ and $Q_{\mathrm{broad}}$, which were then included in the analysis of the \ch{TiO2}-bronze data as fixed parameters.
For the refinements of the data for the \ch{TiO2}-bronze nanocrystals, scale factors, unit cell parameters, isotropic atomic displacement parameters, $u_{\mathrm{iso}}$, and quadratic correlated motion parameters, $\delta_{2}$, were refined.
As Li has a low atomic number and therefore a low x-ray scattering power, its contribution to the total scattering signal was expected to be low. In addition, possible disorder of Li would also broaden the scattering signal originating from Li. Therefore, the isotropic ADP of Li was not included in the refinement as a variable but included as a parameter fixed to a value of $u_{\mathrm{Li}}=\SI{0.05}{\angstrom^{2}}$.
The coherent domain size for a spherical model was also refined to take nanosize into account.
The atomic positions were refined as well, using space group symmetry constraints. 
To stabilize the refinement, the atomic positions were refined using restraints of $\pm0.05$ units of the relevant unit cell length.

Single-phase refinements were based on the output from the \strumin app at \pdfitc.
When single-phase refinements were insufficient for the PDF analyses, the PDF fit residual from \diffpy was extracted and saved to a .gr file as if it was an experimental PDF. The extracted fit residual was then uploaded to the \strumin app with the same metadata as the original PDF.
Using the \strumin output once again, two-phase refinements were conducted using \diffpy.
This approach is illustrated in \fig{strumin_diffpy_workflow} and will be described in detail in \sect{results_pdf_exsitu}
\subsection{\textit{Operando} PDF modelling}
\label{sec:method_operando_pdf_modelling}
The results of the PDF modelling of \textit{ex situ} data for the pristine and chemically lithiated materials were used for the PDF modelling of the \textit{operando} data for the Li-poor and Li-rich states, respectively.
To take possible non-subtracted signal from the glassy carbon windows of the AMPIX cell into account together with the signal originating from the conductive carbon in the electrode composite, a modified graphite phase was included in the analysis the \textit{operando} PDF data. 
To take stacking faults and turbostratic disorder into account, an anomalously high value of unity for the atomic displacement parameter along the $c$ axis was used, $u_{\mathrm{33}}=1.0$. 
Graphically speaking, the interlayer atomic pair correlations are broadened so much that only the in-layer atomic pair correlations remain in the model.
Instead of modelling the carbon signal, one could also subtract a scaled carbon signal from a refinement of the first scan of the pristine material. 
However, the modelling approach allows for small adjustments for the scale and in-plane lattice parameter such that the carbon contribution to the fit residual is minimized throughout the sequential refinement~\cite{Christensen2019, Christensen2019b}.
\subsection{Extracting time dependence of chemical components}
\label{method:extracting_time_dependence}
To extract time dependence of chemical components, i.e., the phase evolution during the \textit{operando} experiment, multiple model-free analyses were conducted.
\subsubsection{\simmap (Pearson correlation analysis)} ${}$
\label{sec:method_simmap}

The \simmap app at \pdfitc was used to inspect similarity of the \textit{operando} PDFs.
The Pearson correlation coefficient (PCC) is the measure of similarity. Similar PDFs are expected to represent similar phase content and \textit{vice versa} for dissimilar PDFs. Therefore, \simmap allows to inspect phase evolution during the \textit{operando} experiment, especially when comparing the correlation matrix to the Galvanostatic cyling.
The onset of phase transitions will be evident as sudden dissimilarity between neighboring PDFs. The nature of phase transitions, e.g., two-phase or solid-solution, will be evident as discrete or continuous changes of similarity, respectively. 
Disordering, i.e., shortening of the length of structural coherence, can be probed by conducting the correlation analysis for different $r$-ranges. See Appendix \ref{sec:si_pearson} in the supporting information.
\subsubsection{\nmfmap (Non-negative matrix factorization)} ${}$
\label{sec:method_nmfmap}

The \nmfmap was used to identify the number of components needed to describe the trends in the \textit{operando} data. This was done through the reconstruction error as a function of the number of components.
Due to the non-negative constraint on the matrix decomposition, the behavior of the NMF weights are likely to be physically meaningful.
The NMF weights as a function of time during the Galvanostatic cycling of the \textit{operando} experiment provided information on the phase evolution.
The behavior of the NMF weights was used as guidance for when to include certain phases in the refinement of the \textit{operando} data.
In \sect{method_pca} below, it is described how PCA was used for denoising the \textit{operando} data. NMF could also be used for denoising. 
However, it should be kept in mind that the PDFs have been shifted to be positive for NMF to be applicable. Hence, using NMF for denoising would require a down shift of the components to oscillate around zero as originally. 
This would of course not be a problem for the case of intensity vs. momentum transfer, $I(Q)$, PXRD data, as no shift would had to be introduced prior to the NMF analysis.
Also, the \nmfmap app only returns normalized weights, as these can be interpreted as fractions of the total signal. Therefore, the reconstructed signal from NMF would not be on the same scale as the experimental data, making comparison and validation less straightforward, though, e.g., Pearson correlation analysis would be insensitive to both shift and scale.
If one had access to the non-normalized NMF weights, the NMF reconstruction would be on the same scale as the experimental \textit{operando} data and direct comparison, e.g., through a difference plot, would be more straightforward.
\subsubsection{Principal component analysis (PCA)} ${}$
\label{sec:method_pca}

The \texttt{scikit-learn.decomposition.PCA}~\cite{scikit-learn} Python~\cite{Rossum2009} module was used for the PCA.
The primary purpose was to denoise the experimental \textit{operando} data.
A part of the denoising process is to set the level of noise filtering. This was done by inspection of the (cumulated) explained variance ratio as a function of the number of components. See Appendix \ref{sec:si_pca} in the supporting information.
Thereby, an indication on the number of components needed to describe the trends in the \textit{operando} data was also obtained. This number was compared that obtained through NMF analysis, as described in \sect{method_nmfmap}.
\subsubsection{Putting it all together} ${}$
\label{sec:method_summary}

\begin{figure}
	\center
	\includegraphics[width=\columnwidth]{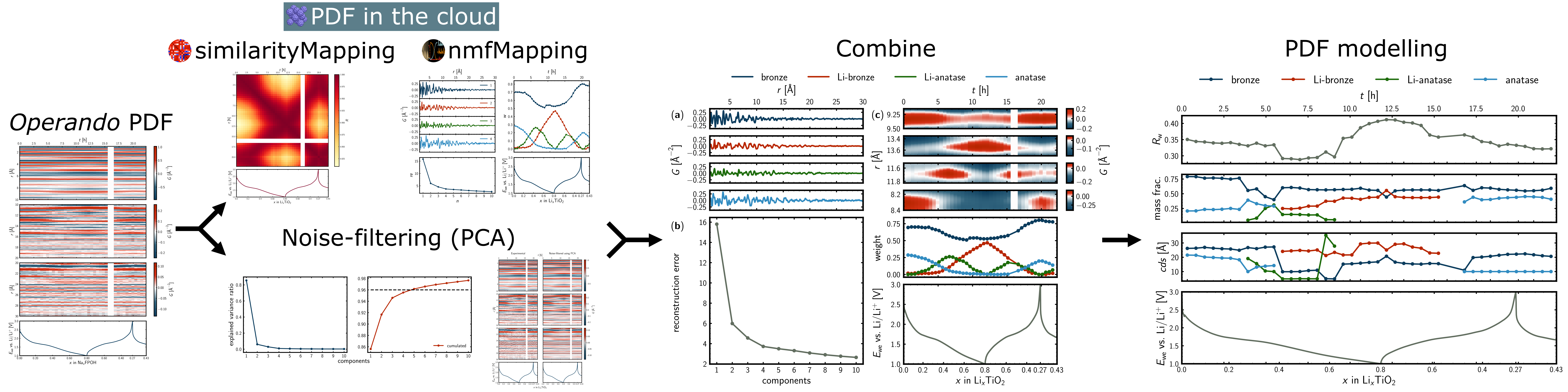}
	\caption{
Modelling of \textit{operando} PDF data based on model-free analyses.
\pdfitc offers model-free analyses through the \simmap and \nmfmap apps.
Experimental noise can be filtered using PCA to improve visual inspection of the data.
The structural modelling is based on the outputs of the model-free analyses.
	}
	\label{fig:simmap_nmfmap_diffpy_workflow}
\end{figure}

Using the denoised data from the PCA, phase-specific atomic pair correlations were identified and their evolution lined up with the behavior of the NMF weights. This underline the physical significance of the NMF output.
The workflow for extracting time dependence of chemical components to use for modelling of the \textit{operando} PDF data is illustrated in \fig{simmap_nmfmap_diffpy_workflow}.
\section{Results}
\label{sec:results}
\subsection{\textit{Ex situ} Rietveld analysis}
\label{sec:results_exsitu_rietveld}
The nanosize crystallites in the materials were readily evident from the very broad reflections in the diffraction patterns. 
Fits from Rietveld analyses of the \textit{ex situ} PXRD data of pristine and chemically lithiated materials of batch one can be found as \fig{pxrd_rietveld} in the supporting information.
The broadening of the peaks limits the amount of structural information extractable through Rietveld analysis, as uncertainties on the refined values are high.
This underlines the need for PDF analysis.
For further discussion of the Rietveld analysis results, please refer to the supporting information.
\subsection{\textit{Ex situ} PDF analysis: \strumin and PDF modelling}
\label{sec:results_pdf_exsitu}
The \strumin app at \pdfitc was used to identify primary and secondary phases for \textit{ex situ} PDF data of the pristine and chemically lithiated materials.
For the pristine materials, the chemical composition was set to \ch{TiO2}.
For the chemically lithiated material, the queries were run for all lithium titanium oxides, \ch{Li_{x}Ti_{y}O_{z}}, by putting \texttt{Li-Ti-O} for the composition on the app. 
Initially, primary phases of the PDFs were identified.
To identify secondary phases, the initial \strumin output was used as input model for the \diffpy software and the model was refined. The \diffpy fit residual was extracted and saved with the same metadata as the original PDF and then uploaded to \strumin app, using the same composition as the initial search query.
The second \strumin output was then incorporated into a two-phase model that was refined using \diffpy. 
The workflow is illustrated in \fig{strumin_diffpy_workflow}.
The top fives results of each \strumin search query are presented in the supporting information.

\begin{figure}
	\center
	\includegraphics[width=\columnwidth]{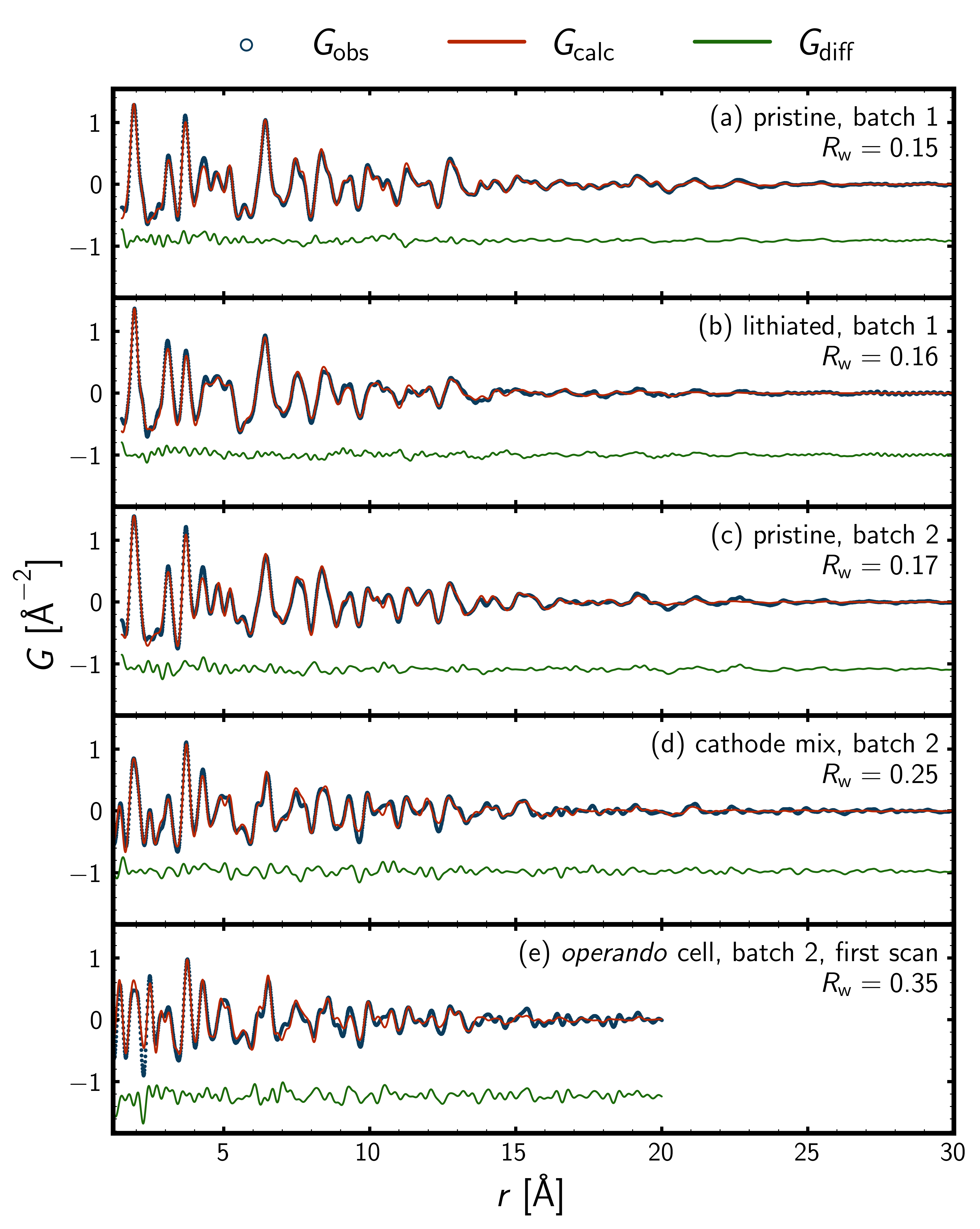}
	\caption{
\textit{Ex situ} PDF refinements. The observed PDF, \gobs, is shown as blue circles, the calculated PDF, \gcalc, is shown as a red line, and the difference between the observed and calculated PDFs, \gdiff, is shown as a green line. The reduced atomic pair distribution function, $G$, in inverse Ångström squared, $\mathrm{\AA}^{-2}$, is shown as a function of the interatomic distance, $r$, in Ångström, Å. The sample label and the weighted residual value, \rw , are displayed to the right in each subplot.
	}
	\label{fig:pdf_exsitu_stack}
\end{figure}

For the pristine materials, the topmost candidate returned by \strumin was the \ch{TiO2}-bronze structure (space group $C2/m$)~\cite{Feist1992}.
\fig{pdf_exsitu_stack}a displays a single-phase fit of the pristine material of batch one, where atomic positions were included in the refinement using space group constraints and restraints of $\pm0.05$ of the relevant unit cell side length. A weighted residual value of \rwvalue{0.15} was obtained. The coherent spherical domain size was estimated to \SI{30\pm6}{\angstrom}.

For the pristine material of batch two in \fig{pdf_exsitu_stack}c, the residual from a single-phase PDF refinement using the \ch{TiO2}-bronze phase was extracted and uploaded to the \strumin app. The topmost candidate for the residual was the \ch{TiO2}-anatase structure (space group $I4_{1}/amd$)~\cite{Horn1972}. A weighted residual value of \rwvalue{0.17} was obtained, when including atomic position subject to space group constraints of the two phases. The estimated weights fractions of the \ch{TiO2}-bronze and anatase phases were 0.85 and 0.15, and the estimated spherical coherent domain size were 25(6) and \SI{40(40)}{\angstrom}, respectively.
The high uncertainty for coherent domain size of the secondary anatase phase should be noted. It is common for minor phases to display high uncertainties for scale factors and coherent domain sizes, due to the minor contributions to the total scattering signal. Also, scale factors and coherent domain sizes are often highly correlated, especially, for nanosized and minor phases. This should be kept in mind for the estimated weight fractions, as these are based on the scale factors of the refinement.

For the chemically litihated material of batch one in \fig{pdf_exsitu_stack}b, the topmost candidate was a lithiated version of the \ch{TiO2}-bronze structure (space group $C2$), with the formula \ch{LiTi4O8}, i.e., \ch{Li_{0.25}TiO2}. The fit residual of a single-phase refinement was extracted and uploaded to the \strumin app. The topmost candidate was a lithiated version of the \ch{TiO2}-anatase structure with composition \ch{Li7Ti8O16} i.e., \ch{Li_{0.875}TiO2} (space group $I\bar{4}2m$). For a two-phase refinement, a weighted residual value of \mbox{\rwvalue{0.16}} was obtained together with mass fractions of 0.85 and 0.15, and coherent domain sizes of 26(7) and \SI{20(20)}{\angstrom} for the lithiated bronze and anatase phases, respectively.

For the cathode composite (containing PVDF binder and conductive carbon) and the first \textit{operando} frame in \fig{pdf_exsitu_stack}d-e, a modified graphite phase was included in addition to the \ch{TiO2}-bronze and anatase phases used for the pristine material.
The most evident difference to the PDF of the pristine material in \fig{pdf_exsitu_stack}c was the additional atomic pair correlations, e.g., the C-C correlation at \SI{1,4}{\angstrom} together with the 'dilution' of the \ch{TiO2} signal, due to the additives.
When including atomic position subject to space group constraints for the \ch{TiO2}-bronze and anatase phases, a weighted residual value of \rwvalue{0.25} was obtained.
The estimates for the coherent domain sizes of the bronze and anatase phases were similar to those obtained for the pristine material in \fig{pdf_exsitu_stack}a, however, the estimated weight fractions were a little different, as they were estimated to 0.8 and 0.2, respectively.

\fig{pdf_exsitu_stack}e displays the PDF fit for the first frame of the \textit{operando} experiment. The increase of, e.g., the C-C atomic pair correlation at \SI{1.4}{\angstrom} reflects the residual signal of the glassy carbon windows of the AMPIX cell, i.e., remaining background signal. 
The modified graphite phase can account for some of the additional residual signal, however, the weighted residual value of \rwvalue{0.35} reflects the increased complexity of the system from which the experimental PDF originates.
The spherical coherent domain sizes for the \ch{TiO2}-bronze and anatase phases were refined to 30(12) and \SI{22(14)}{\angstrom}, respectively, which is a little different but comparable to the \textit{ex situ} estimates in \fig{pdf_exsitu_stack}c-d.
The estimated weight fractions were 0.8 and 0.2, respectively, which was also comparable to the \textit{ex situ} estimates in \fig{pdf_exsitu_stack}c-d.
Tables with refinement results can be found in the supporting information. 
\subsection{\textit{Operando} PDFs}
\label{sec:results_operando}
\begin{figure}
	\center
	\includegraphics[width=0.95\columnwidth]{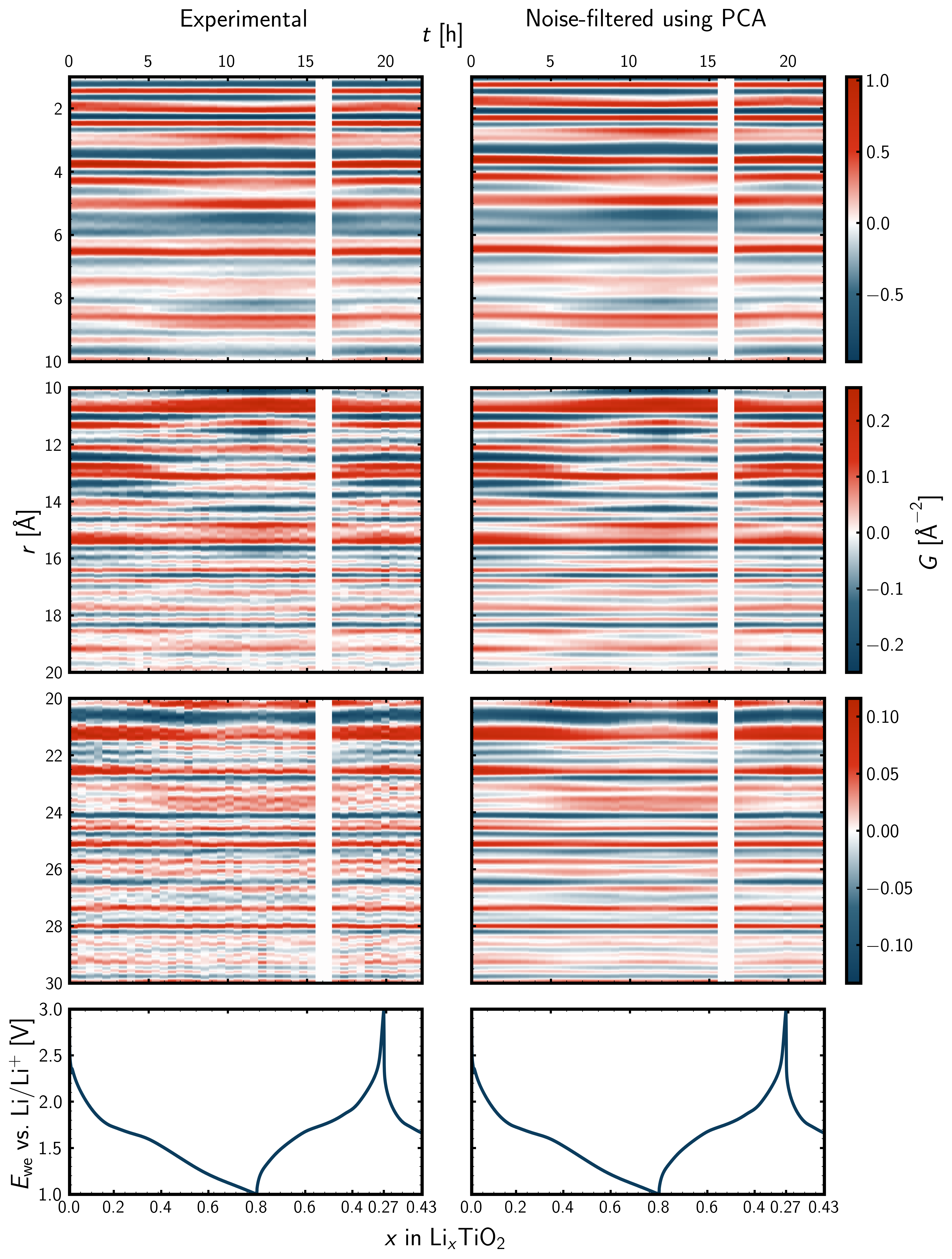}
	\caption{
	Experimental (left) and noise-filtered (PCA-reconstructed, right) \textit{operando} pair distribution function data together with Galvanostatic cycling.
	The reduced atomic pair distribution function, $G$, as a function of the interatomic distance, $r$, and time, $t$, during the \textit{operando} experiment.
	Please note the different scales for the color bars of the different $r$-ranges.
	The white areas represent absence of synchrotron x-rays during the \textit{operando} experiment.
	For the voltage profile, the electrochemical potential of the working electrode, \ewe~vs. Li/\ch{Li+}, is shown as a function of the state of charge, $x$, in \ch{Li_{x}TiO2}.
	}
	\label{fig:gr_gr_gr_echem}
\end{figure}

A real space representation of the \textit{operando} data for the \SI{3}{nm} material of batch two is shown in the left part of \fig{gr_gr_gr_echem}.
A reciprocal space representation of the \textit{operando} can be found in the supporting information as \fig{fq_fq-pca_echem}.
From the low-$r$ part of \fig{gr_gr_gr_echem}, it is evident that the very local structure of the electrode material changes very little during the \textit{operando} experiment.
This was expected, as the polymorphs to be encountered are anticipated all to be build from \ch{TiO6}-octahedra.
Due to the low x-ray scattering power of Li, the appearance of Li correlations in the PDF is not expected. 
However, the structural response upon Li-intercalation resulting in, e.g., an increase in Ti-O distances due to reduction of Ti is expected to be observable, especially for correlations further out in $r$.
From eye-inspection of the \textit{operando} PDF data and the voltage profile of the Galvanostatic cycling in \fig{gr_gr_gr_echem}, the charged state of the electrode at the end of the experiment appears similar to the pristine charged state of the electrode at the beginning.
During the initial discharge, some atomic pair correlations are observed to fade, e.g., around $r=\SI{9}{\angstrom}$, while other are observed to emerge, e.g., around $r=\SI{8}{\angstrom}$.
The fading and the emerging of the atomic pair correlations appear to be reversible.
During the discharge, the fading and emerging of atomic pair correlations can be linked to the kink of the voltage profile around $x=0.4$, for $x$ in \ch{Li_{x}TiO2}. During the charge, the related kink is less pronounced, though, a delicate kink appears around $x=0.6$.
Similar observations can be done in reciprocal space for the reduced total scattering structure function in the supporting information (\fig{fq_fq-pca_echem}).
Thus, from visual inspection of the \textit{operando} data, a reversible two-phase transformation appears to occur.
As the discharge capacity of the Galvanostatic cycling is larger than the charge capacity, the phase evolution might only be partly reversible.
\subsection{\simmap for \textit{operando} PDF data}
\label{sec:results_simmap}
\begin{figure}
	\center
	\includegraphics[width=\columnwidth]{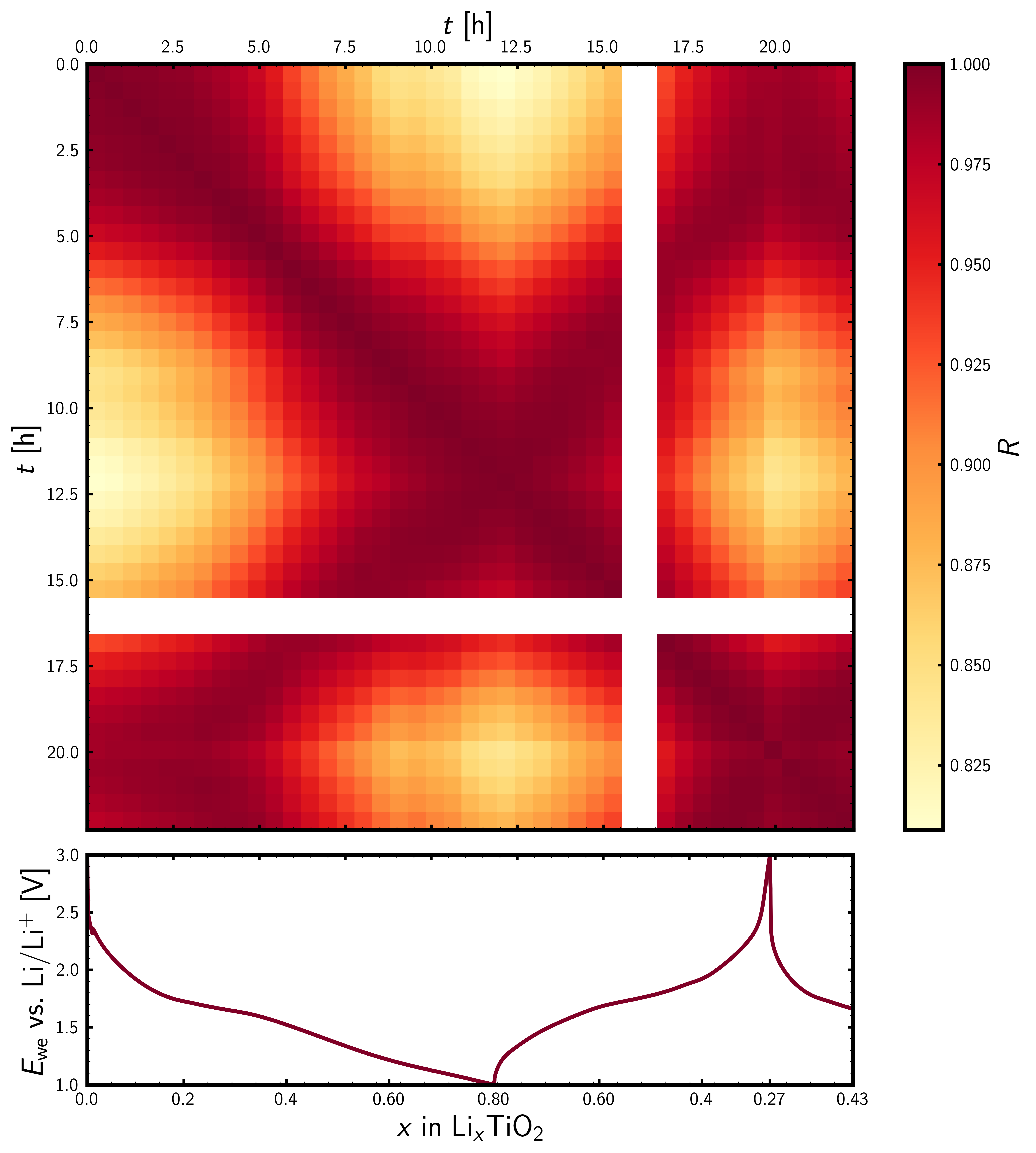}
	\caption{
Pearson cross-correlation matrix for the \textit{operando} PDF data. The corresponding time, $t$, is displayed on the axes. The correlation analysis was conducted for the $r$-range from 0 to \SI{30}{\angstrom}. The white columns are due to absence of synchrotron x-rays during the \textit{operando} experiment. Below, the voltage profile is shown. The electrochemical potential of the working electrode, \ewe~Li/\ch{Li+}, is shown as a function of state of charge, $x$, in \ch{LixTiO2}.
}
	\label{fig:pearson_echem_r=0-30}
\end{figure}

To probe similarity between the individual \textit{operando} PDFs, the \simmap app at \pdfitc was used.
The output of the Pearson correlation analysis is shown in \fig{pearson_echem_r=0-30}, which displays the correlation matrix of the \textit{operando} PDFs together with the voltage profile of the Galvanostatic cycling.
The correlation analysis was conducted for the $r$-range from 0 to \SI{30}{\angstrom}.
From the scale of the colorbar, it is seen that all Pearson correlation coefficients are above 0.8, indicating no severe structural transformations or reconstructions, which is in correspondence with the real space overview plot in \fig{gr_gr_gr_echem} and the reciprocal space overview plot in \fig{fq_fq-pca_echem} in the supporting information.
The aforementioned domains of the voltage profile and reversibility identified from \figs{gr_gr_gr_echem} are also apparent from the correlation matrix.
From the correlation matrix, it is seen that the pristine material is highly similar to the Li-poor state upon charge, just as the Li-rich states of the initial discharge are highly similar to the Li-rich state of the charge. 
Hence, the qualitative interpretations from \fig{gr_gr_gr_echem} are supported by the quantitative but model-free Pearson correlation analysis conducted using the \simmap at \pdfitc.
Outputs of correlations analyses for various $r$-ranges from 0-\SI{10}{\angstrom}, 10-\SI{20}{\angstrom}, and 20-\SI{30}{\angstrom} are available in the supporting information (\figs{pearson_echem_r=0-10}-\ref{fig:pearson_echem_r=20-30}). 
The trends of the correlation matrices do not appear to be $r$-dependent, though the sensitivity towards similarity is highest for the intermediate $r$-range from 10 to \SI{20}{\angstrom} in \fig{pearson_echem_r=10-20}. 
At lower $r$-values, the encountered phases are expected to be similar due to common \ch{TiO6}-octahedra building blocks. At higher $r$-ranges, the signal-to-noise ratio is so low that the noise level hampers the sensitivity towards structural dissimilarities, just as atomic pair correlations also are expected to overlap more and more with increasing $r$.
\subsection{\nmfmap for \textit{operando} PDF data}
\label{sec:results_nmf}
The \nmfmap app at \pdfitc was used to obtain insights into the \textit{operando} PDF data through NMF analysis.
Running the NMF analysis using four components results in the physically interpretable output in \fig{nmf_echem_n=4_gr}.
The first step in the protocol to arrive at the result in \fig{nmf_echem_n=4_gr} was to determine the number of components needed to describe the trends in the \textit{operando} data. This was done by running a query on the \nmfmap app, where no number was stated for the number of components to use during the matrix decomposition. 
Due to the complex and noisy nature of the \textit{operando} PDF data, the default maximum number of components, which is currently ten, was encountered. 
The reconstruction error as a function of the number of components provided was used to determine the number components needed to describe the trends in the data to a sufficient extent, as displayed in \fig{nmf_echem_n=4_gr}b.
A distinct change for the slope of the reconstruction error was observed for two, four, and five components. Going from four to five components barely changed the reconstruction error, whereas a constant slope was observed beyond five components.
The interpretation was that it was meaningful to include up to four components in the analysis. The monotonic decrease from five to ten components was expected to arise from the additional degree of freedom that came with every additional component allowed during the matrix decomposition.
Despite having the indication that the number of components needed to describe the trends in the \textit{operando} PDF data was four, the NMF analysis was not immediately redone using four components, as presented in \fig{nmf_echem_n=4_gr}.
Instead, to have a simpler output to interpret, the NMF analysis was done using two components. Then, the NMF analysis and interpretation was redone using three components, four components, and finally five components, as five components result in an NMF output that cannot be interpreted in a physically meaningful way, as the components look like the PDFs are 'capped' at top and bottom, which was unexpected. Also, it was not possible to make sense of the behavior of the NMF weights anymore.
Plots showing the output of NMF analyses for various number of components in both real and reciprocal space are presented in the supporting information (\figs{nmf_echem_n=2_gr}-\ref{fig:nmf_echem_n=5_fq}).
For the NMF analyses using two and three components, the components appeared fairly reasonable and it was possible to make sense of the behavior of the NMF weights. However, in both analyses, only one initial component was present. From the analysis of the \textit{ex situ} PDF data of the \SI{3}{nm} material of batch two, it was evident that two \ch{TiO2} phases were present initially, i.e., the \ch{TiO2}-bronze and anatase phases.

The second initial component showed up, when the NMF analysis was run with four components, as can be seen in \fig{nmf_echem_n=4_gr}. The components in \fig{nmf_echem_n=4_gr}a look fairly reasonable, even though they do not resemble actual PDFs, which can probably be explained by the multicomponent nature of the electrochemical cell, which was probed in transmission mode by x-rays, resulting in lots of scattering contributions other than those originating from the active electrode material of interest. 
However, the trends in the data described by the NMF weights in \fig{nmf_echem_n=4_gr}c lined nicely up with the voltage profile in \fig{nmf_echem_n=4_gr}d.
Interestingly, even though four phases seem to be present at intermediate discharge and charge, only two phases seem to be present by the end of the discharge and at the end of the charge. The latter feature, together with the overview plot in \fig{gr_gr_gr_echem} and the \simmap output in \fig{pearson_echem_r=0-30}, indicates the partly reversible nature of the phase behavior.
From the NMF analysis, it appears that two Li-poor phases exists for the pristine material, which are reformed upon charge in a reversible manner. From the \textit{ex situ} PDF analysis (\fig{pdf_exsitu_stack}), the two NMF components are expected to represent the \ch{TiO2}-bronze and \ch{TiO2}-anatase phases. Their reversible formation is also expected from the correlation analysis (\fig{pearson_echem_r=0-30}).
At deep discharge, the NMF analysis indicates the presence of two NMF components, which are the major pristine component together with another component that emerges during the discharge and fades upon charge. From the \textit{ex situ} PDF analyses (\fig{pdf_exsitu_stack}), the two components are interpreted as representing the Li-poor and Li-rich \ch{TiO2}-bronze phases.
Finally, the most exciting outcome of the NMF analysis is the NMF component representing an intermediate, which emerges and fades during both the discharge and the charge, where a maximum of the corresponding NMF weight occurs at the kinks of the voltage profile. Its presence also explains the more gradual changes observed for the correlation matrix (\fig{pearson_echem_r=0-30}) at intermediate state of charge.
From the correlation matrix, it can be seen that the PDF, where the weight of the intermediate component is at maximum, is more similar to the pristine and charged states than the discharged state, which might indicate a structural similarity to the \ch{TiO2}-anatase phase present at these states of charge. This would be somewhat in line with the \textit{ex situ} PDF analysis of the chemically lithiated material (\fig{pdf_exsitu_stack}b), which was modelled using lithiated bronze and anatase phases.
\subsection{Denoising using PCA}
\label{sec:result_pca}
In the \textit{operando} PDF data in the left part of \fig{gr_gr_gr_echem}, the signal-to-noise ratio quickly decreases with $r$. Already from around \SI{15}{\angstrom}, the signal it affected significantly by noise.
Therefore, the noise was filtered from experimental \textit{operando} PDF data in \fig{gr_gr_gr_echem} using PCA. When trying to filter noise from the signal of interest, it is key to be highly aware of what is captured be the filter. Filtering too little will not yield the desired signal of interest, whereas filtering too much will hamper signal of interest, which will result in improper interpretation of the data.
To arrive at a proper level of noise filtering, PCA was conducted iteratively using one to ten principal components for the matrix decomposition.
\fig{pca_evr_cevr_gr} in the supporting information shows the explained variance ratio and the cumulated explained variance ratio as a function of the number of principal components for the \textit{operando} PDF data. In both cases, a kink at four components is observed. For the cumulated explained variance ratio, a value of 0.96 for the PCA will ensure that enough of the trends in the experimental data are included in the reconstruction.
When instantiating the \texttt{sklearn.decomposition.PCA} class with the \texttt{n\char`_components} set to 0.96, the algorithm will select the number of components such that the amount of variance that needs to be explained is greater than the percentage specified~\cite{scikit-learn}.
As the value of 0.96 for the cumulated explained variance ratio in \fig{pca_evr_cevr_gr} is between four and five components, the algorithm will 'round up' and use five components for the matrix decomposition. Therefore, one could also just set the number of component to five, when instantiating the class. Hence, the filtering ends up being a bit more conservative, as less data is excluded or filtered, compared to using four components. However, this approach should prevent undesired 'over filtering' of the data.

The right part of \fig{gr_gr_gr_echem} shows the PCA reconstruction of the \textit{operando} PDF data together with the galvanostatic cycling data. That the PCA serves as a noise filter is clearly seen for the high-$r$ region, as the unfiltered signal suffers from a lower signal-to-noise ratio, resulting from the sample and instrumental dampening of the PDFs.
The difference between the experimental data and the PCA reconstruction, i.e., the filtered noise, of the reduced atomic pair distribution function, $G(r)$, can be found in the supporting information as \fig{gr_gr_gr_pca-recon-diff_echem}.
As desired, the part of the signal that is excluded or captured by the filter appears structureless, i.e, behaves as noise.
From the relative trends within each subplot and comparing the color scales of the subplots, the noise-level appears to be relatively constant with $r$, as should also be expected for noise.
Another way of comparing experimental and PCA-reconstructed data is through Pearson correlation analysis, as done in \fig{00_gr_obs_pca-recon_diff_pearson} in the supporting information, where it is seen that all the reconstructed PDFs are highly similar to the experimental ones.
A direct comparison for the first \textit{operando} PDF it found as \fig{00_gr_obs_pca-recon_diff_pearson} in the supporting information, where it is seen that the difference curve behaves as noise, as it should.
The noise-filtering using PCA was also done in reciprocal space for the \textit{operando} $F(Q)$ data, which is presented as \fig{fq_fq-pca_echem} in the supporting information. 
The explained variance ratio as a function of the number of principal components is presented in \fig{pca_evr_cevr_fq}.
The difference between the experimental and PCA-reconstructed $F(Q)$ is presented in \fig{fq_fq_fq_pca-recon-diff_echem}, whereas a comparison of experimental and PCA-reconstructed $F(Q)$ data through Pearson correlation analysis is presented as \fig{pearson_pca_recon_gr}. A direct comparison of the experimental and PCA-reconstructed $F(Q)$ for the first \textit{operando} frame is presented as \fig{00_fq_obs_pca-recon_diff_pearson} in the supporting information.

\begin{figure}
	\center
	\includegraphics[width=\columnwidth]{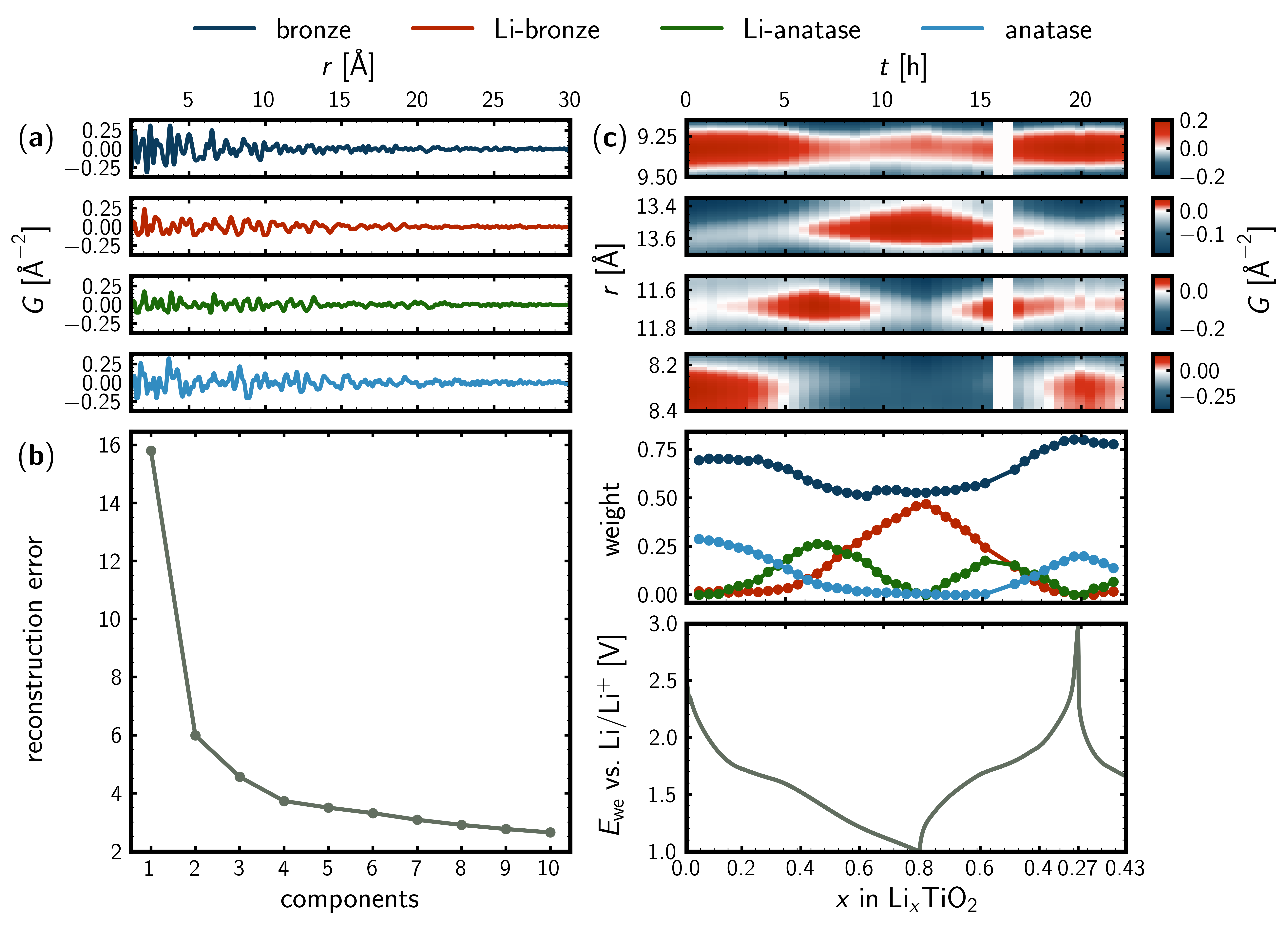}
	\caption{
Output from \nmfmap at \pdfitc when setting the threshold for the number of components to four.
(a) Component PDFs ($G$ vs. $r$).
(b) Reconstruction error as a function of the number of components.
(c) Top: phase-specific atomic pair correlations of the bronze, lithiated bronze, lithiated anatase, and anatase phases, respectively, from noise-filtered \textit{operando} PDF data.
Middle: NMF weights. 
Bottom: voltage profile. The electrochemical potential is shown as a function of the lithiation degree (state of charge).
	}
	\label{fig:nmf_echem_n=4_gr}	
\end{figure}

Atomic pair correlations that follow the behavior of the NMF weights have been identified from the noise-filtered \textit{operando} PDF data for each NMF component. These are plotted together with the NMF weights and the voltage profile in \fig{nmf_echem_n=4_gr}c to emphasize on the physical relevance of the NMF output.
\subsection{Modelling of \textit{operando} PDF data}
\label{sec:results_pdf_modelling}
\begin{figure}
	\center
	\includegraphics[width=\columnwidth]{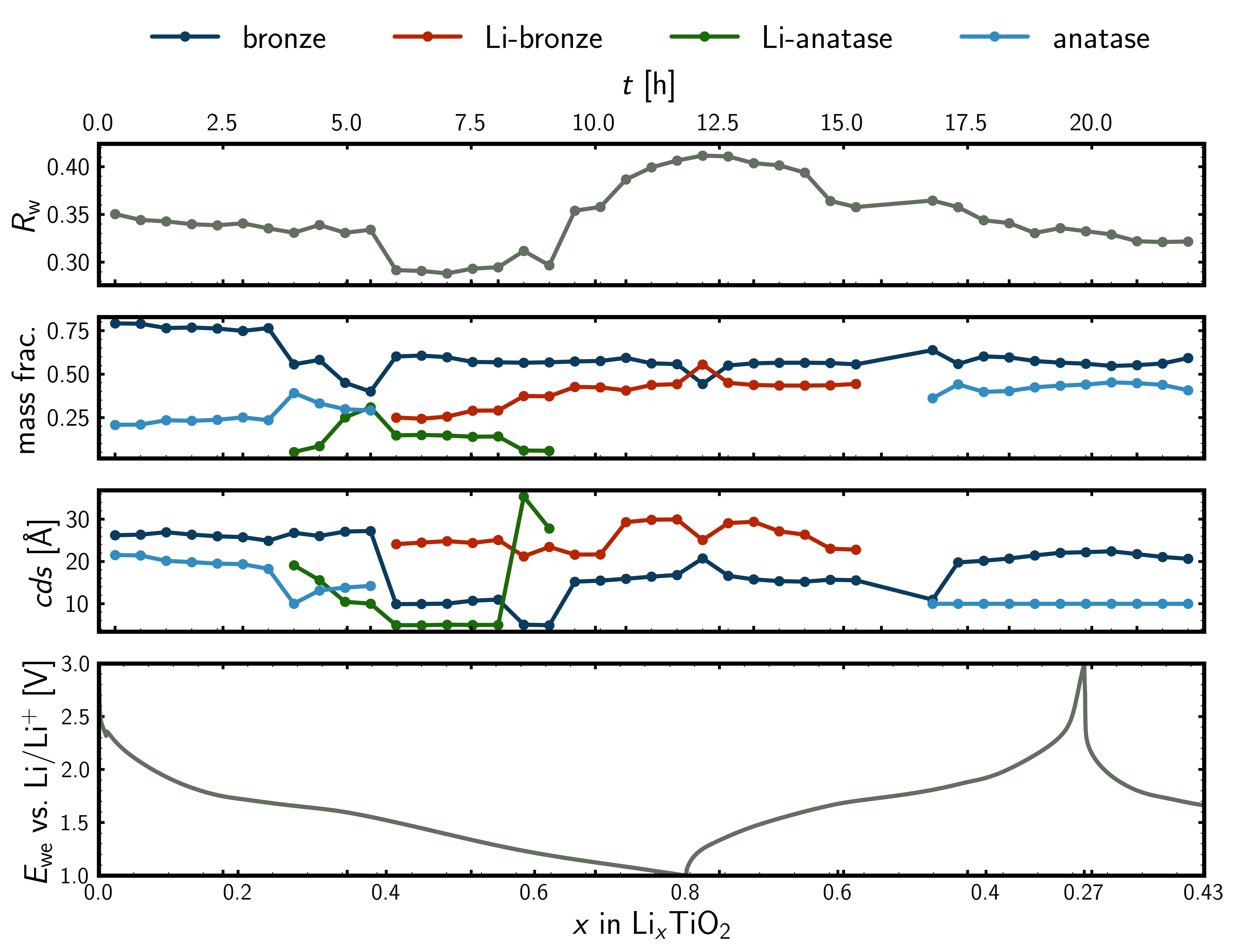}
	\caption{
	Results from PDF modelling using \diffpy.
	The weighted resdiual values for the PDF refinements, \rw, the mass fractions and the coherent domain sizes, $cds$, assuming a spherical domain, for each of the four phases included during the refinement of the \textit{operando} dataset. 
	The values are plotted for each frame of the \textit{operando} experiment as a function of time, $t$, together with the Galvanostatic cycling, showing the working electrode potential, \ewe, as a function of the Li-content of the positive electrode, $x$ in \ch{Li_{x}TiO2}.
	}
	\label{fig:rw_fracs_psize_echem}
\end{figure}
\fig{rw_fracs_psize_echem} displays refinement results for the \textit{operando} PDF analysis.
The nature of the phases appearing during the \textit{operando} analysis was assisted by the \textit{ex situ} analyses (\fig{pdf_exsitu_stack}).
The inclusion of the various phases was greatly guided by the NMF analyses in both real and reciprocal space (\fig{nmf_echem_n=4_gr} and \fig{nmf_echem_n=4_fq} in the supporting information).
The output of the NMF analyses indicated the presence of four components (phases) during the \textit{operando} experiment of the \ch{TiO2} nanocrystals.
Keeping the similarity of the \ch{TiO2} and \ch{Li_{x}TiO2} polymorphs from the Pearson correlation analysis in \fig{pearson_echem_r=0-30} in mind, together with the rather limited range of data to refine ($r$-range from 1.2 to \SI{20}{\angstrom}), due to sample and instrumental dampening, the PDF analysis was highly challenging.
The weighted residual values ranging from a little below 0.3 to a little above 0.4 in \fig{rw_fracs_psize_echem} indicate rather reasonable fits, taking the \textit{operando} nature of the PDF data in mind.
The reversible nature of the phase evolution was reflected by comparable descriptors of the pristine state and charged states, though the coherent domain sizes are observed to decrease a little for the charged states compared to the pristine state, which probably also explains the discrepancy for the estimated weight fractions. Especially the rather low coherent domain size estimated for the anatase phase in light blue is expected to hamper the weight fraction estimates.

A clear challenge for the \textit{operando} analysis was the detection limits, especially for the regions where three or four-phase systems were expected from the NMF analyses. For instance, this was seen during the charge, where it was not possible to include the intermediate \ch{Li_{x}TiO2}-anatase phase, even though it was expected to be present, both from the NMF analyses but also from inspection of the \textit{operando} data, as highlighted in \fig{nmf_echem_n=4_gr}.
Of course, the absence of minor phases in the structural modelling are to be kept in mind when evaluating the modelling results, as it always goes for incomplete (i.e., all) models.
However, being able to compare such modelling results to those of other types of analyses, e.g., Pearson correlation analysis and NMF analysis, is of immeasurable value, as this provides a measure of trustworthiness, increasing the value of the modelling results significantly.
Refined unit cell parameters for each of the phases, including the modified graphite phase, can be found as \figs{bronze_latpars_echem}-\ref{fig:graphite_latpars_echem} in the supporting information.

From the NMF analysis of the \textit{operando} data in \fig{rw_fracs_psize_echem}, the \ch{TiO2}-anatase (light blue) phase appears to transform before the \ch{TiO2}-bronze phase (navy) upon Li-ion intercalation during the discharge.
The \ch{TiO2}-anatase phase transforms into a lithiated analogue, \ch{Li_{x}TiO2}-anatase (green), which is an intermediate, as the phase is absent at deep discharge,
\begin{equation}
	x\ch{Li+} + x\ch{e-} + \ch{TiO2(anatase)} \rightarrow \ch{Li}_{x}\ch{TiO2(anatase)}.
\end{equation}
The phase transition of \ch{TiO2}-anatase into \ch{Li_{0.5}TiO2}-anatase has previously been reported~\cite{Murphy1983, Lafont2010}.
When the formation of the intermediate \ch{Li_{x}TiO2}-anatase is complete, around $x=0.4$ in \ch{Li_{x}TiO2}, the lihiated bronze phase, \ch{Li_{x}TiO2}-bronze (red) forms from the \ch{TiO2}-bronze as well as from the \ch{Li_{x}TiO2}-anatase,
\begin{align}
	x\ch{Li+} + x\ch{e-} + \ch{TiO2(bronze)} &\rightarrow \ch{Li}_{x}\ch{TiO2(bronze)}, \\
	x\ch{Li+} + x\ch{e-} + \ch{Li}_{y}\ch{TiO2(anatase)} &\rightarrow \ch{Li}_{x+y}\ch{TiO2(bronze)}.
\end{align}
At deep discharge, a biphasic mixture of \ch{TiO2}-bronze and \ch{Li_{x}TiO2}-bronze is present: due to reaction (2).

Upon recharge, the reverse behavior is observed and the phase behavior appears to be somewhat reversible, as also indicated by Pearson correlation analysis in \fig{pearson_echem_r=0-30}
\begin{align}
	\ch{Li}_{x+y}\ch{TiO2(bronze)} &\rightarrow x\ch{Li+} + x\ch{e-} + \ch{Li}_{y}\ch{TiO2(anatase)}, \\
	\ch{Li}_{x}\ch{TiO2(bronze)} &\rightarrow x\ch{Li+} + x\ch{e-} + \ch{TiO2(bronze)}, \\
	\ch{Li}_{x}\ch{TiO2(anatase)} &\rightarrow x\ch{Li+} + x\ch{e-} + \ch{TiO2(anatase)}.
\end{align}
A possible interpretation of these observations would be that the minor and more disordered \ch{TiO2}-anatase phase is related to the surface of the nanocrystals. The surface would be lithiated before the 'bulk' of the nanocrystals, such that the lithiated anatase phase forms before the lithiated bronze phase.
Another possibility would be individual bronze and anatase nanoparticles, where the anatase particles were lithiated before the bronze particles.

The transformation of \ch{Li_{x}TiO2}-anatase to \ch{Li_{x}TiO2}-bronze at $x \gtrsim 0.4$ may be enabled by the small domain size of the anatase phase, as the anatase phase usually is not lithiated beyond $x\sim 0.5$~\cite{Murphy1983, Lafont2010}. However, if this would also occur in single-phase anatase materials remains to be investigated.
\section{Conclusions}
\label{sec:conclusions}
Our work demonstrates various approaches to obtain insights into structural transformations of nanocrystalline functional materials through \textit{operando} pair distribution function analysis.
The methods applied constitute general toolbox, especially when dealing with highly complex time-series of PDF data, i.e., \textit{operando} or \textit{in situ} data.
The case for which the toolbox has been demonstrated was \ch{TiO2}-bronze nanocrystals, which were studied during Galvanostatic cycling when incorporated into the positive electrode of a rechargeable Li-ion battery.
The nanosize limited the value of conventional powder x-ray diffraction and Rietveld analysis but information on the atomic structure was accessible through x-ray total scattering and pair distribution analysis. 
The multi-component nature of a battery resulted in highly complex \textit{operando} pair distribution function data. 
The data analysis was greatly assisted through the novel tools at \pdfitc (\url{pdfitc.org}), which have proven to be of enormous value for the later quantitative refinement of the PDF data.
The \strumin app was used for phase identification for pristine and chemically lithiated materials, which were utilized for the \textit{operando} modelling.
The \simmap app was used for simiarlaity measure of the \textit{operando} PDFs to probe phase transformations and reversibility.
The \nmfmap app was used to obtain invaluable insights into the number of components (phases) and the behavior of their weights (fractions) during the \textit{operando} experiment.
To interpret the \nmfmap output, noise-filtering of the \textit{operando} PDF data using PCA greatly improved the visibility of unique atomic pair correlations for each of the phases (NMF components) encountered during the \textit{operando} experiment.
All of these model-free tools made it possible to model the \textit{operando} PDF data for this highly complex system, which involve two pristine nanocrystalline phases of \ch{TiO2}-bronze and anatase phases together with lithiated versions of them.
Without the use of the various tools of the toolbox presented herein, the structural modelling would have been immensely complicated to complete but careful use of the tools enabled the completion of the analyses presented here.
%
     %
%
\newpage
\appendix
\setcounter{equation}{0}
\setcounter{figure}{0}
\setcounter{table}{0}
\renewcommand{\theequation}{A\arabic{equation}}
\renewcommand{\thefigure}{A\arabic{figure}}
\renewcommand{\thetable}{A\arabic{table}}
\section{Rietveld analysis of \textit{ex situ} data for batch one}
\label{sec:si_rietveld}
\begin{figure}
	\center
	\includegraphics[width=\columnwidth]{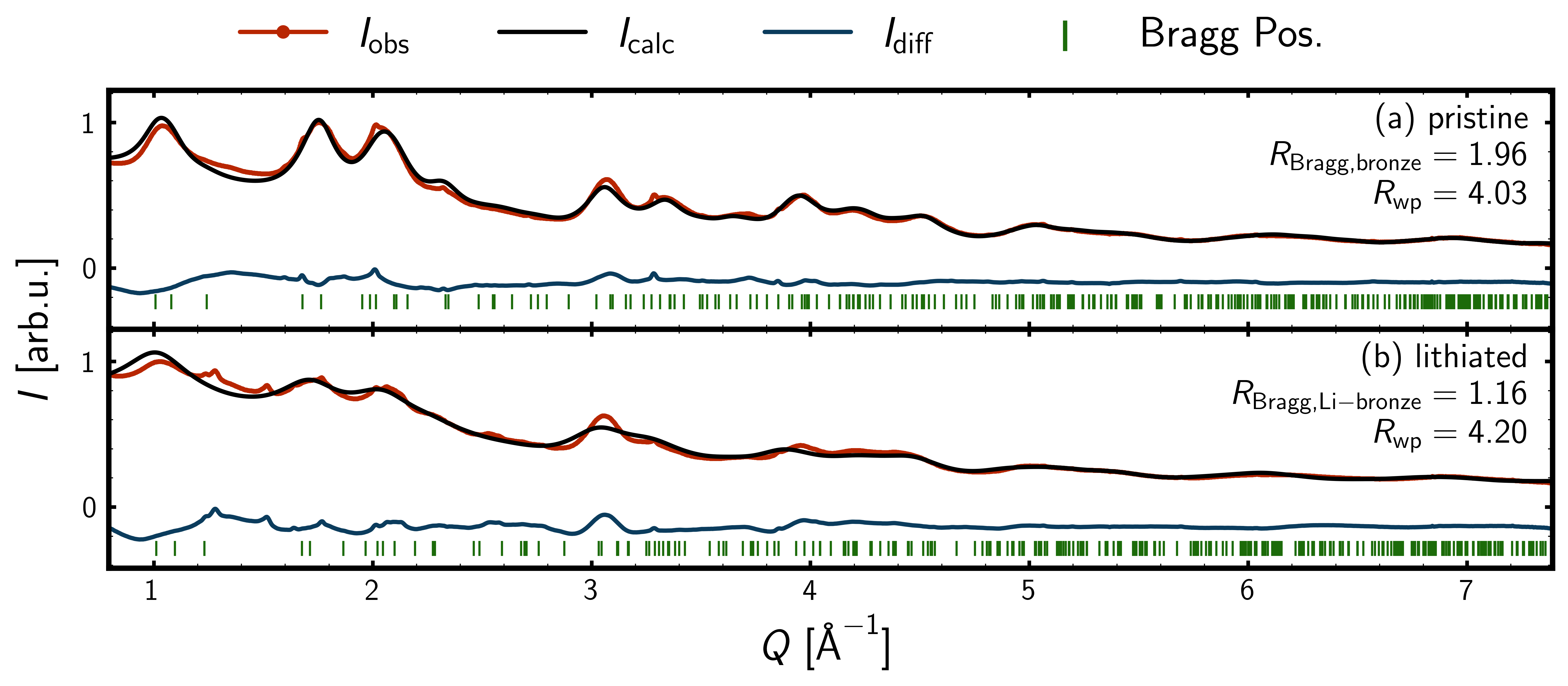}
	\caption{
The observed intensity, $I_{\mathrm{obs}}$, is displayed as red dotted lines,
the calculated intensity, $I_{\mathrm{calc}}$, is shown as a black line,
and the difference between the observed and calculated intensities, $I_{\mathrm{diff}}$, is displayed as a blue line.
The Bragg positions of the phases included in the refinements are indicate by vertical green lines.
The Bragg residual values, \rbragg, and the weighted profile residuals, \rwp, are shown to the right in each fit.
(a) The single-phase fit of the pristine material.
(b) The single-phase fit of the chemically lithiated material.
	}
	\label{fig:pxrd_rietveld}
\end{figure}
\fig{pxrd_rietveld} displays the fits from Rietveld analysis of the \textit{ex situ} data for the pristine and chemically lithiated material of batch one.
For the pristine \SI{3}{nm} material in \fig{pxrd_rietveld}a, a crystallite size of \SI{3.36}{nm} was estimated, well in line with expected size of \SI{3}{nm}.
Peaks that could be described by the \ch{TiO2}-bronze phase were observed, indicating that at least one secondary phase was present.
The fit from the Rietveld analysis of the chemically lithiated material is shown in \fig{pxrd_rietveld}b.
A single phase of lithiated bronze was used.
The estimated crystallite size was \SI{1.51}{nm}.
Peaks that could not be described by the \ch{Li_{0.5}TiO2} phase were observed, indicating that at least one secondary phase was present.
Due to the very broad nature of the features not described by the \ch{TiO2}-bronze and \ch{Li_{0.5}TiO2} phases, it was not attempted to include and a secondary phase in the Rietveld analyses of the materials.
Comparing the data of the chemically lithiated material to the data of the pristine counterpart, additional peak broadening is observed for the chemically lithiated material, indicative of domain size decrease and/or increase in microstrain upon chemical lithiation.
\newpage
\setcounter{equation}{0}
\setcounter{figure}{0}
\setcounter{table}{0}
\renewcommand{\theequation}{B\arabic{equation}}
\renewcommand{\thefigure}{B\arabic{figure}}
\renewcommand{\thetable}{B\arabic{table}}
\section{\strumin outputs}
\label{sec:si_strumin}
\subsection*{\normalfont\textbf{Batch 1}}
\label{sec:si_strumin_b1}
\textbf{3 nm: pristine}
\begin{table}
\caption{\normalfont
\strumin output for the \textit{ex situ} PDF data of the batch one pristine material, when setting the composition to \ch{TiO2}. The weighted residual, \rw , the space group, the database from which the CIF originated, the database ID, and the reference in which the CIF was published.
}
\begin{center}
\begin{tabular}{|l|l|l|l|l|}
\hline
$\boldsymbol{R_{\mathrm{w}}}$&\textbf{Space group}&\textbf{Database}&\textbf{ID}&\textbf{Reference}	\\ \Xhline{3\arrayrulewidth}
0.26	& $C2/m$	& MPD	& 554278	& \cite{Feist1992}		\\ \hline
0.61 	& $I4/m$	& MPD 	& 1101022 	& \cite{Jain2013}		\\ \hline
0.74 	& $C2/m$	& COD 	& 1528778 	& \cite{Ouhenia2006}	\\ \hline
0.74	& $P1$		& MPD	& 1245308	& \cite{Aykol2018}		\\ \hline
0.78	& $P1$		& MPD	& 1245134	& \cite{Aykol2018} 		\\ \hline
\end{tabular}
\end{center}
\end{table}
\textbf{3 nm: lithiated}
\begin{table}
\caption{\normalfont
\strumin output for the \textit{ex situ} PDF data of the batch one chemically lithiated material, when setting the composition to Li-Ti-O. The weighted residual, \rw , the chemical formula, the space group, the database from which the CIF originated, the database ID, and the reference in which the CIF was published.
}
\label{table:strumin_3nm_Li}
\begin{center}
\begin{tabular}{|l|l|l|l|l|l|}
\hline
$\boldsymbol{R_{\mathrm{w}}}$&\textbf{Formula}&\textbf{Space group}&\textbf{Database}&\textbf{ID}&\textbf{Reference}	\\ \Xhline{3\arrayrulewidth}
0.37	& \ch{LiTi4O8}	& $C2$		& MPD	& 554278	& \cite{Jain2013} 		\\ \hline
0.46 	& \ch{Li2Ti6O13}& $C2/m$	& COD	& 7206075 	& \cite{Kataoka2011}	\\ \hline
0.63 	& \ch{Li8Ti2O7}	& $P2_{1}/c$& MPD 	& 1526931 	& \cite{Jain2013}		\\ \hline
0.63	& \ch{LiTi8O13} & $R\bar{3}$& MPD	& 2310710	& \cite{Jain2013}		\\ \hline
0.71	& \ch{LiTi2O4} 	& $C2/m$	& MPD	& 9008213	& \cite{Armstrong2010}	\\ \hline
\end{tabular}
\end{center}
\end{table}
\newpage
\textbf{3 nm: lithiated, residual}
\begin{table}
\caption{\normalfont
\strumin output for the \diffpy fit residual for the \textit{ex situ} PDF data of the residual of the batch one chemically lithiated material, when setting the composition to Li-Ti-O. The weighted residual, \rw , the chemical formula, the space group, the database from which the CIF originated, the database ID, and the reference in which the CIF was published.
}
\label{table:strumin_3nm_Li_res}
\begin{center}
\begin{tabular}{|l|l|l|l|l|l|}
\hline
$\boldsymbol{R_{\mathrm{w}}}$&\textbf{Formula}&\textbf{Space group}&\textbf{Database}&\textbf{ID}&\textbf{Reference}	\\ \Xhline{3\arrayrulewidth}
0.89	& \ch{Li7Ti16O32}	& $I\bar{4}2m$	& MPD	& 530141	& \cite{Jain2013}	\\ \hline
0.89	& \ch{LiTi3O4}		& $Cmmm$  		& MPD	& 867744	& \cite{Jain2013} 	\\ \hline
0.90 	& \ch{Li4Ti3O8}		& $C2/m$		& MPD 	& 755266 	& \cite{Jain2013}	\\ \hline
0.90 	& \ch{LiTi8O16}		& $P\bar{4}m2$	& MPD 	& 1222545 	& \cite{Jain2013}	\\ \hline
0.90	& \ch{Li2TiO3} 		& $P\bar{1}$	& MPD	& 760017	& \cite{Jain2013}	\\ \hline
\end{tabular}
\end{center}
\end{table}
\subsection*{\normalfont\textbf{Batch 2}}
\label{sec:si_strumin_b2}
\textbf{3 nm: pristine}
\begin{table}
\caption{\normalfont
\strumin output for the \textit{ex situ} PDF data of the batch two pristine material, when setting the composition to \ch{TiO2}. The weighted residual, \rw , the space group, the database from which the CIF originated, the database ID, and the reference in which the CIF was published.
}
\label{table:strumin_b2_3nm}
\begin{center}
\begin{tabular}{|l|l|l|l|l|}
\hline
$\boldsymbol{R_{\mathrm{w}}}$&\textbf{Space group}&\textbf{Database}&\textbf{ID}&\textbf{Reference}	\\ \Xhline{3\arrayrulewidth}
0.37	& $C2/m$	& MPD	& 554278	& \cite{Feist1992}		\\ \hline
0.71 	& $I4/m$	& MPD 	& 1101022 	& \cite{Jain2013}		\\ \hline
0.72 	& $C2/m$	& COD 	& 1528778 	& \cite{Ouhenia2006}	\\ \hline
0.74	& $P1$		& MPD	& 1245308	& \cite{Aykol2018}		\\ \hline
0.75	& $Pbca$	& COD	& 8104269	& \cite{Pauling1928} 	\\ \hline
\end{tabular}
\end{center}
\end{table}
\textbf{3 nm: pristine, residual}
\begin{table}
\caption{\normalfont
\strumin output for the \diffpy fit residual for the \textit{ex situ} PDF data of the residual of the batch two pristine material, when setting the composition to \ch{TiO2}. The weighted residual, \rw , the space group, the database from which the CIF originated, the database ID, and the reference in which the CIF was published.
}
\label{table:strumin_b2_3nm_res}
\begin{center}
\begin{tabular}{|l|l|l|l|l|}
\hline
$\boldsymbol{R_{\mathrm{w}}}$&\textbf{Space group}&\textbf{Database}&\textbf{ID}&\textbf{Reference}	\\ \Xhline{3\arrayrulewidth}
0.88			& $I4_{1}/amd$	  & COD	 	& 1530151	& \cite{Khitrova1977} 	\\ \hline
0.90 			& $I4_{1}/amd$	  & COD 	& 1010942 	& \cite{Parker1924}		\\ \hline
0.90 			& $I4_{1}/amd$	  & MPD 	& 390	 	& \cite{Jain2013}		\\ \hline
0.90			& $I4_{1}/amd$	  & COD		& 9009086	& \cite{Wyckoff1963}	\\ \hline
0.90			& $I4_{1}/amd$	  & COD		& 9008216	& \cite{Horn1972}		\\ \hline
\end{tabular}
\end{center}
\end{table}
\newpage
\setcounter{equation}{0}
\setcounter{figure}{0}
\setcounter{table}{0}
\renewcommand{\theequation}{C\arabic{equation}}
\renewcommand{\thefigure}{C\arabic{figure}}
\renewcommand{\thetable}{C\arabic{table}}
\section{\textit{Ex situ} PDF refinements}
\label{sec:si_exsitu_pdf}
\subsection*{\normalfont\textbf{Batch one: pristine material}}
\label{sec:si_exsitu_pdf_b1_pristine}
\begin{table}
\makegapedcells
\caption{\normalfont
Results from one-phase refinement of \textit{ex situ} PDF data for the pristine material of batch one.
}
\label{table:si_exsitu_pdf_b1_3nm_1phase}
\begin{center}
\begin{tabular}{|l|l|}
\hline
\textbf{Variable [unit]} & \textbf{Value~$\pm$~esd}	\\ \Xhline{3\arrayrulewidth}
Scale						& $0.35		\pm	0.06$	\\ \hline
\latonea					& $12.17	\pm	0.06$	\\ \hline
\latoneb					& $3.74		\pm	0.02$ 	\\ \hline
\latonec					& $6.49		\pm	0.04$ 	\\ \hline
\latonebeta					& $107.1	\pm	0.5$	\\ \hline
\uisoone{Ti}				& $0.006	\pm	0.002$	\\ \hline
\uisoone{O}					& $0.015	\pm	0.009$	\\ \hline
\deltatwoone				& $2.4		\pm	1.1$	\\ \hline
\cdsone						& $30		\pm	6$		\\ \hline
\atomposone{x}{Ti}{1}{a}	& $0.101	\pm 0.004$	\\ \hline
\atomposone{x}{Ti}{2}{a}	& $0.197	\pm 0.004$	\\ \hline
\atomposone{x}{O}{1}{a}		& $0.06		\pm 0.02$	\\ \hline
\atomposone{x}{O}{2}{a}		& $0.138	\pm 0.009$	\\ \hline
\atomposone{x}{O}{3}{a}		& $0.132	\pm 0.012$	\\ \hline
\atomposone{x}{O}{4}{a}		& $0.240	\pm 0.015$	\\ \hline
\atomposone{z}{Ti}{1}{c}	& $0.707	\pm 0.007$	\\ \hline
\atomposone{z}{Ti}{2}{c}	& $0.288	\pm 0.005$	\\ \hline
\atomposone{z}{O}{1}{c}		& $0.37		\pm 0.02$	\\ \hline
\atomposone{z}{O}{2}{c}		& $0.011	\pm 0.001$	\\ \hline
\atomposone{z}{O}{3}{c}		& $0.71		\pm 0.02$	\\ \hline
\atomposone{z}{O}{4}{c}		& $0.35		\pm 0.02$	\\ \Xhline{3\arrayrulewidth}
\rw							& $0.15$				\\ \hline
\end{tabular}
\end{center}
\end{table}
\newpage
\subsection*{\normalfont\textbf{Batch one: lithiated material (two \ch{Li_{x}TiO2} phases)}}
\label{sec:si_exsitu_pdf_b1_lithiated}
\textbf{\ch{Li_{x}TiO2}-bronze}
\begin{table}
\makegapedcells
\caption{\normalfont
Results for the lithiated bronze phase from two-phase refinement of \textit{ex situ} PDF data for the chemically lithiated material of batch one. Please see \tabl{si_exsitu_pdf_b1_3nmLi_anatase} for results of the lithiated anatase phase and the weighted residual.
}
\label{table:si_exsitu_pdf_b1_3nmLi_bronze}
\begin{center}
\begin{tabular}{|l|l|}
\hline
\textbf{Variable [unit]}	& \textbf{Value~$\pm$~esd} 	\\ \Xhline{3\arrayrulewidth} 
Scale factor				& $0.32		\pm	0.09$ 	\\ \hline
\latonea					& $12.3		\pm	0.1$	\\ \hline 	
\latoneb					& $3.79		\pm	0.03$ 	\\ \hline
\latonec					& $6.45		\pm	0.08$ 	\\ \hline
\latonebeta					& $107		\pm 1$	\\ \hline
\uisoone{Ti}				& $0.009	\pm	0.005$	\\ \hline
\uisoone{O}					& $0.02		\pm	0.03$	\\ \hline
\deltatwoone				& $2		\pm	2$		\\ \hline
\cdsone						& $26		\pm	7$		\\ \hline
\atomposone{x}{Ti}{1}{a}	& $0.103	\pm 0.007$	\\ \hline
\atomposone{x}{Ti}{2}{a}	& $0.201	\pm 0.007$	\\ \hline
\atomposone{x}{O}{1}{a}		& $0.06		\pm 0.04$	\\ \hline
\atomposone{x}{O}{2}{a}		& $0.11		\pm 0.02$	\\ \hline
\atomposone{x}{O}{3}{a}		& $0.16		\pm 0.04$	\\ \hline
\atomposone{x}{O}{4}{a}		& $0.25		\pm 0.02$	\\ \hline
\atomposone{y}{Ti}{1}{a}	& $0.54		\pm 0.08$	\\ \hline
\atomposone{y}{Ti}{2}{a}	& $0.54		\pm 0.10$	\\ \hline
\atomposone{y}{O}{1}{a}		& $0.5355	\pm 0.0010$	\\ \hline
\atomposone{y}{O}{2}{a}		& $0.050	\pm 0.002$	\\ \hline
\atomposone{y}{O}{3}{a}		& $0.4458	\pm 0.0010$	\\ \hline
\atomposone{y}{O}{4}{a}		& $0.02		\pm 0.1$	\\ \hline
\atomposone{z}{Ti}{1}{c}	& $0.708	\pm 0.007$	\\ \hline
\atomposone{z}{Ti}{2}{c}	& $0.287	\pm 0.006$	\\ \hline
\atomposone{z}{O}{1}{c}		& $0.35		\pm 0.03$	\\ \hline
\atomposone{z}{O}{2}{c}		& $0.01		\pm 0.03$	\\ \hline
\atomposone{z}{O}{3}{c}		& $0.17		\pm 0.03$	\\ \hline
\atomposone{z}{O}{4}{c}		& $0.36		\pm 0.03$	\\ \hline
Weight frac.				& $0.85$				\\ \hline
\end{tabular}
\end{center}
\end{table}
\newpage
\textbf{\ch{Li_{x}TiO2}-anatase}
\begin{table}
\makegapedcells
\caption{\normalfont
Results for the lithiated anatase phase from two-phase refinement of \textit{ex situ} PDF data for the chemically lithiated material of batch one. Please see \tabl{si_exsitu_pdf_b1_3nmLi_bronze} for results of the lithiated bronze phase.}
\label{table:si_exsitu_pdf_b1_3nmLi_anatase}
\begin{center}
\begin{tabular}{|l|l|}
\hline
\textbf{Variable [unit]}	& \textbf{Value~$\pm$~esd} 	\\ \Xhline{3\arrayrulewidth}
Scale factor	& $0.06		\pm	0.06$	\\ \hline
\latonea		& $8.2		\pm	0.2$	\\ \hline
\latonec		& $17.7		\pm	0.6$ 	\\ \hline
\uisoone{Ti}	& $0.01		\pm	0.03$	\\ \hline
\uisoone{O}		& $0.005	\pm	0.031$	\\ \hline
\deltatwoone	& $3.8		\pm	0.2$	\\ \hline
\cdsone			& $20		\pm	19$		\\ \hline
Weight frac.	& $0.15$				\\ \Xhline{3\arrayrulewidth}
\rw				& 0.16					\\ \hline
\end{tabular}
\end{center}
\end{table}
\newpage
\subsection*{\normalfont\textbf{Batch two: pristine (two \ch{TiO2} phases)}}
\label{sec:si_exsitu_pdf_b2_3nm}
\textbf{\ch{TiO2}-bronze}
\begin{table}
\makegapedcells
\caption{\normalfont
Results for the bronze phase from two-phase refinement of \textit{ex situ} PDF data for the pristine sample of batch two. Please see \tabl{si_exsitu_pdf_b2_2phase_anatase} for results of the anatase phase.
}
\label{table:si_exsitu_pdf_b2_2phase_bronze}
\begin{center}
\begin{tabular}{|l|l|}
\hline
\textbf{Variable [unit]} & \textbf{Value~$\pm$~esd}	\\ \Xhline{3\arrayrulewidth}
Scale						& $0.32		\pm	0.07$	\\ \hline
\latonea					& $12.14	\pm	0.08$	\\ \hline
\latoneb					& $3.75		\pm	0.02$ 	\\ \hline
\latonec					& $6.51		\pm	0.05$ 	\\ \hline
\latonebeta					& $107.0	\pm	0.7$	\\ \hline
\uisoone{Ti}				& $0.005	\pm	0.002$	\\ \hline
\uisoone{O}					& $0.012	\pm	0.009$	\\ \hline
\deltatwoone				& $3		\pm	3$	\\ \hline
\cdsone						& $25		\pm	6$		\\ \hline
\atomposone{x}{Ti}{1}{a}	& $0.100	\pm 0.005$	\\ \hline
\atomposone{x}{Ti}{2}{a}	& $0.197	\pm 0.004$	\\ \hline
\atomposone{x}{O}{1}{a}		& $0.06		\pm 0.02$	\\ \hline
\atomposone{x}{O}{2}{a}		& $0.138	\pm 0.012$	\\ \hline
\atomposone{x}{O}{3}{a}		& $0.131	\pm 0.013$	\\ \hline
\atomposone{x}{O}{4}{a}		& $0.241	\pm 0.014$	\\ \hline
\atomposone{z}{Ti}{1}{c}	& $0.706	\pm 0.009$	\\ \hline
\atomposone{z}{Ti}{2}{c}	& $0.289	\pm 0.006$	\\ \hline
\atomposone{z}{O}{1}{c}		& $0.37		\pm 0.02$	\\ \hline
\atomposone{z}{O}{2}{c}		& $0.01		\pm 0.03$	\\ \hline
\atomposone{z}{O}{3}{c}		& $0.71		\pm 0.02$	\\ \hline
\atomposone{z}{O}{4}{c}		& $0.36		\pm 0.03$	\\ \Xhline{3\arrayrulewidth}
Weight frac.				& $0.85$				\\ \Xhline{3\arrayrulewidth}
\rw							& $0.17$				\\ \hline
\end{tabular}
\end{center}
\end{table}
\newpage
\textbf{\ch{TiO2}-anatase}
\begin{table}
\makegapedcells
\caption{\normalfont
Results for the anatase phase from two-phase refinement of \textit{ex situ} PDF data for the pristine material of batch two. Please see \tabl{si_exsitu_pdf_b2_2phase_bronze} for results of the bronze phase.
}
\label{table:si_exsitu_pdf_b2_2phase_anatase}
\begin{center}
\begin{tabular}{|l|l|}
\hline
\textbf{Variable [unit]}	& \textbf{Value~$\pm$~esd} 	\\ \Xhline{3\arrayrulewidth}
Scale factor	& $0.06		\pm	0.05$	\\ \hline
\latonea		& $3.8		\pm	0.5$	\\ \hline
\latonec		& $9.6		\pm	0.2$ 	\\ \hline
\uisoone{Ti}	& $0.01		\pm	0.01$	\\ \hline
\uisoone{O}		& $0.04		\pm	0.08$	\\ \hline
\deltatwoone	& $4		\pm	2$		\\ \hline
\cdsone			& $40		\pm	40$		\\ \hline
Weight frac.	& $0.15$				\\ \Xhline{3\arrayrulewidth}
\rw				& 0.17					\\ \hline
\end{tabular}
\end{center}
\end{table}
\newpage
\subsection*{\normalfont\textbf{Batch two: cathode composite (two \ch{TiO2} phases and a modified graphite phase)}}
\label{sec:si_exsitu_pdf_b2_3nm_pm}
\textbf{\ch{TiO2}-bronze}
\begin{table}
\makegapedcells
\caption{\normalfont
Results for the bronze phase from the refinement of the \textit{ex situ} PDF data for the cathode composite. Please see \tabl{si_exsitu_pdf_b2_2phase_pm_anatase} and \tabl{si_exsitu_pdf_b2_2phase_pm_graphite} for the results of the anatase and modified graphite phases, respectively.
}
\label{table:si_exsitu_pdf_b2_2phase_pm_bronze}
\begin{center}
\begin{tabular}{|l|l|}
\hline
\textbf{Variable [unit]} & \textbf{Value~$\pm$~esd}	\\ \Xhline{3\arrayrulewidth}
Scale						& $0.24		\pm	0.07$	\\ \hline
\latonea					& $12.14	\pm	0.09$	\\ \hline
\latoneb					& $3.75		\pm	0.03$ 	\\ \hline
\latonec					& $6.50		\pm	0.07$ 	\\ \hline
\latonebeta					& $107.0	\pm	0.8$	\\ \hline
\uisoone{Ti}				& $0.008	\pm	0.005$	\\ \hline
\uisoone{O}					& $0.002	\pm	0.005$	\\ \hline
\deltatwoone				& $1		\pm	5$		\\ \hline
\cdsone						& $26		\pm	8$		\\ \hline
\atomposone{x}{Ti}{1}{a}	& $0.100	\pm 0.006$	\\ \hline
\atomposone{x}{Ti}{2}{a}	& $0.192	\pm 0.006$	\\ \hline
\atomposone{x}{O}{1}{a}		& $0.058	\pm 0.010$	\\ \hline
\atomposone{x}{O}{2}{a}		& $0.116	\pm 0.011$	\\ \hline
\atomposone{x}{O}{3}{a}		& $0.125	\pm 0.010$	\\ \hline
\atomposone{x}{O}{4}{a}		& $0.235	\pm 0.010$	\\ \hline
\atomposone{z}{Ti}{1}{c}	& $0.692	\pm 0.014$	\\ \hline
\atomposone{z}{Ti}{2}{c}	& $0.282	\pm 0.010$	\\ \hline
\atomposone{z}{O}{1}{c}		& $0.39		\pm 0.02$	\\ \hline
\atomposone{z}{O}{2}{c}		& $-0.3		\pm 0.02$	\\ \hline
\atomposone{z}{O}{3}{c}		& $0.71		\pm 0.02$	\\ \hline
\atomposone{z}{O}{4}{c}		& $0.35		\pm 0.02$	\\ \Xhline{3\arrayrulewidth}
Weight frac.				& $0.80$				\\ \Xhline{3\arrayrulewidth}
\rw							& $0.25$				\\ \hline
\end{tabular}
\end{center}
\end{table}
\newpage
\textbf{\ch{TiO2}-anatase}
\begin{table}
\makegapedcells
\caption{\normalfont
Results for the anatase phase from the refinement of the \textit{ex situ} PDF data for the cathode composite. Please see \tabl{si_exsitu_pdf_b2_2phase_pm_bronze} and \tabl{si_exsitu_pdf_b2_2phase_pm_graphite} for the results of the bronze and modified graphite phases, respectively.
}
\label{table:si_exsitu_pdf_b2_2phase_pm_anatase}
\begin{center}
\begin{tabular}{|l|l|}
\hline
\textbf{Variable [unit]}	& \textbf{Value~$\pm$~esd} 	\\ \Xhline{3\arrayrulewidth}
Scale factor	& $0.06		\pm	0.05$	\\ \hline
\latonea		& $3.8		\pm	0.5$	\\ \hline
\latonec		& $9.6		\pm	0.2$ 	\\ \hline
\uisoone{Ti}	& $0.01		\pm	0.02$	\\ \hline
\uisoone{O}		& $0.2		\pm	0.2$	\\ \hline
\deltatwoone	& $4		\pm	1$		\\ \hline
\cdsone			& $40		\pm	60$		\\ \hline
Weight frac.	& $0.15$				\\ \Xhline{3\arrayrulewidth}
\rw				& 0.17					\\ \hline
\end{tabular}
\end{center}
\end{table}
\textbf{Modified graphite}
\begin{table}
\makegapedcells
\caption{\normalfont
Results for the modified graphite phase from the refinement of the \textit{ex situ} PDF data for the cathode composite. Please see \tabl{si_exsitu_pdf_b2_2phase_pm_bronze} and \tabl{si_exsitu_pdf_b2_2phase_pm_anatase} for results of the bronze and anatase phases, respectively.
}
\label{table:si_exsitu_pdf_b2_2phase_pm_graphite}
\begin{center}
\begin{tabular}{|l|l|}
\hline
\textbf{Variable [unit]}	& \textbf{Value~$\pm$~esd} 	\\ \Xhline{3\arrayrulewidth}
Scale factor	& $0.04		\pm	0.02$	\\ \hline
\latonea		& $2.457	\pm	0.013$	\\ \hline
\uisoone{C}		& $0.001	\pm	0.003$	\\ \hline
\deltatwoone	& $2.01		\pm	0.03$	\\ \hline
\cdsone			& $40		\pm	40$		\\ \hline
\rw				& 0.17					\\ \hline
\end{tabular}
\end{center}
\end{table}
\newpage
\subsection*{\normalfont\textbf{3 nm: \textit{operando} first frame (two \ch{TiO2} phases and a modified graphite phase)}}
\label{sec:si_operando_pdf_b2_3nm_first_frame}
\textbf{\ch{TiO2}-bronze}
\begin{table}
\makegapedcells
\caption{\normalfont
Results for the bronze phase from the refinement of the first \textit{operando} frame for the batch two material. Please see \tabl{si_exsitu_pdf_b2_2phase_operando_anatase} and \tabl{si_exsitu_pdf_b2_2phase_operando_graphite} for the results of the anatase and modified graphite phases, respectively.
}
\label{table:si_exsitu_pdf_b2_2phase_operando_bronze}
\begin{center}
\begin{tabular}{|l|l|}
\hline
\textbf{Variable [unit]} & \textbf{Value~$\pm$~esd}	\\ \Xhline{3\arrayrulewidth}
Scale						& $0.17		\pm	0.06$	\\ \hline
\latonea					& $12.16	\pm	0.14$	\\ \hline
\latoneb					& $3.78		\pm	0.04$ 	\\ \hline
\latonec					& $6.54		\pm	0.08$ 	\\ \hline
\latonebeta					& $107.3	\pm	1.2$	\\ \hline
\uisoone{Ti}				& $0.006	\pm	0.005$	\\ \hline
\uisoone{O}					& $0.01	\pm	0.02$	\\ \hline
\deltatwoone				& $1		\pm	5$		\\ \hline
\cdsone						& $26		\pm	12$		\\ \Xhline{3\arrayrulewidth}
Weight frac.				& $0.80$				\\ \Xhline{3\arrayrulewidth}
\rw							& $0.35$				\\ \hline
\end{tabular}
\end{center}
\end{table}
\textbf{\ch{TiO2}-anatase}
\begin{table}
\makegapedcells
\caption{\normalfont
Results for the anatase phase from the refinement of the first \textit{operando} frame for the batch two material. Please see \tabl{si_exsitu_pdf_b2_2phase_operando_bronze} and \tabl{si_exsitu_pdf_b2_2phase_operando_graphite} for the results of the bronze and modified graphite phases, respectively.
}
\label{table:si_exsitu_pdf_b2_2phase_operando_anatase}
\begin{center}
\begin{tabular}{|l|l|}
\hline
\textbf{Variable [unit]}	& \textbf{Value~$\pm$~esd} 	\\ \Xhline{3\arrayrulewidth}
Scale factor	& $0.06		\pm	0.05$	\\ \hline
\latonea		& $3.8		\pm	0.5$	\\ \hline
\latonec		& $9.6		\pm	0.2$ 	\\ \hline
\uisoone{Ti}	& $0.01		\pm	0.02$	\\ \hline
\uisoone{O}		& $0.2		\pm	0.2$	\\ \hline
\deltatwoone	& $4		\pm	1$		\\ \hline
\cdsone			& $40		\pm	60$		\\ \hline
Weight frac.	& $0.15$				\\ \Xhline{3\arrayrulewidth}
\rw				& 0.17					\\ \hline
\end{tabular}
\end{center}
\end{table}
\newpage
\textbf{Modified graphite}
\begin{table}
\makegapedcells
\caption{\normalfont
Results for the modified graphite phase from the refinement of the first \textit{operando} frame for the batch two material. Please see \tabl{si_exsitu_pdf_b2_2phase_operando_bronze} and \tabl{si_exsitu_pdf_b2_2phase_operando_anatase} for the results of the bronze and anatase phases, respectively.
}
\label{table:si_exsitu_pdf_b2_2phase_operando_graphite}
\begin{center}
\begin{tabular}{|l|l|}
\hline
\textbf{Variable [unit]}	& \textbf{Value~$\pm$~esd} 	\\ \Xhline{3\arrayrulewidth}
Scale factor	& $0.04		\pm	0.02$	\\ \hline
\latonea		& $2.457	\pm	0.013$	\\ \hline
\uisoone{C}		& $0.001	\pm	0.003$	\\ \hline
\deltatwoone	& $2.01		\pm	0.03$	\\ \hline
\cdsone			& $40		\pm	40$		\\ \hline
\rw				& 0.17					\\ \hline
\end{tabular}
\end{center}
\end{table}
\newpage
\setcounter{equation}{0}
\setcounter{figure}{0}
\setcounter{table}{0}
\renewcommand{\theequation}{D\arabic{equation}}
\renewcommand{\thefigure}{D\arabic{figure}}
\renewcommand{\thetable}{D\arabic{table}}
\section{Noise-filtering using principal component analysis (PCA)}
\label{sec:si_pca}
\subsection*{\normalfont\textbf{Real space}}
\textbf{Explained variance ratio for PCA of $\boldsymbol{G(r)}$}

\begin{figure}
	\center
	\includegraphics[width=\columnwidth]{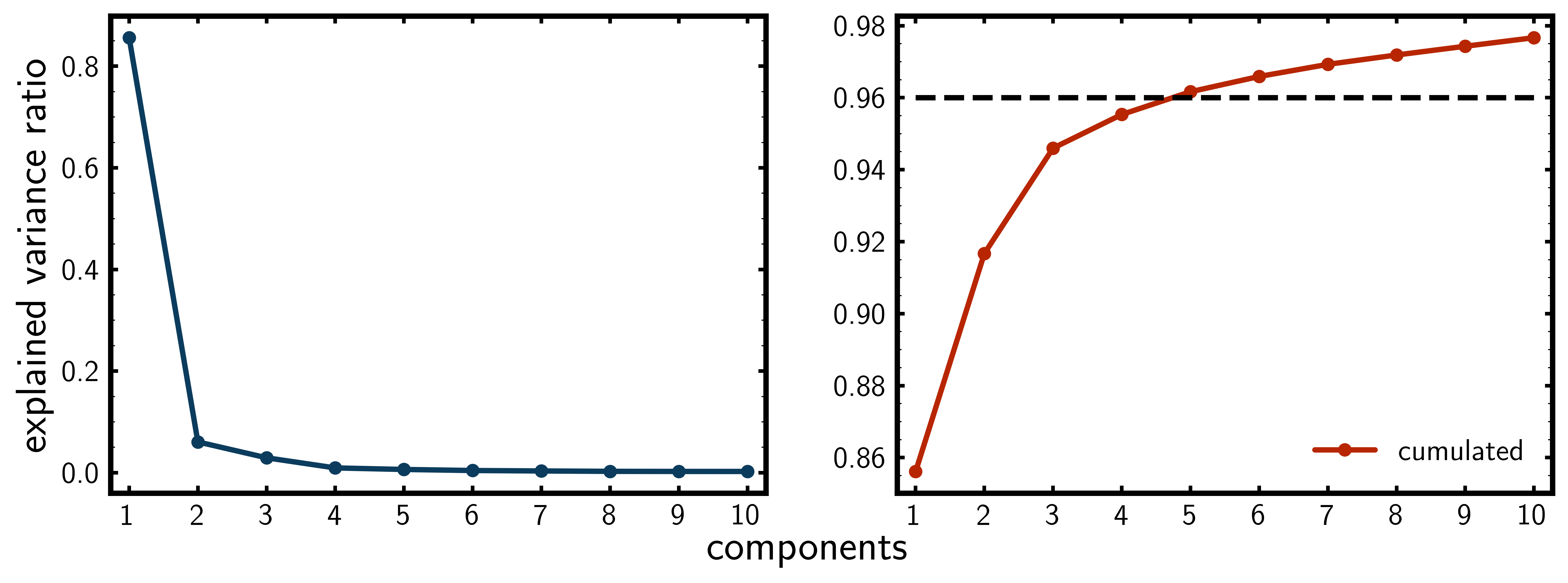}
	\caption{
	Explained variance ratio and cumulated explained variance ratio as a function of the number of principal components. The value of 0.96 used for noise filtering is marked with a black dashed line.
	\label{fig:pca_evr_cevr_gr}
	}
\end{figure}

\fig{pca_evr_cevr_gr} displays the explained variance ratio and its cumulated version for PCA of the \textit{operando} PDF data. A kink at four components is observable, indicating that four components are probably needed to describe the trends in the \textit{operando} PDF data. The dashed line to the right indicates that four components are included if including 0.96 of the cumulated explained variance ratio for the PCA.
\newpage
\textbf{Difference between experimental and PCA-reconstructed $\boldsymbol{G(r)}$}

\begin{figure}
	\center
	\includegraphics[width=0.8\columnwidth]{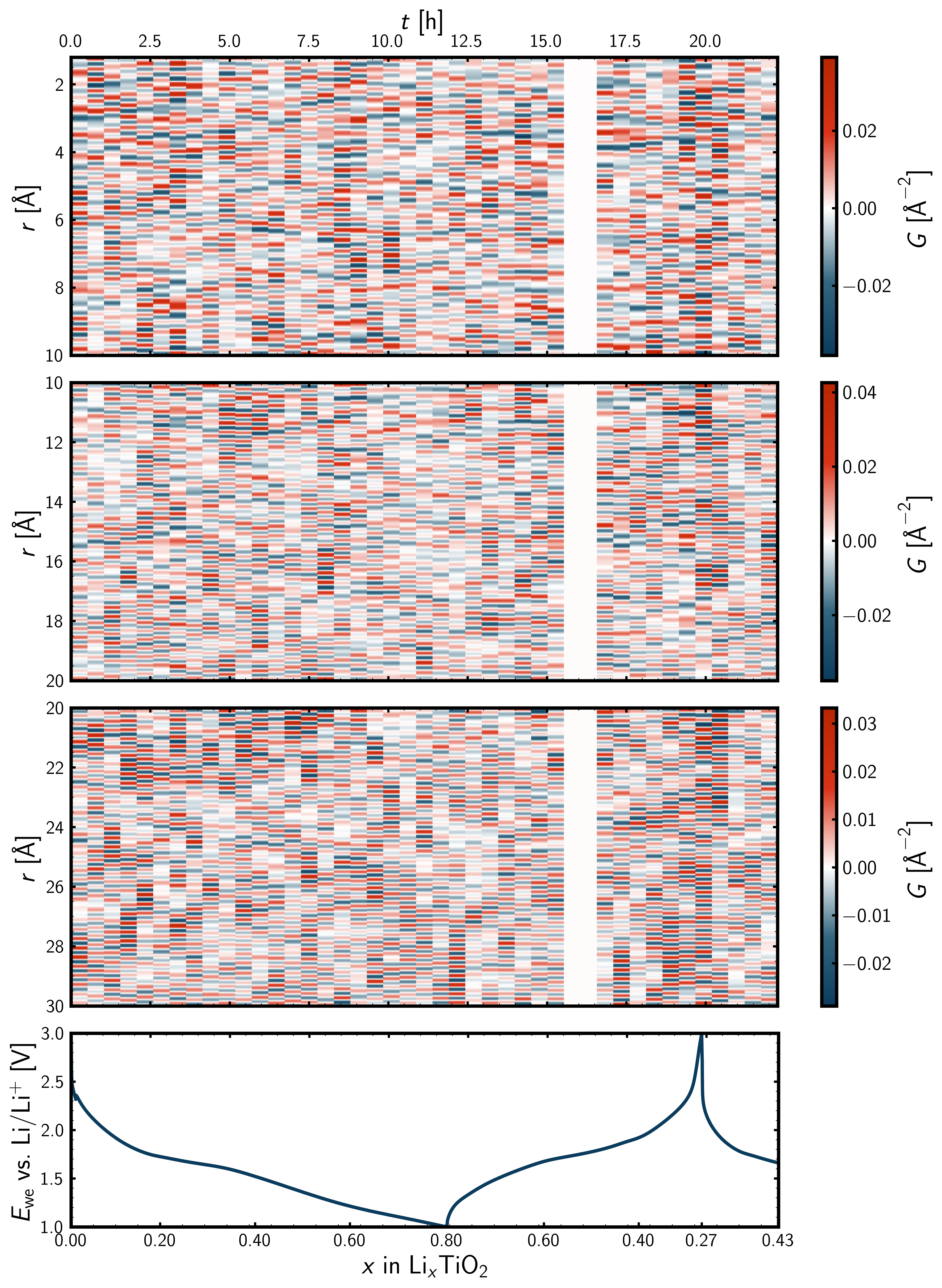}
	\caption{
Difference between the experimental and the PCA reconstruction of the reduced atomic pair distribution function, $G(r)$, together with the galvanostatic cycling.
The $G(r)$ data have been divided into three $r$-regions, each region having its own colorbar to improve the visualization.
The white column is due to absence of synchrotron x-rays during the \textit{operando} experiment.
	}
	\label{fig:gr_gr_gr_pca-recon-diff_echem}	
\end{figure}

\fig{gr_gr_gr_pca-recon-diff_echem} displays the between the experimental \textit{operando} PDF data and the PCA-reconstructed data. As the difference appears structureless and relatively constant with $r$, it does look like noise, as desired.
\newpage
\textbf{Pearson correlation coefficients between experimental and PCA-reconstructed $\boldsymbol{G(r)}$}

\begin{figure}
	\center
	\includegraphics[width=0.9\columnwidth]{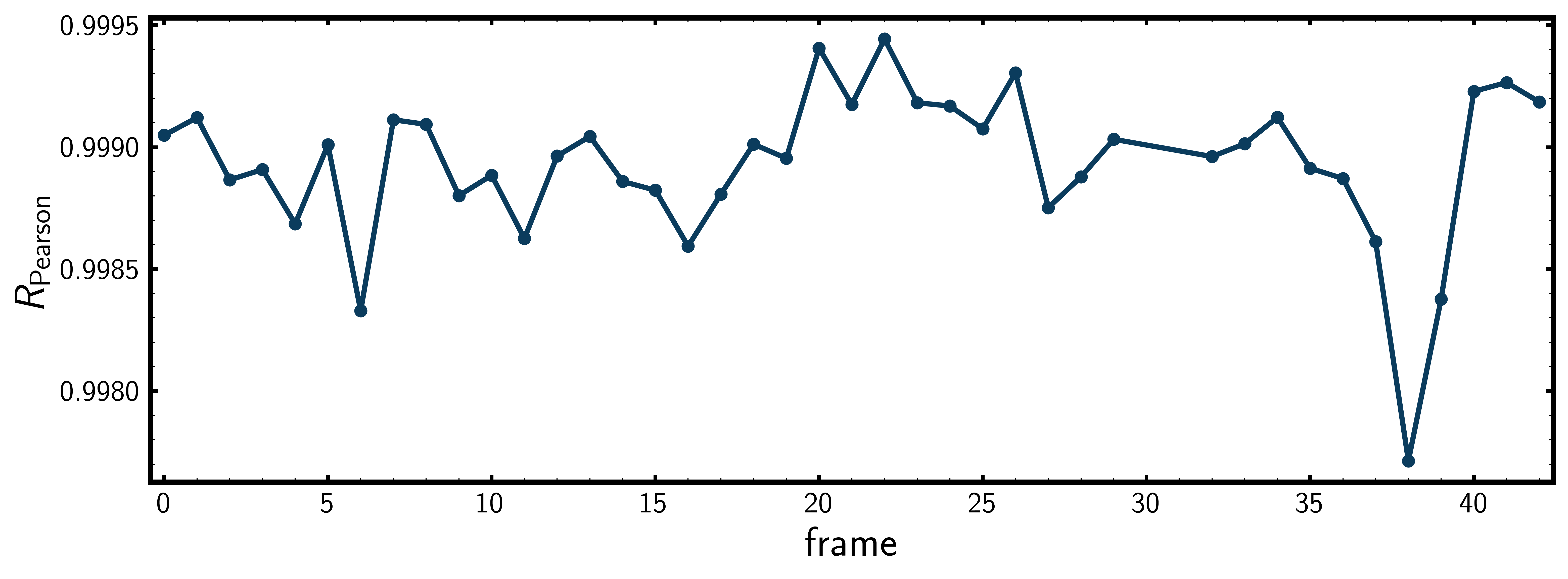}
	\caption{
	Pearson correlation analysis of experimental and PCA-reconstructed reduced atomic pair distribution function data for all frames during the \textit{operando} experiment.
	}
	\label{fig:pearson_pca_recon_gr}
\end{figure}

\fig{pearson_pca_recon_gr} displays the Pearson correlation coefficient between the experimental PDF and the PCA-reconstruction as a function of \textit{operando} PDF frame number. All correlation coefficients are close to unity, revealing highly similarity as desired.
\textbf{Experimental, PCA-reconstructed, and difference $\boldsymbol{G(r)}$}

\begin{figure}
	\center
	\includegraphics[width=0.9\columnwidth]{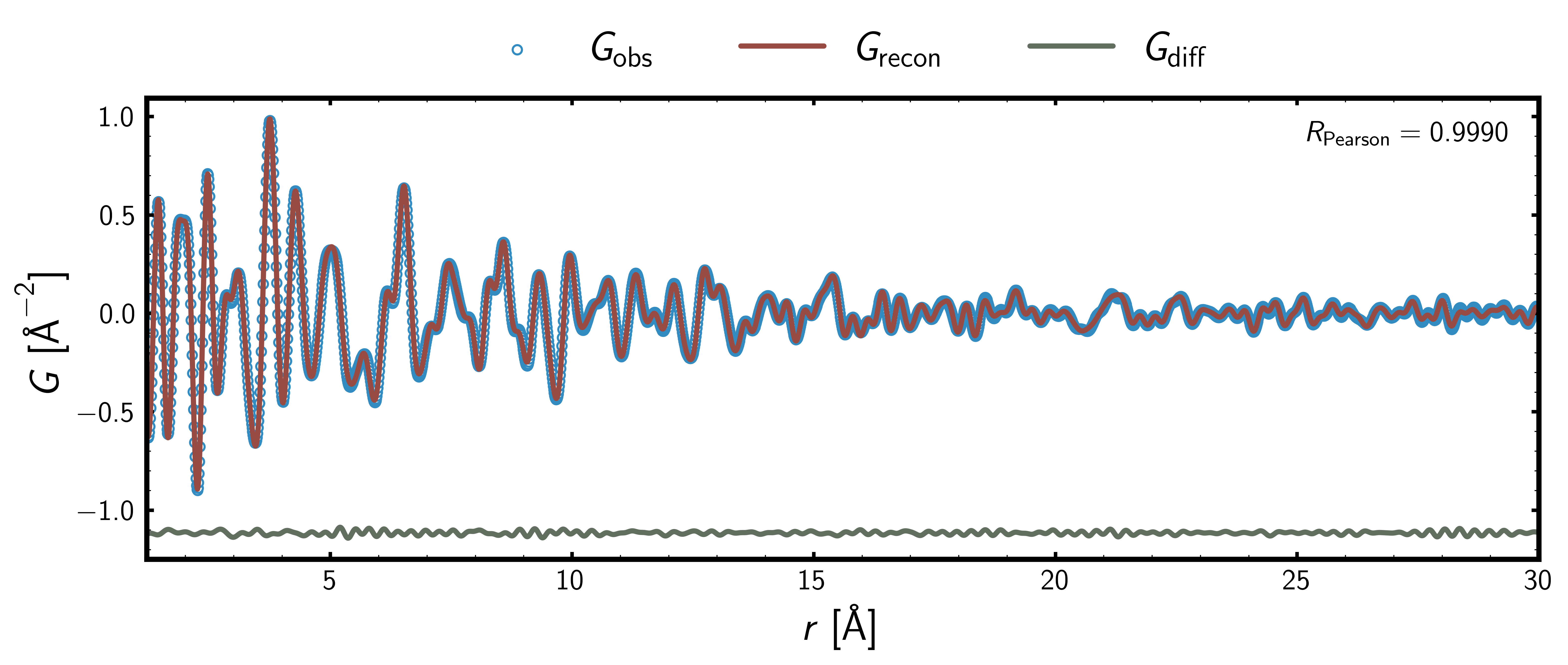}
	\caption{
	Experimental reduced atomic pair distribution function data, PCA-reconstructed data, and the difference of the two, for the first frame of the \textit{operando} experiment.
	The Pearson correlation coefficient is shown in the upper right corner.	
	}
	\label{fig:00_gr_obs_pca-recon_diff_pearson}
\end{figure}

\fig{00_gr_obs_pca-recon_diff_pearson} displays the experimental PDF, the PCA-reconstruction, and their difference for the first \textit{operando} PDF frame.
\newpage
\subsection*{\normalfont\textbf{Reciprocal space}}
\label{sec:si_pca_qspace}
\textbf{Explained variance ratio for PCA of $\boldsymbol{F(Q)}$}

\fig{pca_evr_cevr_gr} shows the explained variance ratio and the cumulated explained variance ratio as a function of the number of principal components for the \textit{operando} reduced total scattering structure function data. In both cases, a kink at four components is observed. For the cumulated explained variance ratio, a value of 0.93 for the PCA will ensure that enough of the trends in the experimental data are included in the reconstruction.
\begin{figure}
	\center
	\includegraphics[width=\columnwidth]{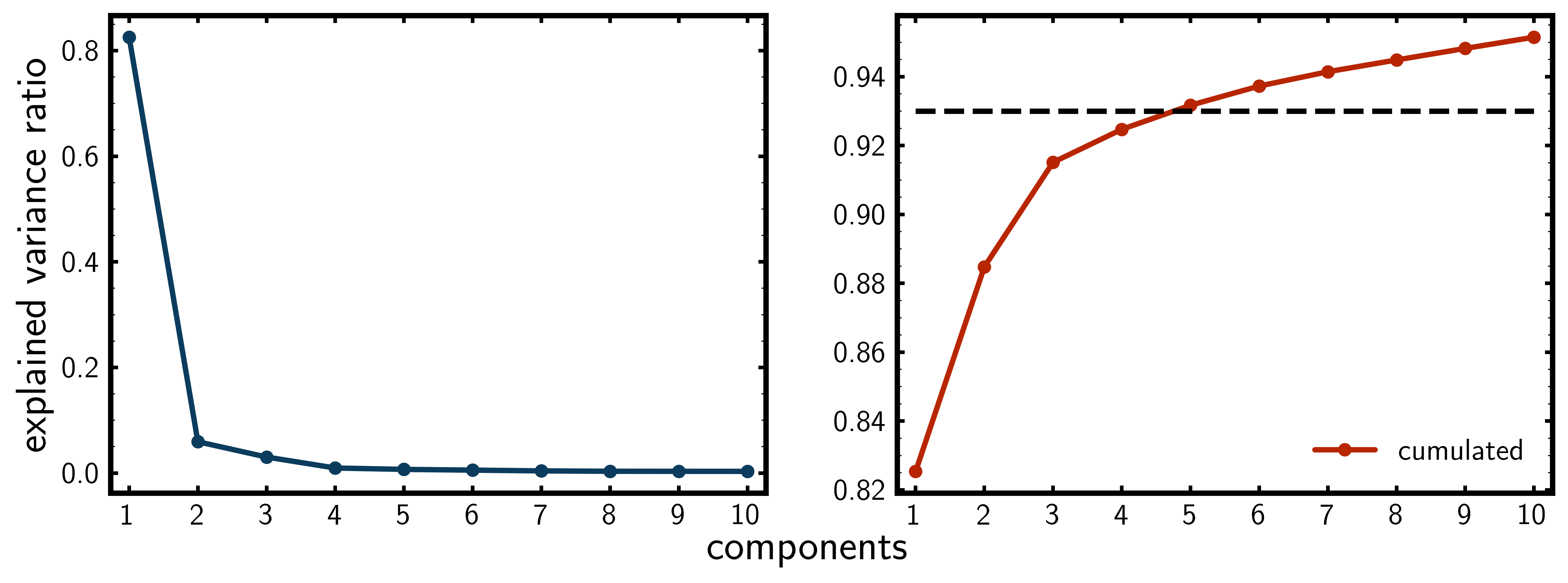}
	\caption{
	Explained variance ratio and cumulated explained variance ratio as a function of the number of principal components. The value of 0.93 used for noise filtering is marked with a black dashed line.
	\label{fig:pca_evr_cevr_fq}
	}
\end{figure}
\newpage
\textbf{Experimental and PCA-reconstructed $\boldsymbol{F(Q)}$}

\begin{figure}
	\center
	\includegraphics[width=0.8\columnwidth]{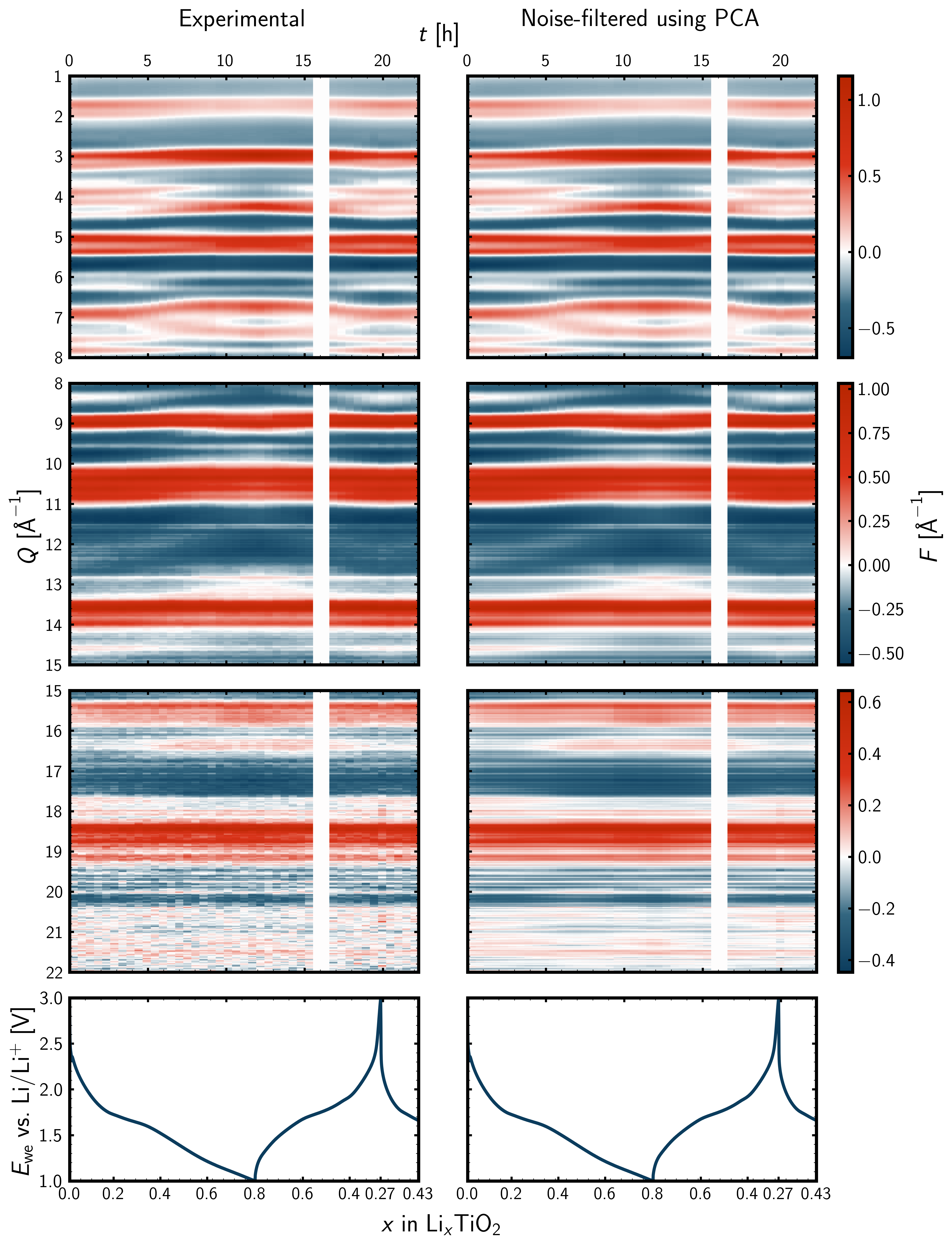}
	\caption{
Experimental (left) and noise-filtered (PCA-reconstructed, right) \textit{operando} reduced total scattering structure function, $F(Q)$, together with the galvanostatic cycling.
The $F(Q)$ data have been divided into three $Q$-regions, each region having its own colorbar to improve the visualization.
The white column is due to absence of synchrotron x-rays during the \textit{operando} experiment.
	}
	\label{fig:fq_fq-pca_echem}	
\end{figure}

\fig{fq_fq-pca_echem} shows the PCA reconstruction of the $F(Q)$ data together with the galvanostatic cycling data.
That the PCA serves as a noise filter is clearly seen for the high-$Q$ region, as the unfiltered signal suffers from a lower signal-to-noise ratio there, resulting from the x-ray atomic form factor.
\newpage
\textbf{Difference between experimental and PCA-reconstructed $\boldsymbol{F(Q)}$}

\begin{figure}
	\center
	\includegraphics[width=0.95\columnwidth]{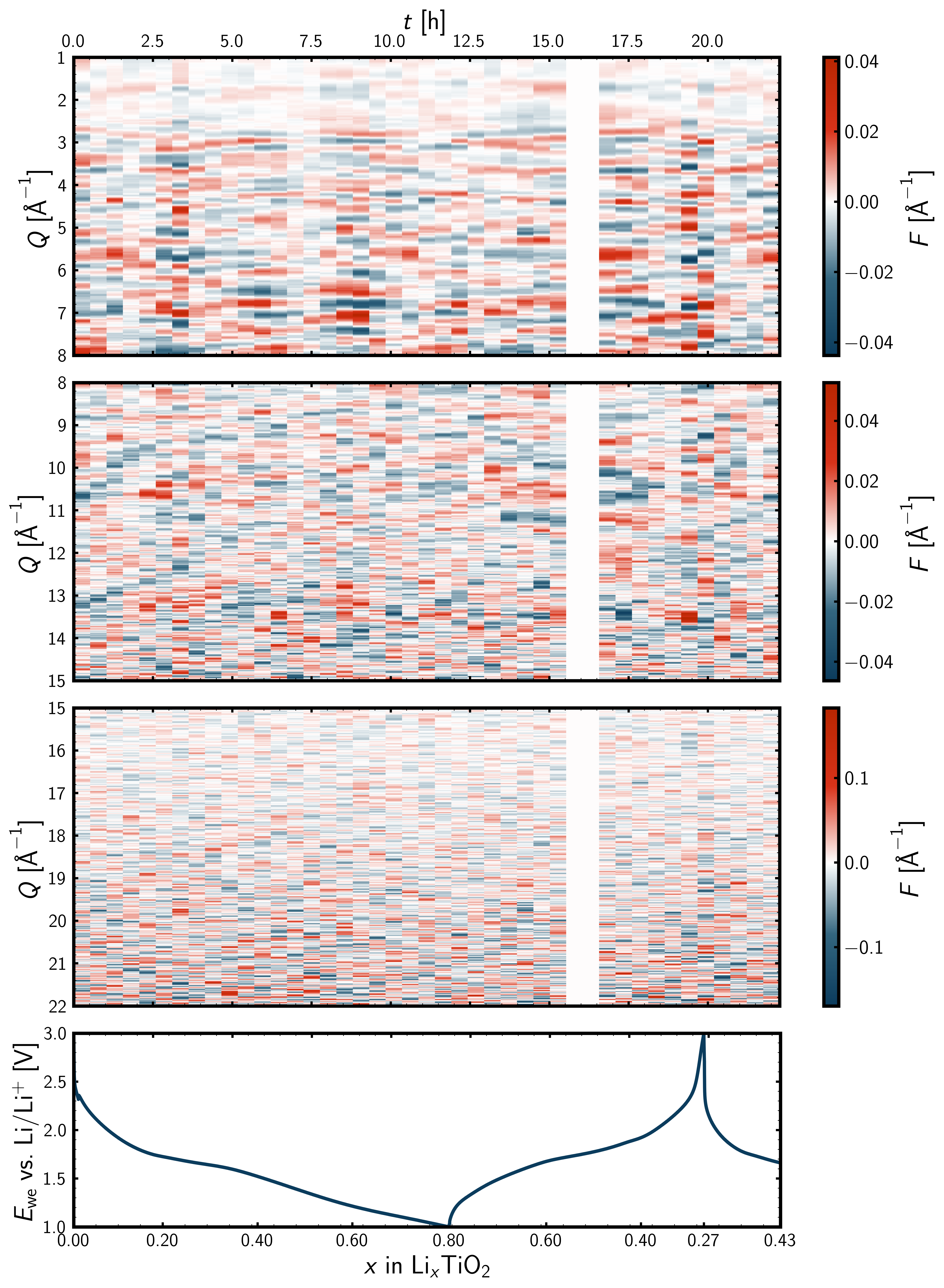}
	\caption{
Difference between the experimental and the PCA reconstruction of the reduced total scattering structure function, $F(Q)$, together with the galvanostatic cycling.
The $F(Q)$ data have been divided into three $Q$-regions, each region having its own colorbar to improve the visualization.
The white column is due to absence of synchrotron x-rays during the \textit{operando} experiment.
	}
	\label{fig:fq_fq_fq_pca-recon-diff_echem}	
\end{figure}

\fig{fq_fq_fq_pca-recon-diff_echem} shows the difference between the experimental data and the PCA reconstruction, i.e., the filtered noise, of the reduced total scattering structure function, $F(Q)$, together with the galvanostatic cycling data.
As desired, the part of the signal that is filtered appears structureless, i.e, behaves as noise.
From the relative trends within each subplot, the noise-level increases momentum transfer, $Q$.
Comparing the color scales of the subplots, the noise-level also increases with momentum transfer in absolute terms.
This arises from the multiplication with $Q$ during the last step in the data processing to obtain the reduced total scattering structure function, $I(Q) \rightarrow S(Q) \rightarrow F(Q)=Q[S(Q)-1]$.
\newpage
\textbf{Pearson correlation coefficients between experimental and PCA-reconstructed $\boldsymbol{F(Q)}$}

\begin{figure}
	\center
	\includegraphics[width=0.85\columnwidth]{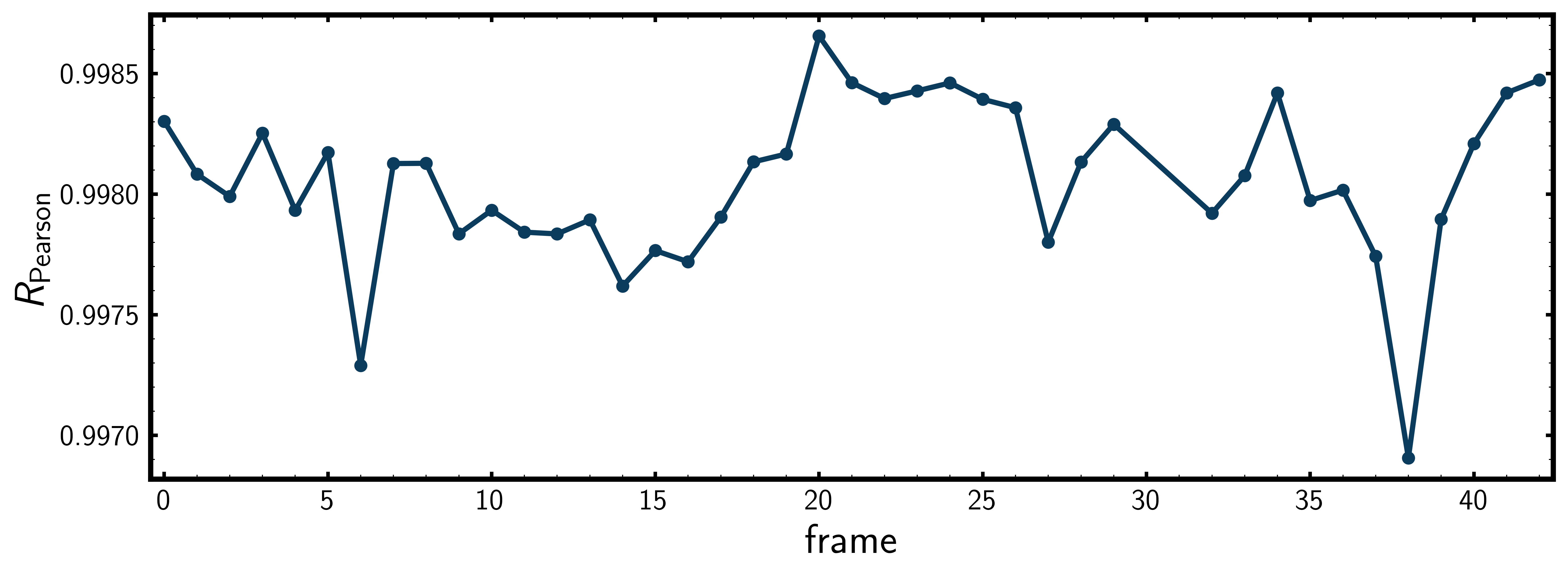}
	\caption{
	Pearson correlation analysis of experimental and PCA-reconstructed reduced total scattering structure function data for all frames during the \textit{operando} experiment.
	}
	\label{fig:pearson_pca_recon_fq}
\end{figure}

\fig{pearson_pca_recon_fq} displays the Pearson correlation coefficient for the experimental and PCA-reconstructed reduced total scattering structure function data as a function of frame number during the \textit{operando} experiment.
All correlation coefficients are above 0.99 and the patterns are practically speaking identical. The minor deviation from a value of unity reflects that the small difference from the filtering of the experimental data.

\begin{figure}
	\center
	\includegraphics[width=0.85\columnwidth]{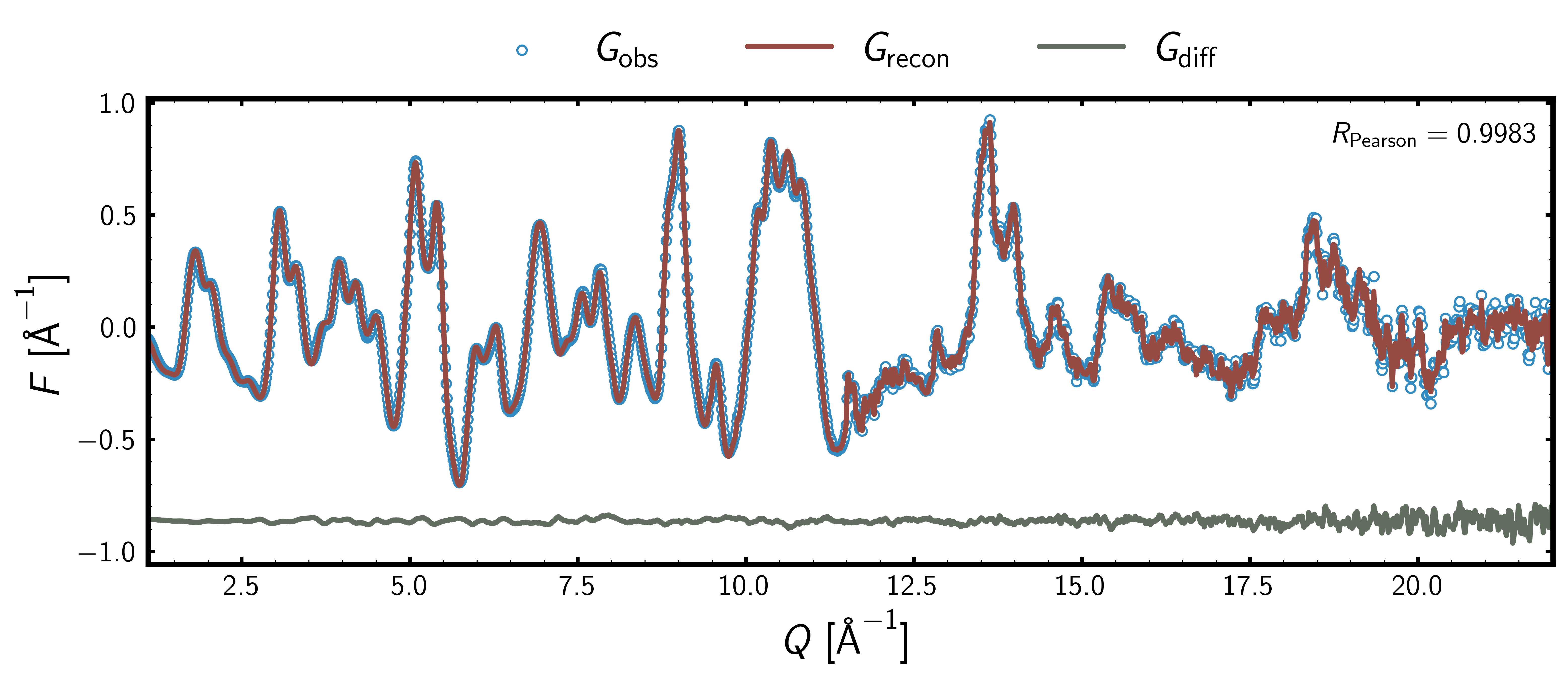}
	\caption{
	Experimental reduced total scattering structure function data, PCA-reconstructed data, and the difference of the two, for the first frame of the \textit{operando} experiment.
	The Pearson correlation coefficient is shown in the upper right corner.
	}
	\label{fig:00_fq_obs_pca-recon_diff_pearson}
\end{figure}

Frame 38 during the last part of the experiment stands out. However, re-inspection of the metadata of the raw data revealed that the exposure time for this frame was only one-fourth of that of the remaining frames, i.e., only one minute instead of four minutes.
\newpage
\textbf{Experimental data, PCA reconstruction, and difference}
\fig{00_fq_obs_pca-recon_diff_pearson} shows a plot of the experimental, the PCA-reconstructed data, and the difference of the two, for the reduced total scattering structure function for the first frame of the \textit{operando} experiment.
The Pearson correlation coefficient is also displayed.
From the difference curve, it is evident that more noise is filtered from the experimental data at high values of momentum transfer, $Q$.
\newpage
\setcounter{equation}{0}
\setcounter{figure}{0}
\setcounter{table}{0}
\renewcommand{\theequation}{E\arabic{equation}}
\renewcommand{\thefigure}{E\arabic{figure}}
\renewcommand{\thetable}{E\arabic{table}}
\section{\simmap: Pearson correlation analysis for \textit{operando} PDF data}
\label{sec:si_pearson}
From the $r$-dependent Pearson correlation analyses in Figs. \ref{fig:pearson_echem_r=0-10}-\ref{fig:pearson_echem_r=20-30}, similar trends are observed as for the full range in \fig{pearson_echem_r=0-30}.
Comparing the values of all the color scales reveals that higher dissimilarity (lower correlation coefficients) for the intermediate $r$-range from \SI{10}{\angstrom} to \SI{20}{\angstrom} in \fig{pearson_echem_r=10-20}. 
In the low-$r$ range in \fig{pearson_echem_r=0-10}, the PDFs are highly similar, as the phases are made from the same building blocks, i.e., \ch{TiO6}-octahedra.
In the high-$r$ range in \fig{pearson_echem_r=20-30}, the signal-noise-ratio is low that the correlation analysis is less sensitive to structural differences.
\newpage
\subsection*{\normalfont\textbf{From 0 to 10 Å}}
\begin{figure}
	\center
	\includegraphics[width=0.85\columnwidth]{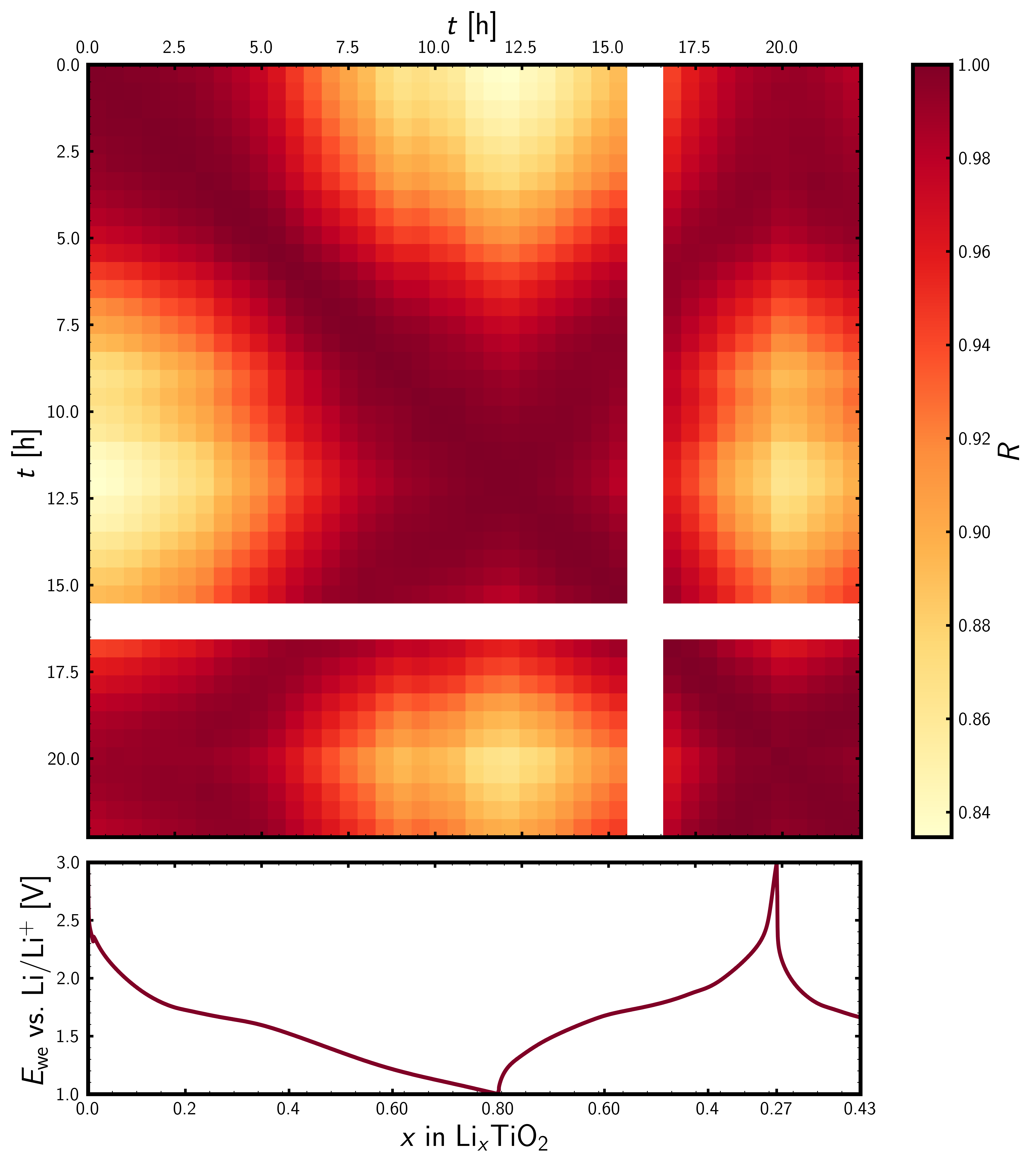}
	\caption{
Top: Pearson cross-correlation matrix for the reduced atomic pair distribution functions, $G(r)$, of the \textit{operando} experiment, with the corresponding time, $t$, on the axes.
The correlation analysis was conducted for the $r$-range from 0 to \SI{10}{\angstrom}.
The value of the Pearson correlation coefficient, $R$, is given by the colorbar to the right.
Bottom: voltage profile with the electrochemical potential of the working electrode, \ewe~Li/\ch{Li+}, as a function of state of charge, $x$, in \ch{Li_{x}TiO2}.
}
	\label{fig:pearson_echem_r=0-10}
\end{figure}

\fig{pearson_echem_r=0-10} shows the result of Pearson correlation analysis for the $r$ range from 0 to \SI{10}{\angstrom}. 
The result is highly similar to \fig{pearson_echem_r=0-30}, only with a little less contrast, i.e., the minimum $R$-value observed is a little higher in \fig{pearson_echem_r=0-10}.
This reflects the high similarity of the PDFs, and therefore the atomic structures, in this $r$-region.
\newpage
\subsection*{\normalfont\textbf{From 10 to 20 Å}}
\begin{figure}
	\center
	\includegraphics[width=0.75\columnwidth]{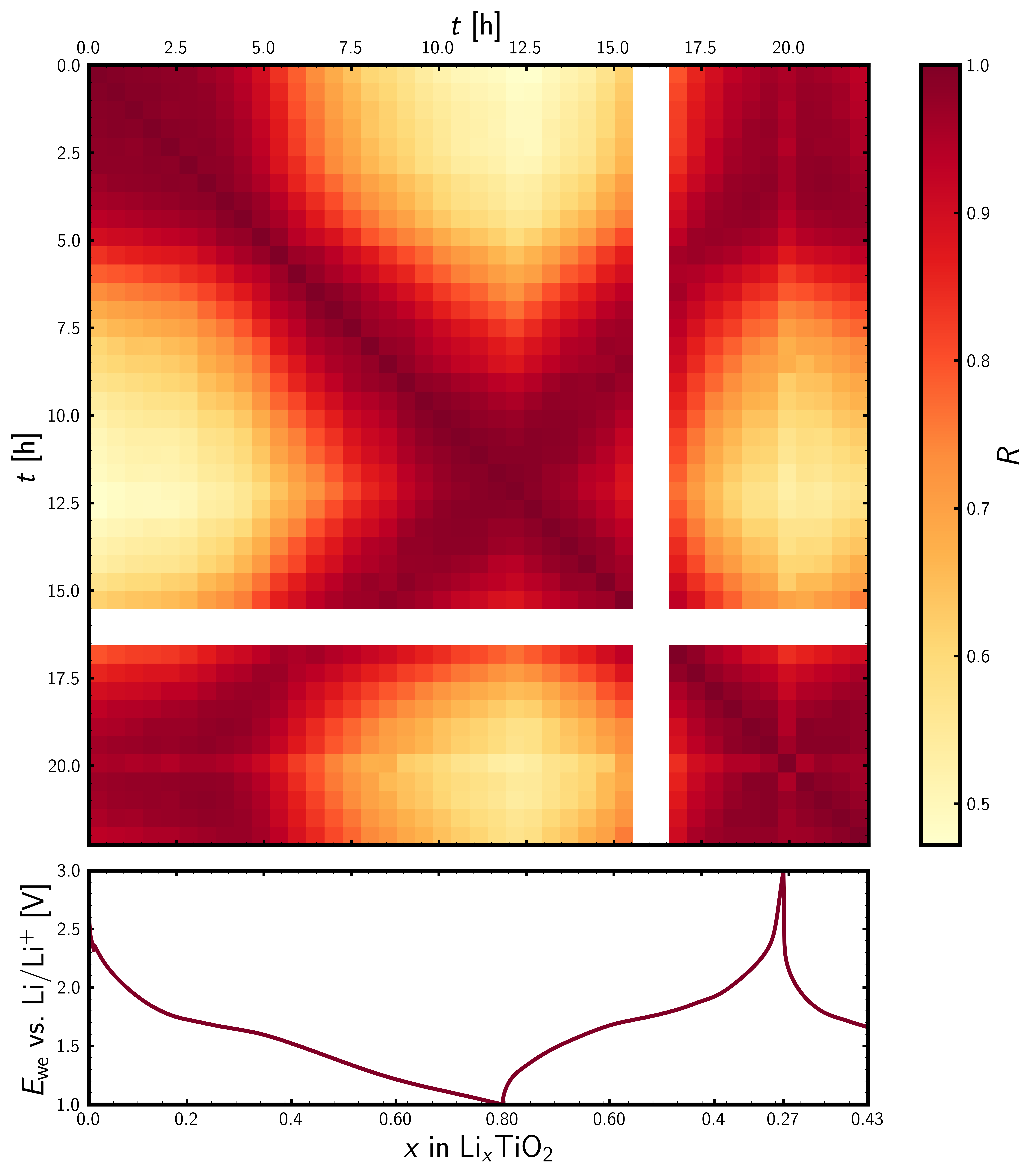}
	\caption{
Top: Pearson cross-correlation matrix for the reduced atomic pair distribution functions, $G(r)$, of the \textit{operando} experiment, with the corresponding time, $t$, on the axes.
The correlation analysis was conducted for the $r$-range from 10 to \SI{20}{\angstrom}.
The value of the Pearson correlation coefficient, $R$, is given by the colorbar to the right.
Bottom: voltage profile with the electrochemical potential of the working electrode, \ewe~Li/\ch{Li+}, as a function of state of charge, $x$, in \ch{Li_{x}TiO2}.
}
	\label{fig:pearson_echem_r=10-20}
\end{figure}

\fig{pearson_echem_r=10-20} shows the result of Pearson correlation analysis for the $r$ range from 10 to \SI{20}{\angstrom}. 
Even though the appearance is highly similar to \fig{pearson_echem_r=0-30}, a significantly higher contrast is observed, i.e., the minimum $R$-value observed is significantly lower in \fig{pearson_echem_r=10-20}.
This reflects the higher dissimilarity of the PDFs, and therefore the atomic structures, for this $r$-range.
This should be exptected, e.g., when comparing to the very local range from 0 to \SI{10}{\angstrom} in \fig{pearson_echem_r=0-10}.
\newpage
\subsection*{\normalfont\textbf{From 20 to 30 Å}}
\begin{figure}
	\center
	\includegraphics[width=0.8\columnwidth]{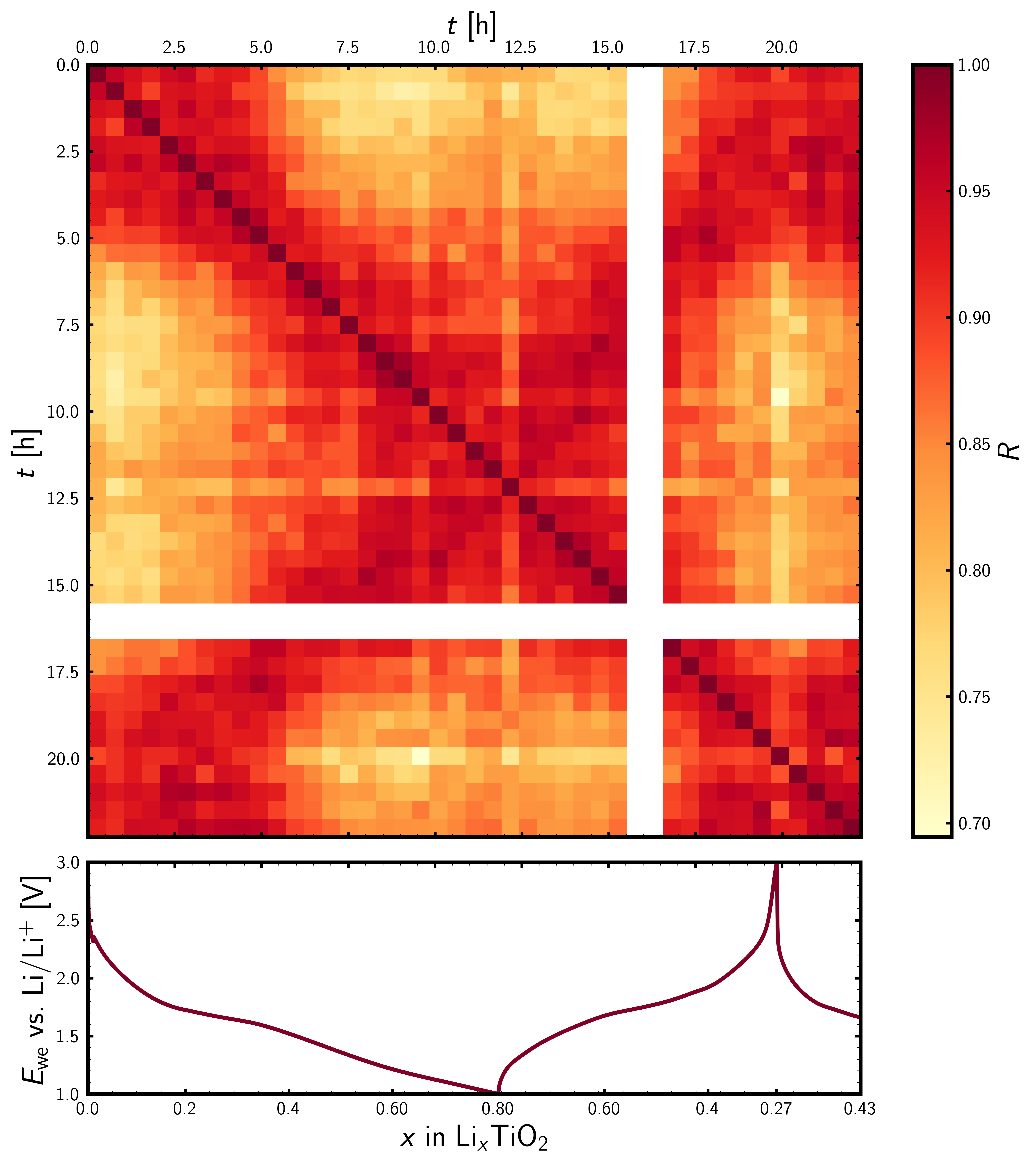}
	\caption{
Top: Pearson cross-correlation matrix for the reduced atomic pair distribution functions, $G(r)$, of the \textit{operando} experiment, with the corresponding time, $t$, on the axes.
The correlation analysis was conducted for the $r$-range from 20 to \SI{30}{\angstrom}.
The value of the Pearson correlation coefficient, $R$, is given by the colorbar to the right.
Bottom: voltage profile with the electrochemical potential of the working electrode, \ewe~Li/\ch{Li+}, as a function of state of charge, $x$, in \ch{Li_{x}TiO2}.
}
	\label{fig:pearson_echem_r=20-30}
\end{figure}

\fig{pearson_echem_r=20-30} shows the result of Pearson correlation analysis for the $r$ range from 20 to \SI{30}{\angstrom}. 
Even though the appearance is sort of similar to \fig{pearson_echem_r=0-30}, the level of noise is much higher.
This reflects the lower signal-to-noise ratio for this $r$-range due to the dampening of the PDFs, together with the increased overlap of correlation peaks with increasing $r$.
\newpage
\setcounter{equation}{0}
\setcounter{figure}{0}
\setcounter{table}{0}
\renewcommand{\theequation}{F\arabic{equation}}
\renewcommand{\thefigure}{F\arabic{figure}}
\renewcommand{\thetable}{F\arabic{table}}
\section{\nmfmap: Non-negative matrix factorization for \textit{operando} data}
\label{sec:si_nmf}
From the various NMF analyses of reciprocal and real space data in \fig{nmf_echem_n=4_gr} and Figs. \ref{fig:nmf_echem_n=2_gr}-\ref{fig:nmf_echem_n=5_fq}, a physical interpretation of the beahvior of the NMF weights is possible up to four components. When using five components, the behavior of the NMF weights cannot be accounted for in a physically meaningful way.
Conducting NMF analysis in reciprocal and real space results in similar behavior of the NMF weights, though the relative size of the weights differ a little. In reciprocal space, the change in the weights is observed to be larger. The interpretation of this is that the phases differ more in reciprocal space than in real space, as should be expected. However, due to the nanosize of the materials, extraction of structural information through modelling must be done in real space through PDF analysis.
\newpage
\subsection*{\normalfont\textbf{Two components: real space}}
\begin{figure}
	\center
	\includegraphics[width=0.9\columnwidth]{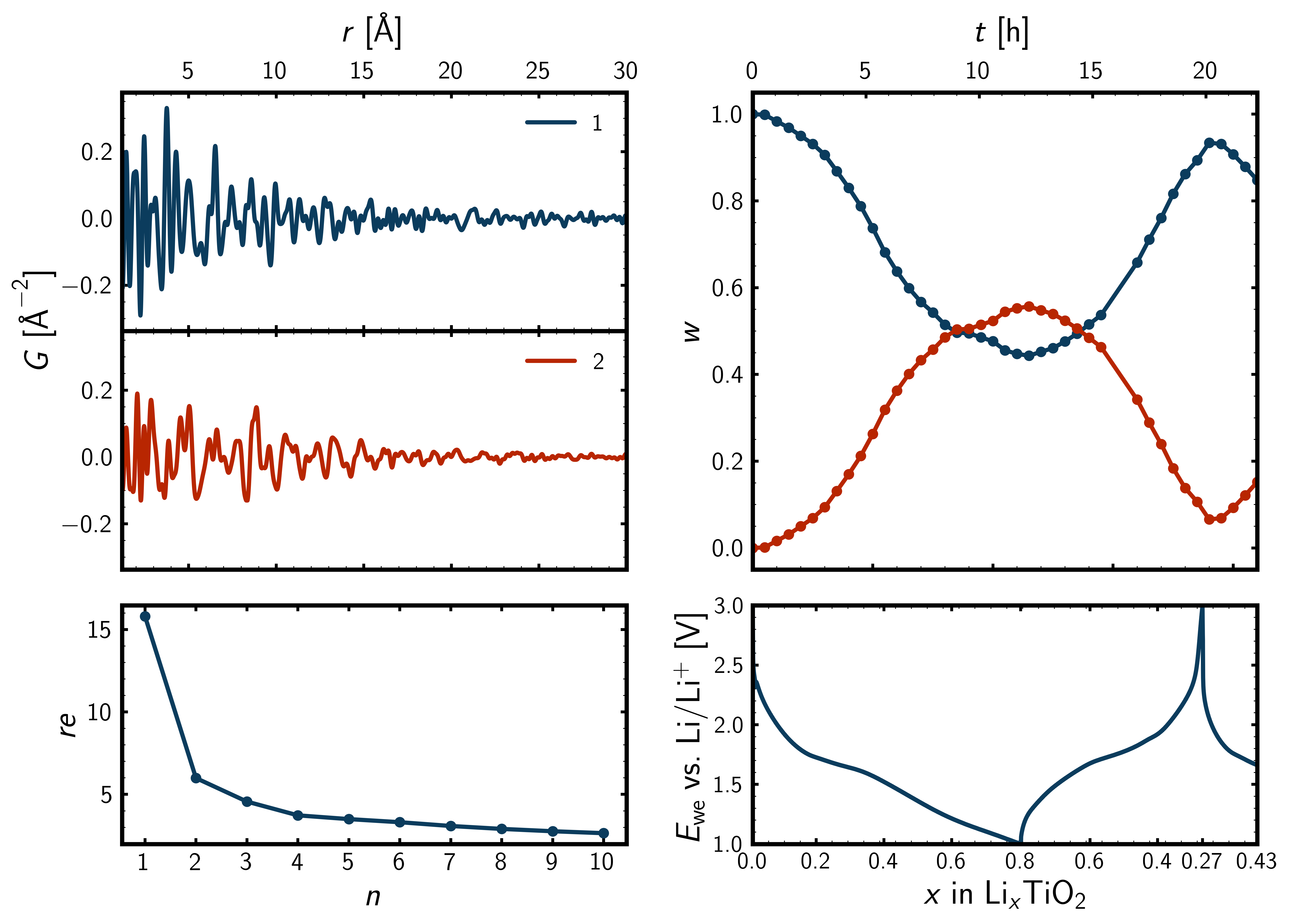}
	\caption{
Output from \nmfmap at \pdfitc when setting the threshold for the number of components to two.
Top left: for each component, the reduced atomic pair distribution function, $G(r)$, is shown.
Bottom left: the reconstruction error, $re$, as a function of the number of components, $n$.
Top right: NMF weights, $w$, as a function of time, $t$, in hours, h.
Bottom right: voltage profile. The electrochemical potential of the working electrode, \ewe~vs. Li/\ch{Li+}, as a function of the state of charge, $x$ in \ch{Li_{x}TiO2}, during the \textit{operando} experiment.  
	}
	\label{fig:nmf_echem_n=2_gr}	
\end{figure}

\fig{nmf_echem_n=2_gr} displays the output from the \nmfmap app using two components for the reduced atomic pair distribution function data, together with the Galvanostatic cycling. 
From the behavior of the weights and the electrochemistry, it can be seen that the two components represent at Li-poor (1, navy) and Li-rich (2, red) states, respectively. 
From the extent of the components, it can be seen that the red components terminates a little earlier, indicating a shorter length of structural coherence for the Li-rich state.
Since only one component is present for the pristine material, the number of components used is too low, as the \textit{ex situ} analysis revealed two phases, \ch{TiO2}-bronze and \ch{TiO2}-anatase.
\newpage
\subsection*{\normalfont\textbf{Two components: reciprocal space}}
\begin{figure}
	\center
	\includegraphics[width=\columnwidth]{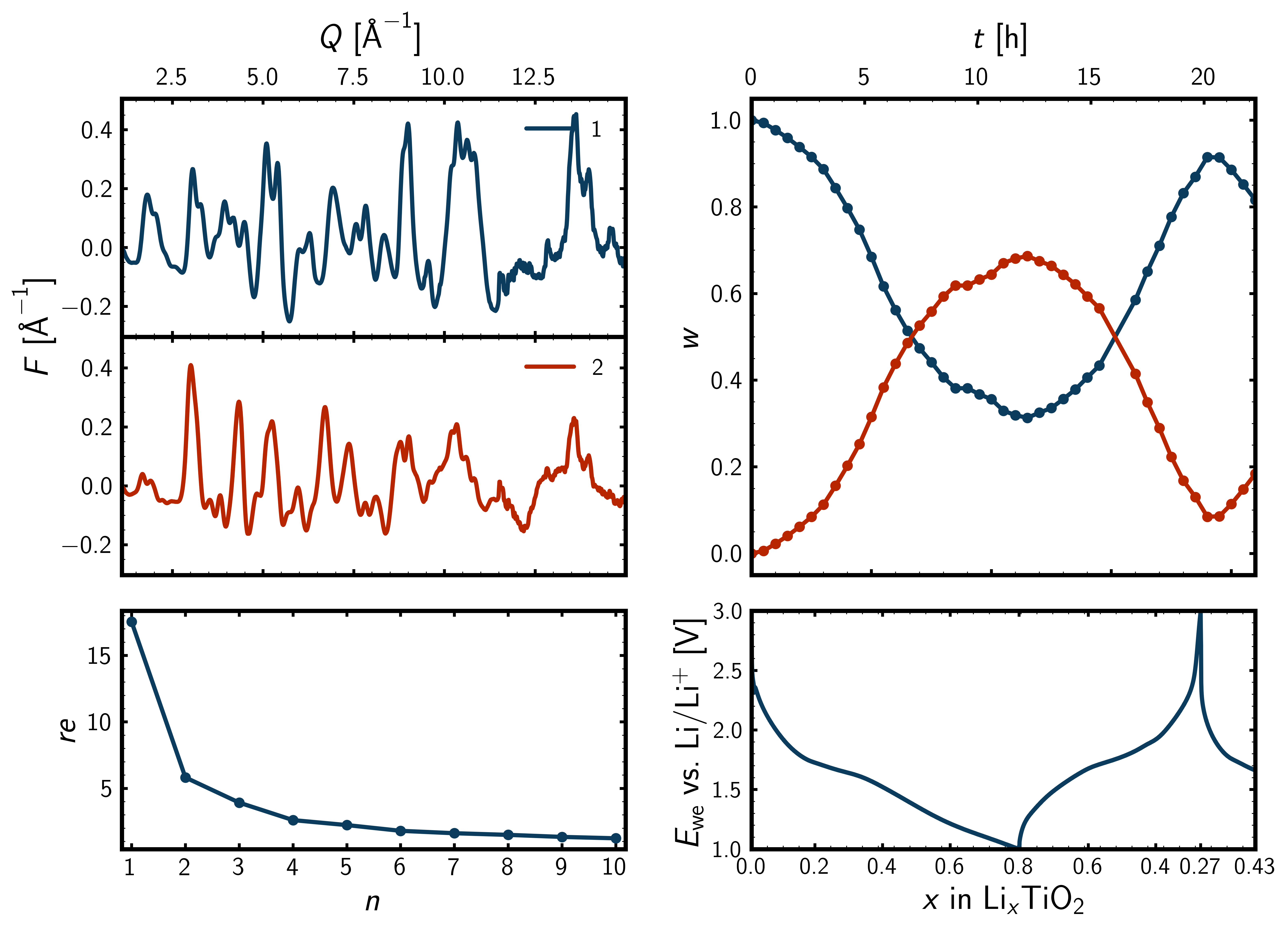}
	\caption{
Output from \nmfmap at \pdfitc when setting the threshold for the number of components to two.
Top left: for each component, the reduced total scattering structure function, $F(Q)$, is shown.
Bottom left: the reconstruction error, $re$, as a function of the number of components, $n$.
Top right: NMF weights, $w$, as a function of time, $t$, in hours, h.
Bottom right: voltage profile. The electrochemical potential of the working electrode, \ewe~vs. Li/\ch{Li+}, as a function of the state of charge, $x$ in \ch{Li_{x}TiO2}, during the \textit{operando} experiment. 
	}
	\label{fig:nmf_echem_n=2_fq}	
\end{figure}

\fig{nmf_echem_n=2_fq} displays the output from the \nmfmap app using two components for the reduced total scattering structure function data, together with the Galvanostatic cycling. As is the case for \fig{nmf_echem_n=2_gr}, the two components represents Li-poor (1, navy) and Li-rich (2, red) states, respectively. The behavior of the weights is comparable to that for the PDF data in \fig{nmf_echem_n=2_gr}, though the changes in weights are a little more pronounced for the $F(Q)$ data here, which might indicate that it is easier to distinguish the two components (phases) in reciprocal space.
\newpage
\subsection*{\normalfont\textbf{Three components: real space}}
\begin{figure}
	\center
	\includegraphics[width=1.0\columnwidth]{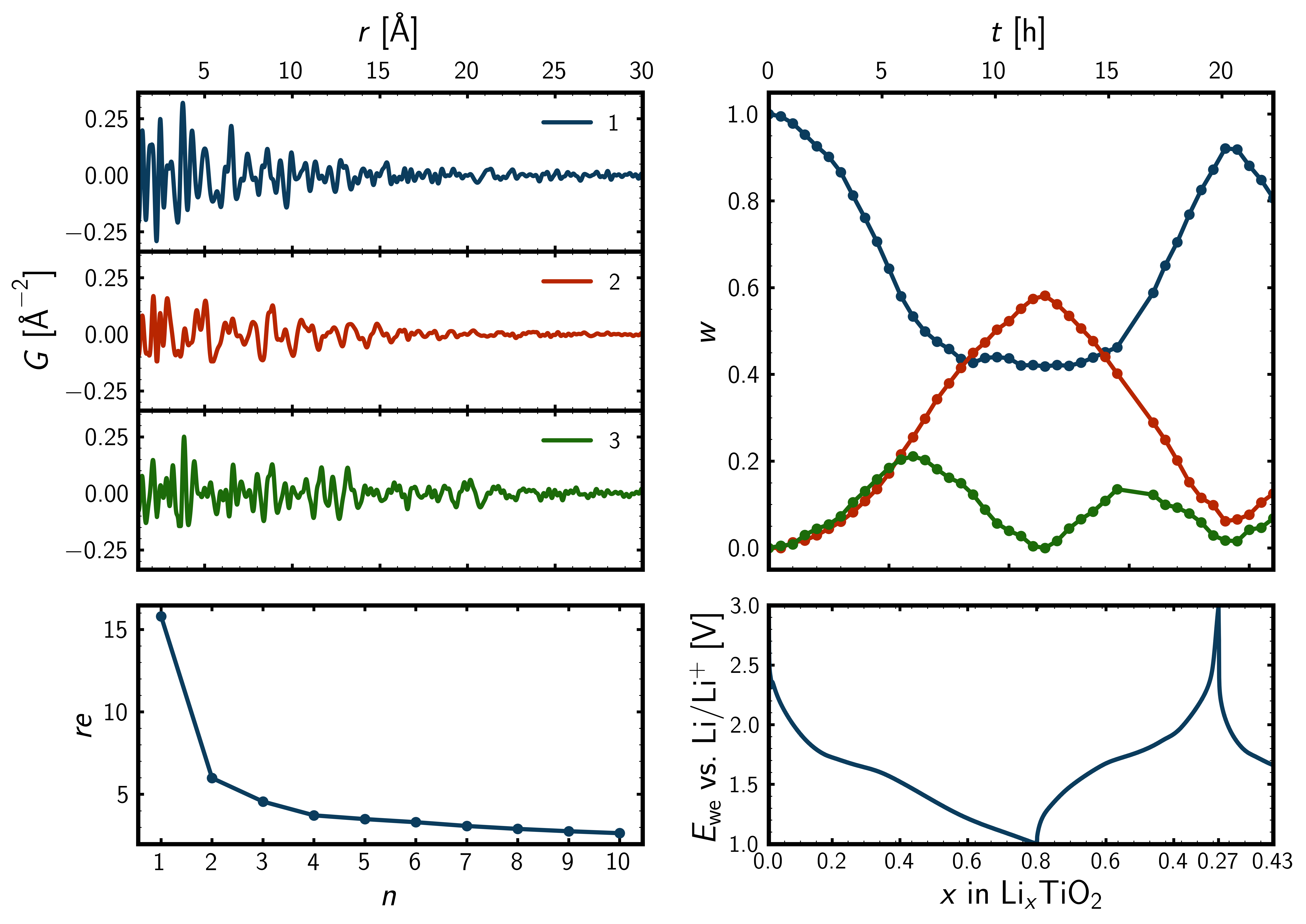}
	\caption{
Output from \nmfmap at \pdfitc when setting the threshold for the number of components to three.
Top left: for each component, the reduced atomic pair distribution function, $G(r)$, is shown.
Bottom left: the reconstruction error, $re$, as a function of the number of components, $n$.
Top right: NMF weights, $w$, as a function of time, $t$, in hours, h.
Bottom right: voltage profile. The electrochemical potential of the working electrode, \ewe~vs. Li/\ch{Li+}, as a function of the state of charge, $x$ in \ch{Li_{x}TiO2}, during the \textit{operando} experiment.  
	}
	\label{fig:nmf_echem_n=3_gr}	
\end{figure}

\fig{nmf_echem_n=3_gr} displays the output from the \nmfmap app using three components for the reduced atomic pair distribution function data, together with the Galvanostatic cycling. In addition to the Li-poor (1, navy) and Li-rich (2, red) components, a component of intermediate degree of lithiation (3, green) appears as an intermediate during both discharge and charge. It is worth to note that the component seems to be absent at complete charged and discharged states and that its evolution appears to be somewhat reversible.
\newpage
\subsection*{\normalfont\textbf{Three components: reciprocal space}}
\begin{figure}
	\center
	\includegraphics[width=\columnwidth]{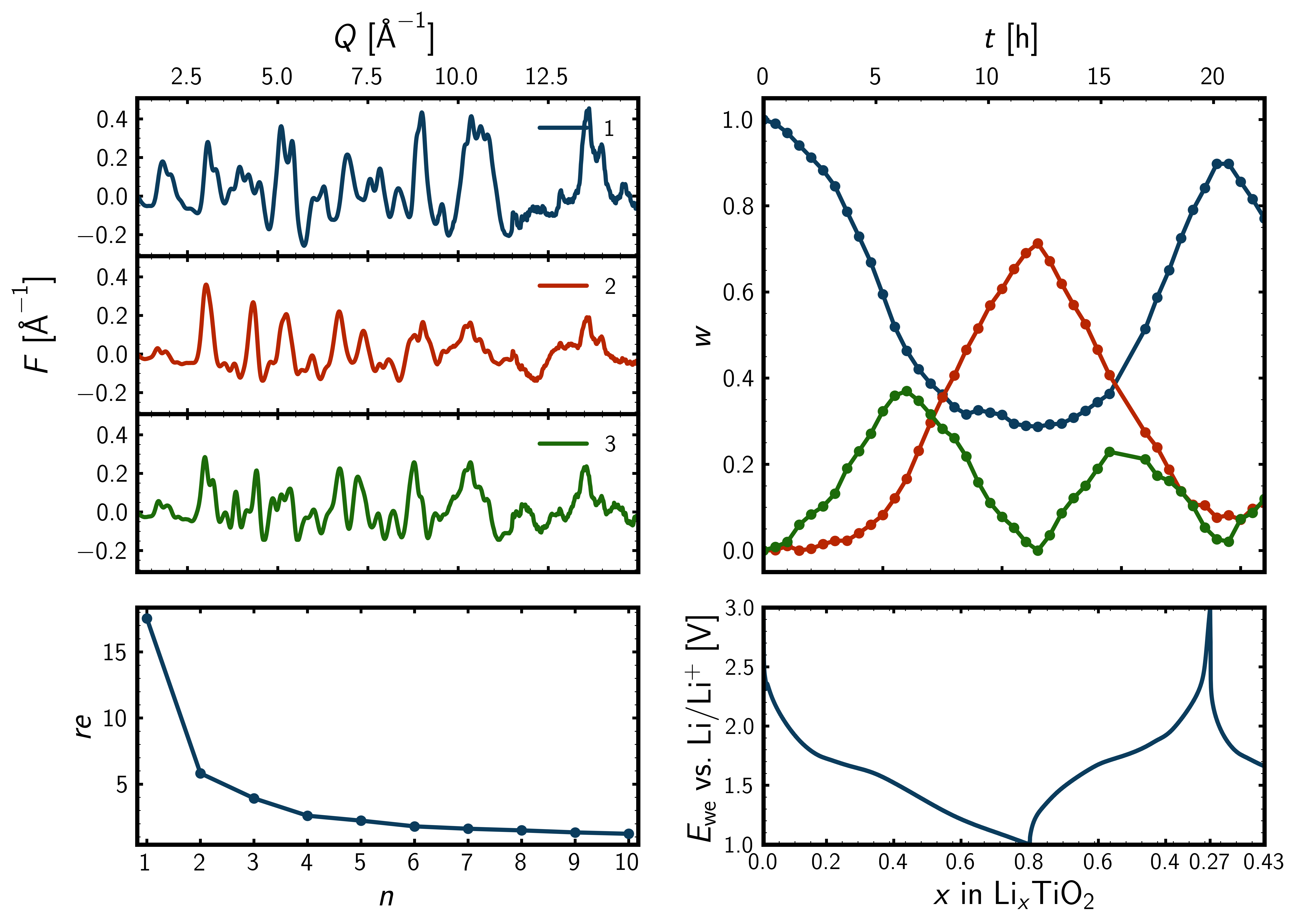}
	\caption{
Output from \nmfmap at \pdfitc when setting the threshold for the number of components to three.
Top left: for each component, the reduced total scattering structure function, $F(Q)$, is shown.
Bottom left: the reconstruction error, $re$, as a function of the number of components, $n$.
Top right: NMF weights, $w$, as a function of time, $t$, in hours, h.
Bottom right: voltage profile. The electrochemical potential of the working electrode, \ewe~vs. Li/\ch{Li+}, as a function of the state of charge, $x$ in \ch{Li_{x}TiO2}, during the \textit{operando} experiment.  
	}
	\label{fig:nmf_echem_n=3_fq}	
\end{figure}

\fig{nmf_echem_n=3_fq} displays the output from the \nmfmap app using three components for the reduced total scattering structure function data, together with the Galvanostatic cycling. The relative behavior of the weights is again comparable to that observe in real space in \fig{nmf_echem_n=3_gr}, however, as observed in \figs{nmf_echem_n=2_gr} and \ref{fig:nmf_echem_n=2_fq}, the changes in the weights are more pronounced for the $F(Q)$ data, indicating that it is easier to distinguish the three components (phases) in reciprocal space.
\newpage
\subsection*{\normalfont\textbf{Four components: reciprocal space}}
\begin{figure}
	\center
	\includegraphics[width=1.0\columnwidth]{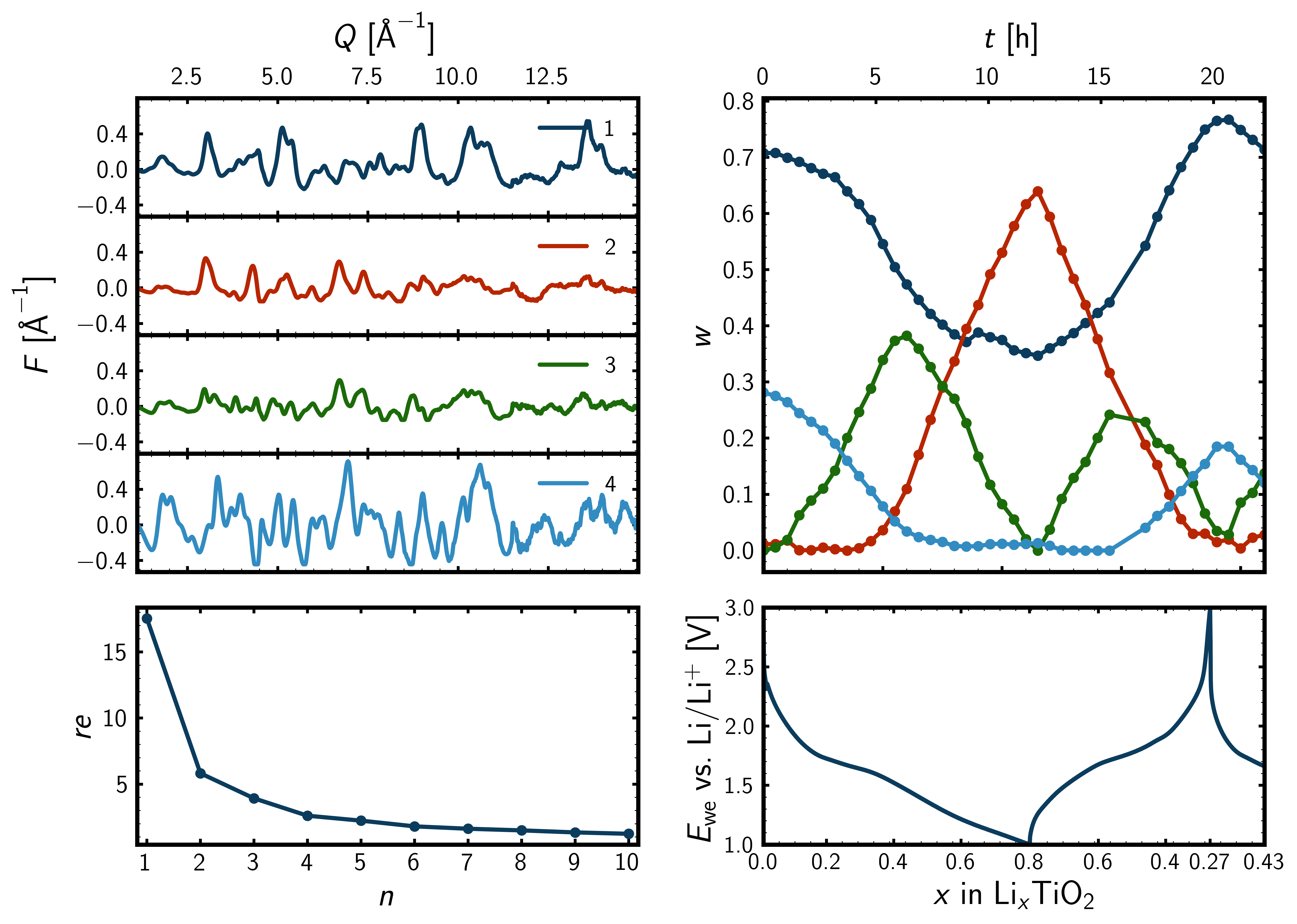}
	\caption{
Output from \nmfmap at \pdfitc when setting the threshold for the number of components to four.
Top left: for each component, the reduced total scattering structure function, $F(Q)$, is shown.
Bottom left: the reconstruction error, $re$, as a function of the number of components, $n$.
Top right: NMF weights, $w$, as a function of time, $t$, in hours, h.
Bottom right: voltage profile. The electrochemical potential of the working electrode, \ewe~vs. Li/\ch{Li+}, as a function of the state of charge, $x$ in \ch{Li_{x}TiO2}, during the \textit{operando} experiment. 
	}
	\label{fig:nmf_echem_n=4_fq}	
\end{figure}

\fig{nmf_echem_n=4_fq} displays the output from the \nmfmap app using four components for the reduced total scattering structure function data, together with the Galvanostatic cycling. Comparing to \fig{nmf_echem_n=4_gr}, the relative beahvior of the weights are similar, though the changes in weights are more pronounced for the $F(Q)$ data here, pointing towards that it is easier to distinguish the four components (phases) in reciprocal space.
\newpage
\subsection*{\normalfont\textbf{Five components: real space}}
\begin{figure}
	\center
	\includegraphics[width=0.9\columnwidth]{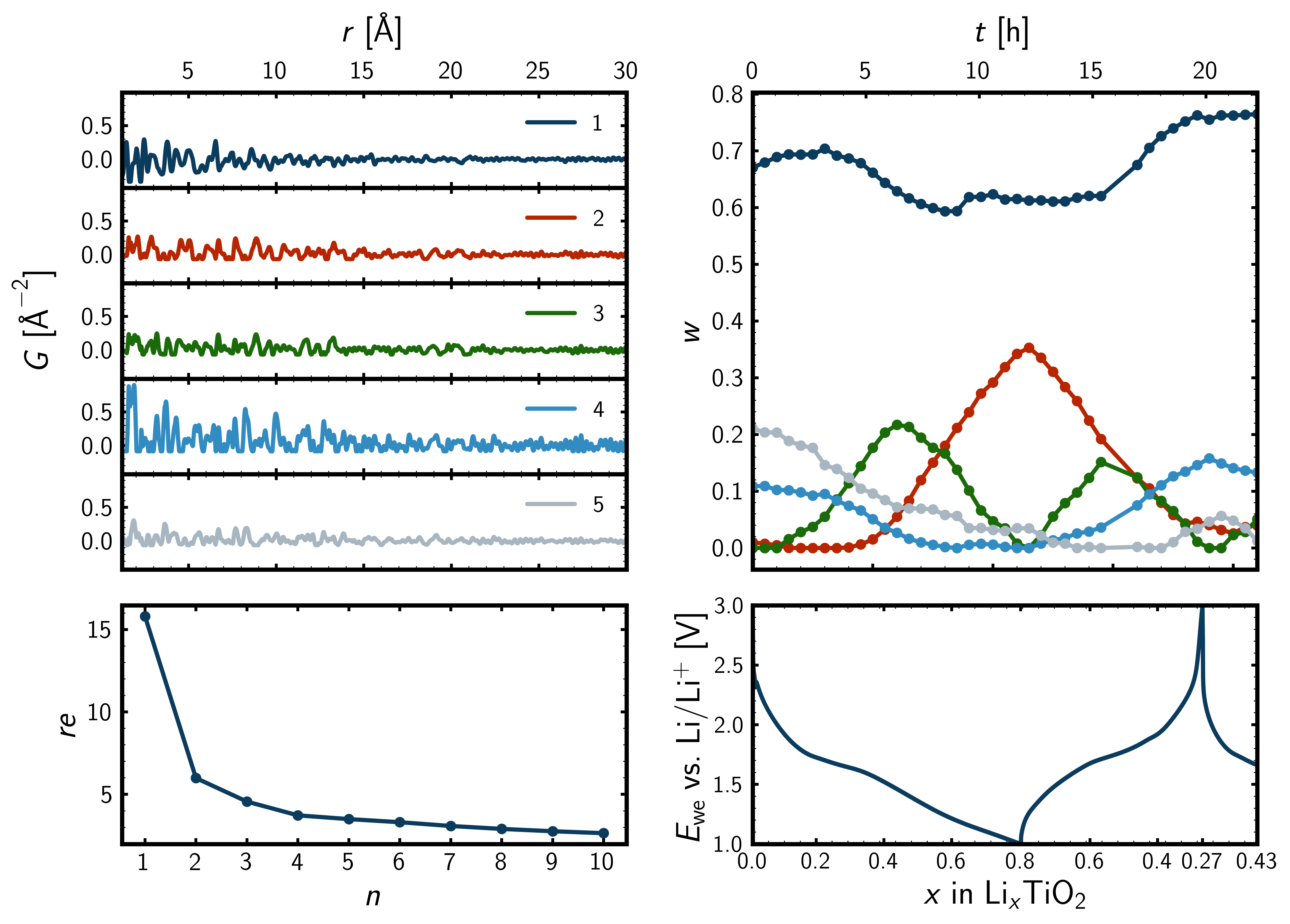}
	\caption{
Output from \nmfmap at \pdfitc when setting the threshold for the number of components to five.
Top left: for each component, the reduced atomic pair distribution function, $G(r)$, is shown.
Bottom left: the reconstruction error, $re$, as a function of the number of components, $n$.
Top right: NMF weights, $w$, as a function of time, $t$, in hours, h.
Bottom right: voltage profile. The electrochemical potential of the working electrode, \ewe~vs. Li/\ch{Li+}, as a function of the state of charge, $x$ in \ch{Li_{x}TiO2}, during the \textit{operando} experiment. 
	}
	\label{fig:nmf_echem_n=5_gr}	
\end{figure}

\fig{nmf_echem_n=5_gr} displays the output from the \nmfmap app using five components for the reduced atomic pair distribution function data, together with the Galvanostatic cycling. From the appearance of the components and the behavior of the weights, it is no longer possible to make sense of the output. Comparing to the weights of the analysis using four components in \fig{nmf_echem_n=4_gr}, it looks like the fifth component in grey accounts for some of the signal that the fourth component in light blue otherwise would do.
Hence, it is not possible to interpret the NMF output in a meaningful way beyond four components, as indicated by the reconstruction error as a function of the number of components in \fig{nmf_echem_n=5_gr} bottom left.
\newpage
\subsection*{\normalfont\textbf{Five components: reciprocal space}}
\begin{figure}
	\center
	\includegraphics[width=0.9\columnwidth]{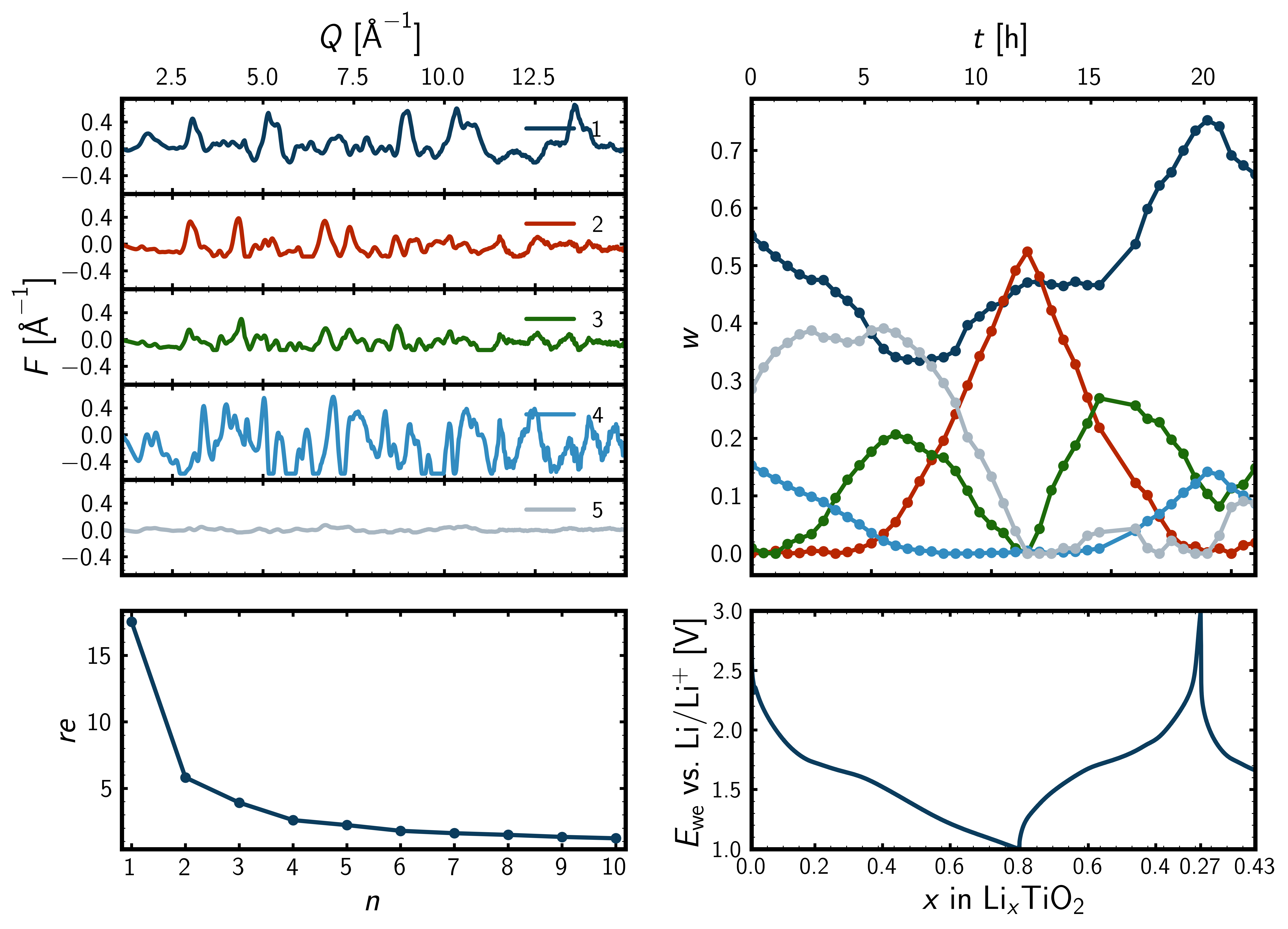}
	\caption{
Output from \nmfmap at \pdfitc when setting the threshold for the number of components to five.
Top left: for each component, the reduced total scattering structure function, $F(Q)$, is shown.
Bottom left: the reconstruction error, $re$, as a function of the number of components, $n$.
Top right: NMF weights, $w$, as a function of time, $t$, in hours, h.
Bottom right: voltage profile. The electrochemical potential of the working electrode, \ewe~vs. Li/\ch{Li+}, as a function of the state of charge, $x$ in \ch{Li_{x}TiO2}, during the \textit{operando} experiment.
	}
	\label{fig:nmf_echem_n=5_fq}	
\end{figure}

\fig{nmf_echem_n=5_fq} displays the output from the \nmfmap app using five components for the reduced total scattering structure function data, together with the Galvanostatic cycling. As is the case in \fig{nmf_echem_n=5_gr}, it is no longer possible to make physical sense of the NMF output. The magnitude of the signal of the fifth component is much smaller than the first four components, indicating that the algorithm might start to include noise into the matrix decompomsition, which is undesirable.
As concluded for the real space data in \fig{nmf_echem_n=5_gr}, it is not possible to interpret the NMF output in a meaningful way beyond four components, as indicated by the reconstruction error as a function of the number of components in \fig{nmf_echem_n=5_fq} bottom left.
\newpage
\setcounter{equation}{0}
\setcounter{figure}{0}
\setcounter{table}{0}
\renewcommand{\theequation}{G\arabic{equation}}
\renewcommand{\thefigure}{G\arabic{figure}}
\renewcommand{\thetable}{G\arabic{table}}
\section{\textit{Operando} PDF modelling}
\subsection*{\normalfont\textbf{First frame}}
\begin{figure}
	\center
	\includegraphics[width=0.95\columnwidth]{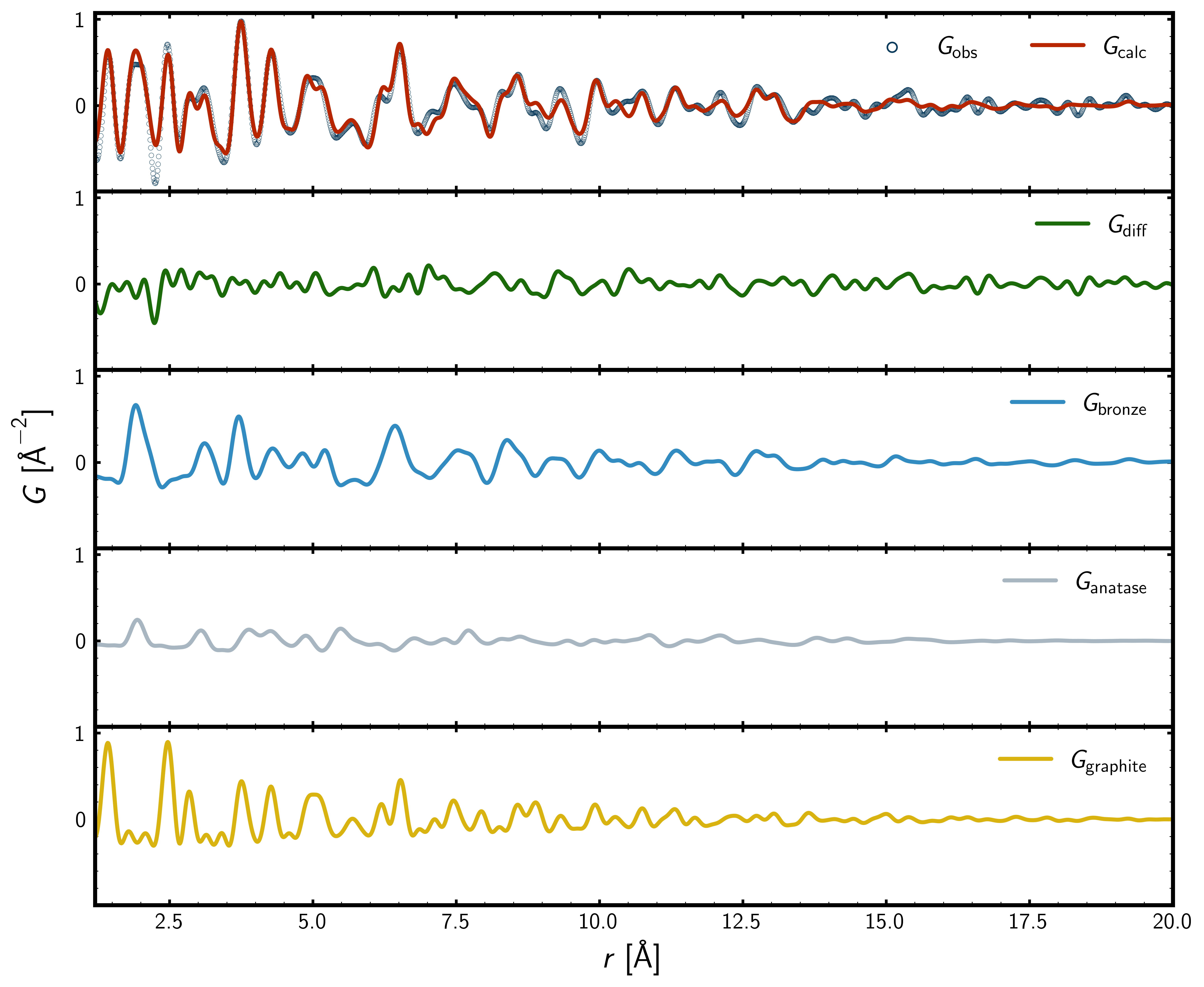}
	\caption{
	PDF fit for first \textit{operando} frame. 
	The observed and calculated PDFs are shown topmost as blue circles and a red line, respectively.
	Below, the difference of the observed and calculated PDFs is shown in green.
	The calculated PDFs of the bronze, anatase, and modified graphite phases are shown in light blue, grey, and yellow, respectively.
	}
	\label{fig:si_pdf_operando_frame1}
\end{figure}

\fig{si_pdf_operando_frame1} displays the PDF fit of the first \textit{operando} frame.
It can be seen that the modified graphite phase accounts for a significant part of the total signal, comparable to that of the bronze phase, whereas the minor anatase phase contributes less.
\newpage
\subsection*{\normalfont\textbf{\ch{TiO2}-bronze unit cell parameters}}
\begin{figure}
	\center
	\includegraphics[width=0.95\columnwidth]{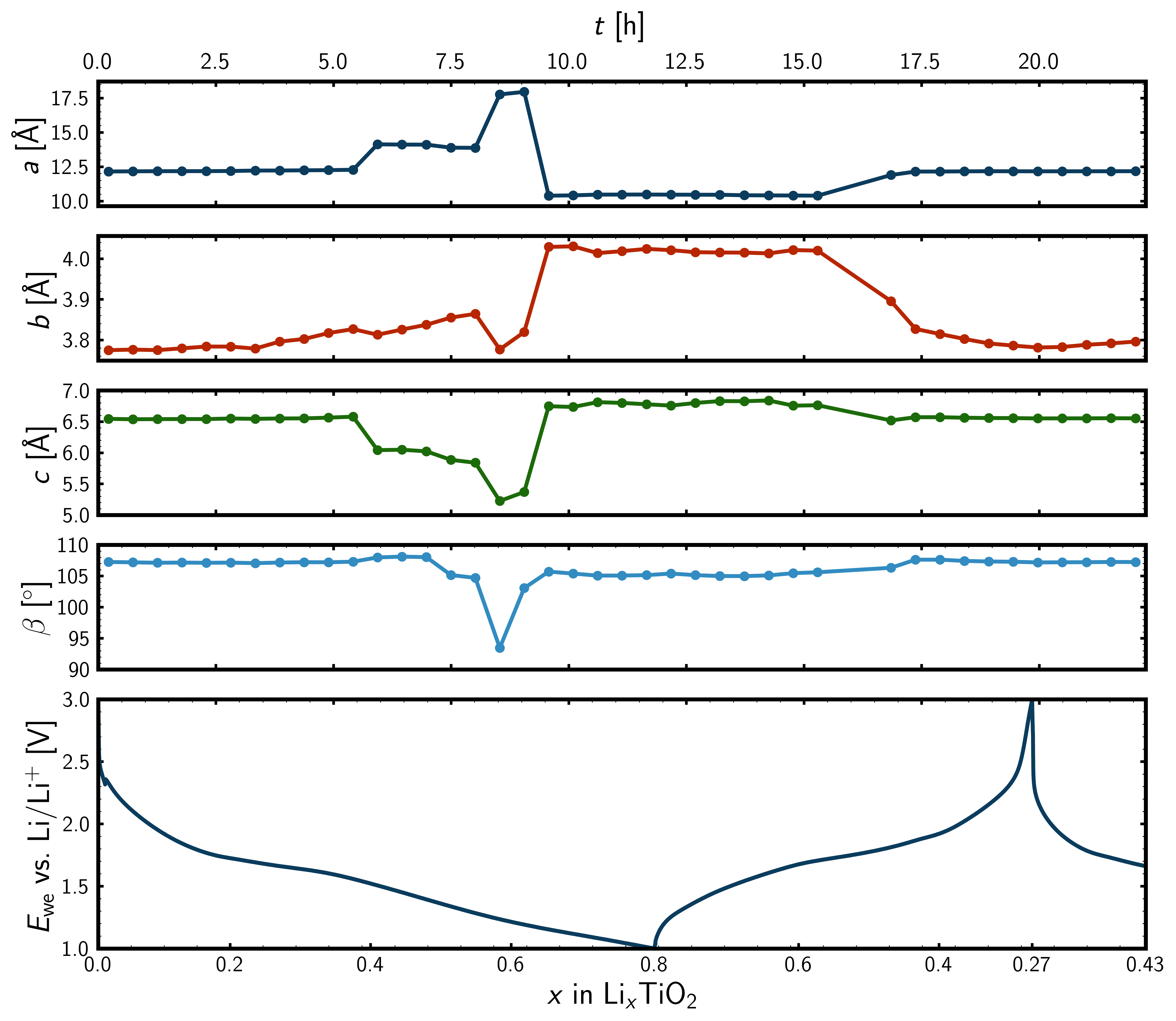}
	\caption{
	Unit cell parameter values for the \ch{TiO2}-bronze phase from PDF modelling of the \textit{operando} data, together with the Galvanostatic cycling.
	}
	\label{fig:bronze_latpars_echem}
\end{figure}

\fig{bronze_latpars_echem} displays the refined lattice parameters for the \ch{TiO2}-bronze phase. 
A discontinuity is observed for the $a$ and $c$ parameters during the initial discharge around $x\approx0.4$, where the \ch{Li_{x}TiO2}-anatase phase is included in the refinement.
Another discontinuity during the initial discharge is observed for all four lattice parameters around $x\approx0.65$, where the \ch{Li_{x}TiO2}-anatase phase is excluded from the refinement again.
In general terms, the $a$-axis appears to shrink for the discharged state compared to the charged states, whereas the $b$- and $c$-axes appear to increase. Only slight changes are observed for the monoclinic angle, $\beta$.
\newpage
\subsection*{\normalfont\textbf{\ch{Li_{x}TiO2}-bronze unit cell parameters}}
\begin{figure}
	\center
	\includegraphics[width=\columnwidth]{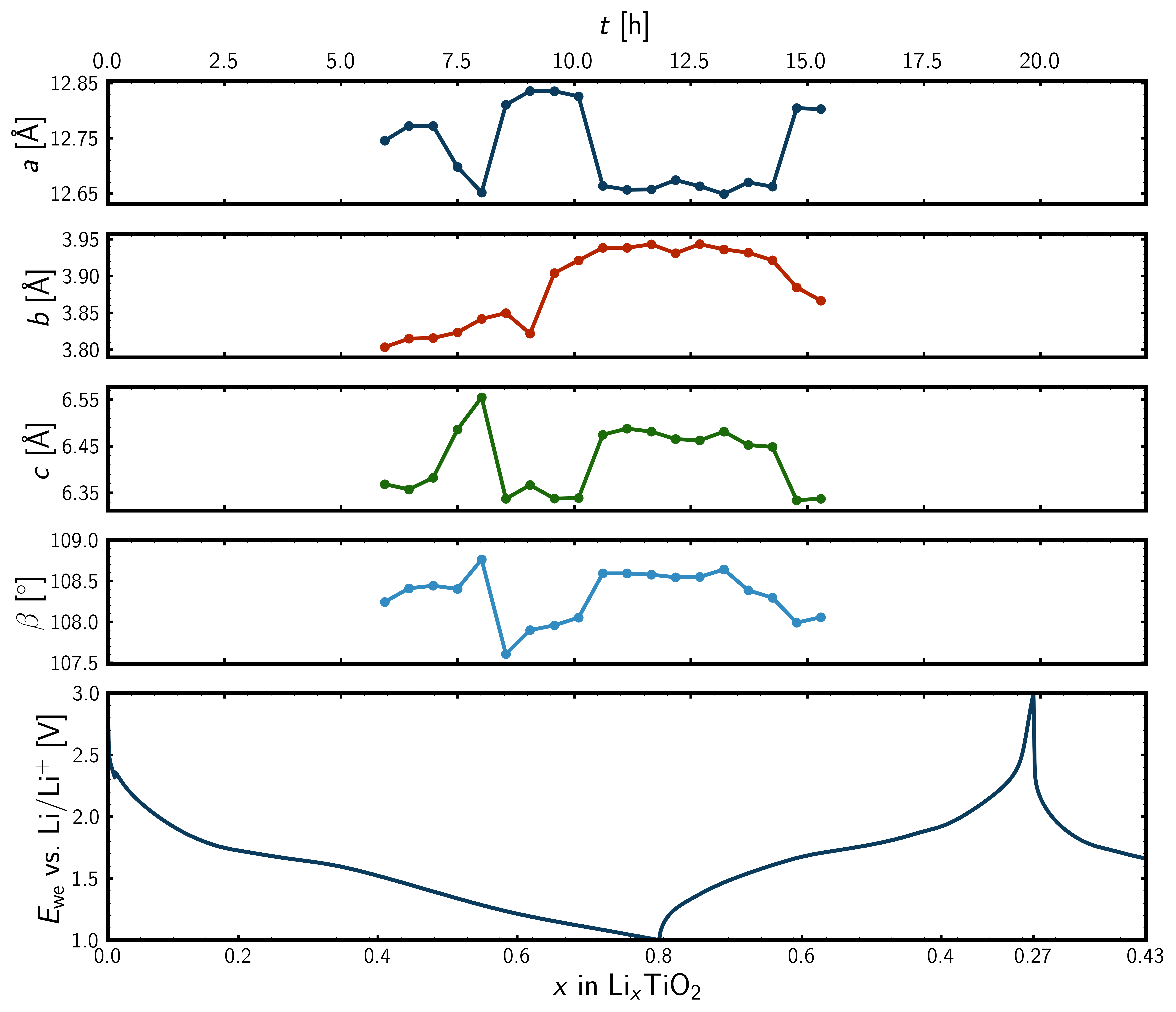}
	\caption{
	Unit cell parameter values for the \ch{Li_{x}TiO2}-bronze phase from PDF modelling of the \textit{operando} data, together with the Galvanostatic cycling.	
	}
	\label{fig:libronze_latpars_echem}
\end{figure}

\fig{libronze_latpars_echem} displays the refined lattice parameters for the \ch{Li_{x}TiO2}-bronze phase. 
The refined values are observed to fluctuate a bit. However, the refined values appear to be stable around deep discharge and the first part of the charge process, where only the \ch{Li_{x}TiO2}-bronze and \ch{TiO2}-bronze phases are included in the refinement.
\newpage
\subsection*{\normalfont\textbf{\ch{Li_{x}TiO2}-anatase unit cell parameters}}
\begin{figure}
	\center
	\includegraphics[width=\columnwidth]{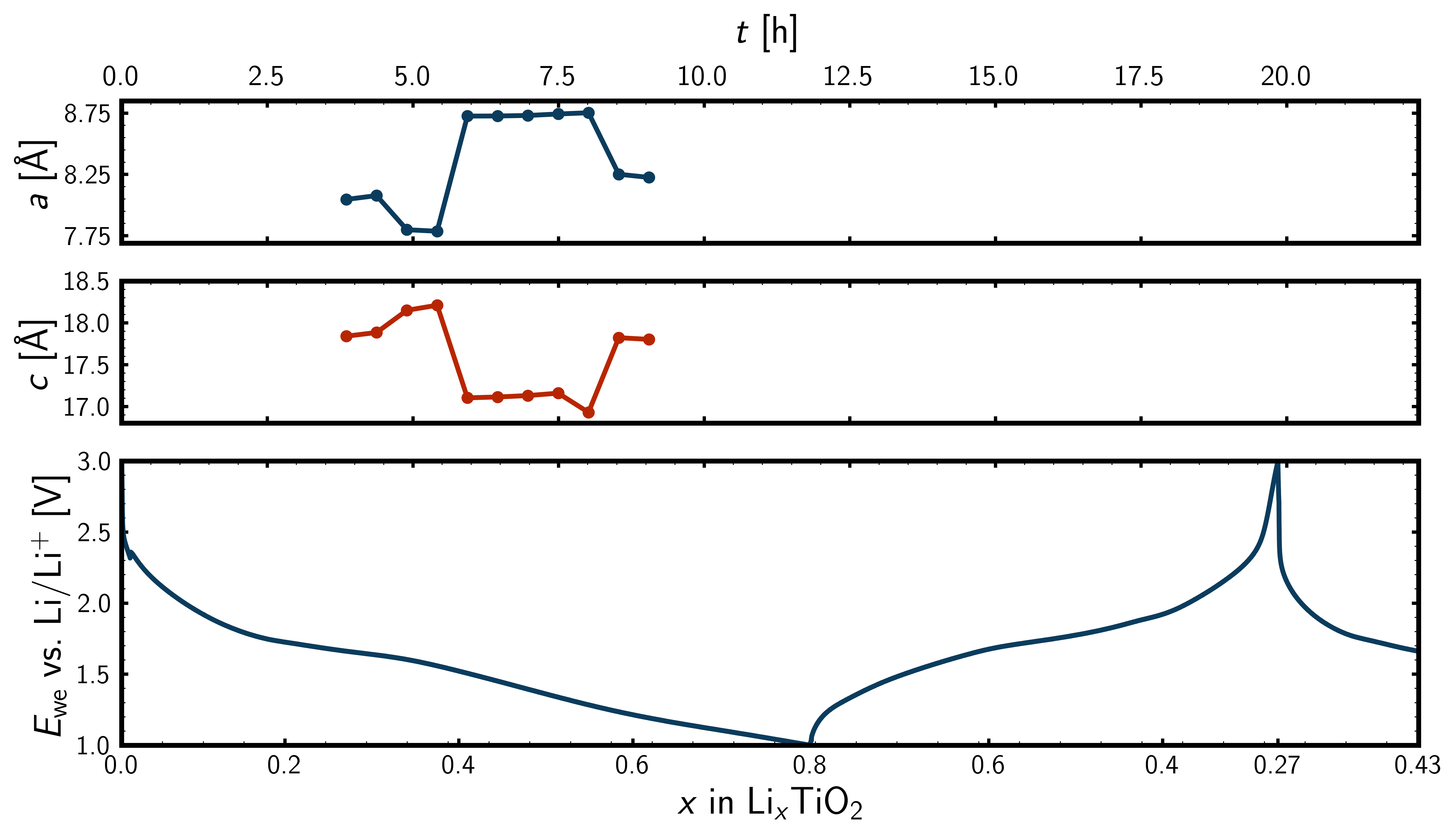}
	\caption{
	Unit cell parameter values for the \ch{Li_{x}TiO2}-anatase phase from PDF modelling of the \textit{operando} data, together with the Galvanostatic cycling.	
	}
	\label{fig:lianatase_latpars_echem}
\end{figure}

\fig{lianatase_latpars_echem} displays the refined lattice parameters for the \ch{Li_{x}TiO2}-anatase phase.
The refined values for the first four frames appear different from the latter. During the refinement of the first four frames, the \ch{TiO2}-anatase is also a part of the refinement. For the last two frames, the phase fraction is low, which results in the sudden jump for the refined values for the last two frames, where the \ch{Li_{x}TiO2}-bronze phase is included in the refinement.
\newpage
\subsection*{\normalfont\textbf{\ch{TiO2}-anatase unit cell parameters}}
\begin{figure}
	\center
	\includegraphics[width=\columnwidth]{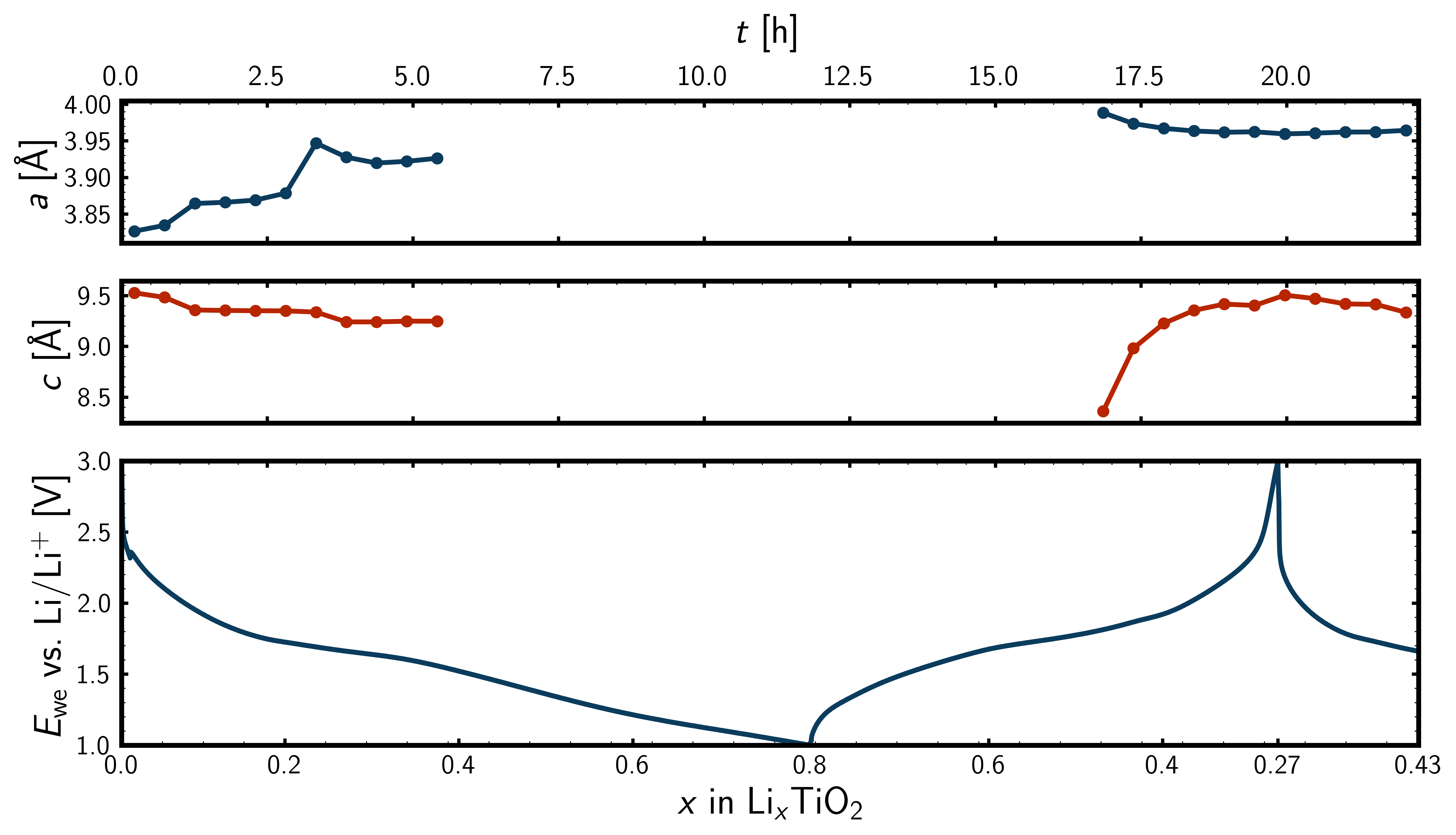}
	\caption{
	Unit cell parameter values for the \ch{TiO2}-anatase phase from PDF modelling of the \textit{operando} data, together with the Galvanostatic cycling.	
	}
	\label{fig:anatase_latpars_echem}
\end{figure}

\fig{anatase_latpars_echem} displays the refined lattice parameters for the \ch{TiO2}-anatase phase.
It is noted that the $a$-axis increases and the $c$-axis decreases monotonically during the initial discharge.
During the last part of the \textit{operando} experiment, the $a$-axis is at the level it ends at during the initial discharged, whereas the $c$-axis is more or less at the same lavel as for the initial discharge, though a decrease it observed for the former part of the charge process when the phase is included in the refinement.
\newpage
\subsection*{\normalfont\textbf{Modified graphite unit cell parameter}}
\begin{figure}
	\center
	\includegraphics[width=\columnwidth]{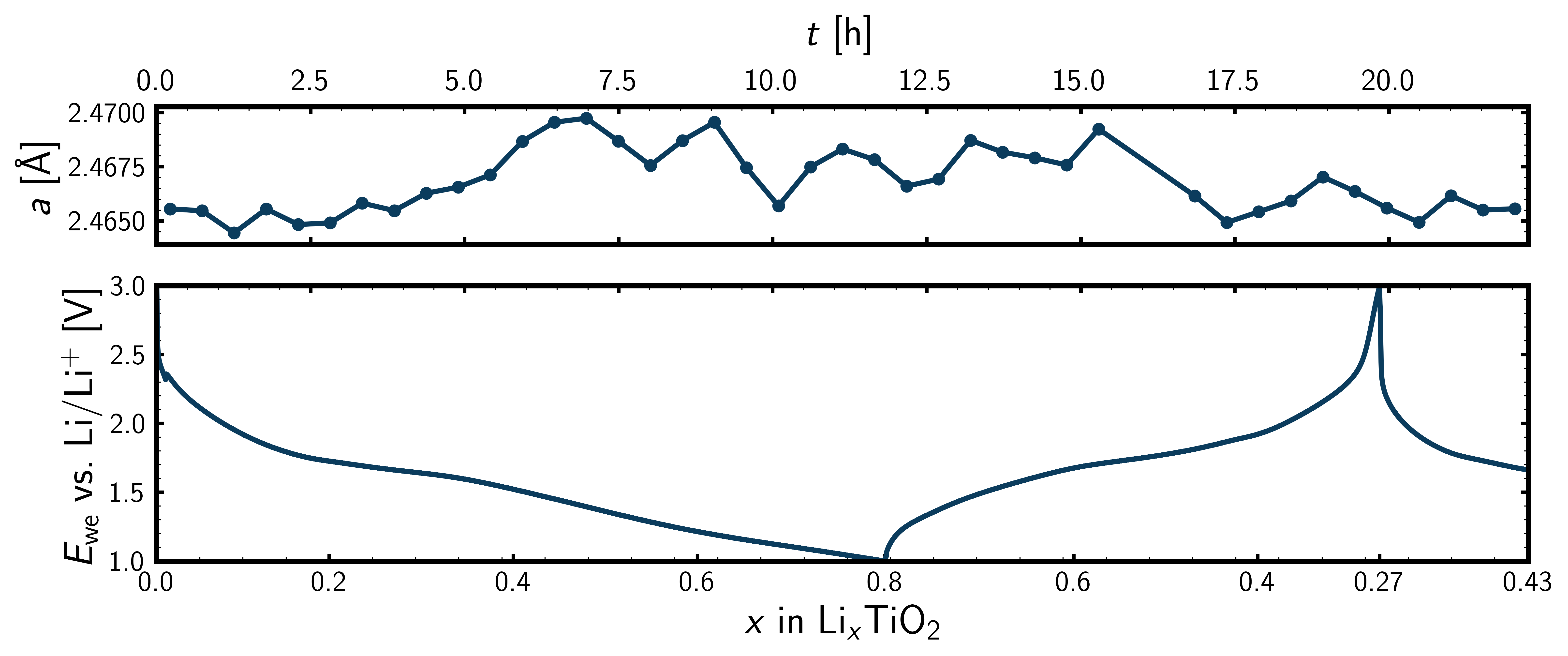}
	\caption{
	Unit cell parameter values for the modified graphite phase from PDF modelling of the \textit{operando} data, together with the Galvanostatic cycling.	
	}
	\label{fig:graphite_latpars_echem}
\end{figure}

\fig{graphite_latpars_echem} displays the refined lattice parameter for the modified graphite phase. As expected, the refined lattice parameter value is almost constant throughout the PDF modelling of the \textit{operando} data.
%

%
\newpage
\ack{Acknowledgements}

We thank the Carlsberg Foundation (grant. no. CF17-0823) and the Novo Nordisk Foundation (grant No. NNF20OC0062068) for supporting this research.
We acknowledge DanScatt for financial support in relation to synchrotron experiments.
We are grateful for access to the facilities and resources at beamline P02.1, PETRA III, DESY, a member of the Helmholtz Association, HGF. We would like to thank Dr. Alexander Schökel for experimental support.
We thank Ghent University for supporting the work conducted at their institution.
Work in the Billinge group was supported as part of GENESIS: A Next Generation Synthesis Center, an Energy Frontier Research Center funded by the U.S. Department of Energy, Office of Science, Basic Energy Sciences under Award Number DE-SC0019212
\newpage
%

%
\newpage
\bibliographystyle{iucr}
\bibliography{tio2_bronze.bib}
\newpage
%






\end{document}